\documentclass[11pt, a4paper]{article}
\usepackage{amsmath, amssymb, amsthm}
\usepackage{graphicx}
\graphicspath{{figures/}}
\usepackage{booktabs}
\usepackage{tabularx}
\usepackage[margin=1in]{geometry}
\usepackage{hyperref}
\hypersetup{bookmarksdepth=1}
\usepackage{bm}
\usepackage{empheq}
\usepackage{float}
\usepackage[section]{placeins}  % auto-FloatBarrier at each \section{}
\usepackage[font=small,labelfont=bf]{subcaption}

\makeatletter
\renewcommand\paragraph{\@startsection{paragraph}{4}{\z@}%
  {3.25ex \@plus1ex \@minus.2ex}%
  {-1em}%
  {\normalfont\normalsize\itshape}}
\makeatother

\newcommand{\TE}{\mathrm{TE}}

\title{\textbf{\Large Are Three Matrices All You Need To Beat the Market? \\
Observable Matrix Dynamics for Portfolio Optimization}}

\author{Igor Halperin\thanks{Opinions expressed here are author's only,
and not of his employer. This paper presents a research exploration, and
is not an investment proposal. All
calculations, numerical analysis, and manuscript
preparation were performed by Claude Code with Opus 4.8
working as an AI assistant under author's supervision. I
would like to thank Miquel Noguer i Alonso, Andres Bagnasco,
Ernest Baver, Eric Berger, Alejandro Rodriguez Dominguez,
Andrey Itkin, Yinsen Miao, and Alexander Vigodner
for very helpful discussions and comments on the manuscript. All remaining errors are my own. All Python
code, analysis scripts, and figures
supporting this paper are available upon request at
\url{https://github.com/ighalp/omd_portfolio}, currently a private repository.}}

\date{\today}

\begin{document}

\emergencystretch=2em

\maketitle

\begin{center}
Email: ighalp@gmail.com
\end{center}

\begin{abstract}
We present a simple framework for dynamic portfolio management that uses
nothing but daily prices, trading volumes, and market capitalizations. Its state
is three fixed-size matrices built from the price history: the
distance matrix of the return correlations and the transition
matrices of two Markov chains that rank the S\&P 500 names monthly by
trailing return and by trailing volatility. These three matrices rest on the
price history alone, the same information Markowitz mean-variance optimization
draws on, but they replace its expected-return vector and covariance matrix. Our method requires no matrix inversion, works on
outlier-robust cross-sectional ranks, and is dynamic rather than
single-period. Empirically the volatility
rank is forecastable one step ahead while the return rank stays close to
unforecastable. A portfolio built on the forecasts, a market-neutral
momentum long-short blended with an opportunistic long-only sleeve, beats
the market on two non-overlapping out-of-sample test sets, January 2022 to
December 2024 and January 2025 to July 2026, at Sharpes of $1.06$ and
$1.32$ against the market's $0.78$ and
$1.14$, respectively, net of a five-basis-point trading cost and marked to
market daily. It also outperforms the classical minimum-variance and
maximum-diversification portfolios. Diversifying the long sleeve by residual
distance adds a further edge on both periods, lifting the Sharpe to $1.08$ and
$1.44$ and the annualized return from $18\%$ to $20\%$ and from $44\%$ to
$56\%$, respectively. A convex information-leader overlay separately
insures the market-neutral sleeve, buying convexity and a shallower drawdown at
a small cost in return, the Sharpe unchanged.
\end{abstract}

\section{Introduction}
\label{sec:intro}

The question whether one can consistently, over a period of years, beat the
market is itself a well-beaten path, in the academic literature and among
practitioners alike. Another well-beaten path, this time predominantly in the AI
literature, is the ``all you need is $\dots$'' motto, which spiked as part of
many AI papers in the wake of the celebrated \emph{Attention Is All You Need} of
Google \cite{vaswani2017}. The present paper carries both terms in its title. We
hope it will prove to be more than an $(N{+}1)$-th entry in either category or in
their intersection. The latter is the easier hope to meet, if only because the
count of papers that carry both themes in a single title is quite sparse, but it
is our hope that the former holds as well. As we set out in detail below, for our
purpose the ``all you need'' part is the claim that a good representation of the
system, the portfolio of $N$ stocks, by three fixed-size matrices is enough, and
that this one representation serves both the statistical description of the
market, as proposed in OMD-Stocks \cite{halperin2026omdstocks}, and
decision-making, that is, trading, in the context of dynamic portfolio
optimization.

The approach is deliberately simple, which is much of its appeal. We use no
predictive signal beyond those built from the price and volume history, keep a
small number of intuitive hyperparameters, and call on no neural network or other
black-box nonlinearity and no long training. Every predictor and every parameter
is motivated in advance and checked for stability, and no advanced numerical
optimization is required beyond standard mathematical software.

Whether such a construction can genuinely beat the market is the old question the
classical theory answers in the negative. The mere existence of the
active-management profession does not on its own support the belief that it is
possible, given survivorship bias and the changing composition of market
participants. The classical theory of finance claims that it is impossible,
because the market is ``efficient'' and current prices already reflect, in a
statistical sense, all valuable information available, so that systematic
overperformance relative to the market itself cannot persist \cite{fama1970}.
This paper takes the question up empirically. We ask whether a systematic
portfolio of S\&P 500 stocks, built from market-based information alone (prices,
sizes, and trading volumes), can outperform the index itself out of sample, and
we find that a disciplined construction does, on data it was never fitted to. We
return in the summary to what this does and does not say about market efficiency,
and to the documented return anomalies our result sits alongside.

Dynamic portfolio optimization amounts to decision-making under uncertainty:
given my current portfolio of $N$ assets (e.g.\ stocks) held today, what
assets should I buy and sell to maximize the performance of my portfolio over
a rolling horizon (e.g.\ a month)? Every portfolio model answers this one
question, and the models differ along three axes: the information set of
predictive signals they draw on, the representation they cast that information
into, and the objective they optimize. Both conceptually and operationally,
dynamic portfolio optimization amounts to methods of stochastic optimal control (SOC), foremost
reinforcement learning (RL) or inverse reinforcement learning (IRL), as
presented for example in \cite{dixon2020ml}. As is well known in
machine-learning research, representation learning plays a critical role in the
performance of RL and SOC models.

For the optimization of financial portfolios, and in the special case of a
one-step optimization, the most classical way of resolving the three choices
formulated above is the Markowitz mean-variance theory \cite{markowitz1952},
which blends all three axes into a single framework. Its input is a vector of expected returns and a
covariance matrix, and its objective trades the portfolio's expected return
against its variance. The covariance half can be estimated from
price history and refined with a factor or a random-matrix model. The
expected-return half is the murky one. Returns are notoriously hard to
forecast, the best alpha models carry very little out-of-sample predictive
power when read as statistical models, and what profit there is comes from a
thin edge spread over many trades rather than from any confident forecast.
Much of the practical response has been to discard the return forecast
altogether: the minimum-variance \cite{clarke2011minvar} and the
most-diversified \cite{choueifaty2008} portfolios keep only the covariance
and optimize the weights on it, trading return estimation for robustness.

The recent OMD-Stocks framework proposed a new, simple yet powerful
representation of the whole dynamics of a market of $N$ stocks by three
fixed-size matrices \cite{halperin2026omdstocks}. The first is the geodesic
distance matrix of the return correlations, the arccos of the correlation,
which measures how far apart the names sit on the correlation manifold. It
is inherited from Observable Matrix Dynamics, the framework OMD-Stocks carries
into finance, which builds dynamic spectral measures of such a distance
matrix to describe learning in neural systems and is rooted in random matrix
theory \cite{halperin2026omd}. The other
two are the transition matrices of two Markov chains that rank each name
monthly by its trailing return and, separately, by its trailing volatility.
That study found the representation simple but rich: from a purely
descriptive, statistical reading of historical equity prices it draws enough
structure to tell the story of how the major market crises unravel.

The present paper develops OMD-Portfolio, which explores this three-matrix
formalism in the context of dynamic portfolio optimization, a problem that,
as explained above, is tightly connected to the prediction problem. We carry
the same three simple matrices of OMD-Stocks into portfolio selection. Like Markowitz mean-variance theory, we draw on the
return distribution only up to its second moments, the mean and the
covariance. But we apply well-motivated non-linear transformations on top of
them: the arccos geometry that turns the correlation matrix into the distance
matrix, and the ranking that turns trailing returns and volatilities into the
states of two Markov chains. The construction uses no predictive signals beyond
market ones built from price and volume data, inverts no covariance matrix, works on
outlier-robust cross-sectional ranks rather than raw returns, and employs a
dynamic distance matrix rather than its single-period snapshot (which is
equivalent to a snapshot correlation matrix).

The three matrices inform one another. The two transition matrices supply the
forecasts and, through their blend, the score that ranks the universe. The
distance matrix enters three times. Its leading eigenvector, centrality, and a
transfer-entropy lead-lag score condition the chains. Its residual form, the
market mode removed, diversifies the long book. An information-leader long-short
from its transfer-entropy graph insures the market-neutral sleeve.

The OMD-Stocks framework treated the two chains as memoryless and unconditional.
Each transition depended only on the current rank, not on anything else known
about the firm. This is a first approximation, and the obvious next step is to
let transitions depend on covariates, the firm characteristics practitioners use
as selection signals. The present paper
develops this extension. We model both ranking chains as Markov chains with
covariates, formulate the conditioning, and show that the discrete construction
keeps the information-theoretic machinery of the unconditional chain intact.

The covariates that enter this conditioning are, in the tradeable strategy,
market indicators constructed from daily prices, trading volumes, and market
capitalizations alone, the same raw data as the state itself. We select the
market capitalization, the market beta, the Amihud illiquidity, the trailing
momentum and short-term reversal, and the realized volatility over several
windows, together with the three signals read from the distance matrix. Both the
state variables, the three matrices, and the indicators that condition them thus
draw on price and volume data alone, with no forward-looking or non-market
signal. We test firms' fundamentals as potential covariates within an
ablation study in Appendix~\ref{sec:fundsection}, which finds over half a century
that the informative ones carry a value-and-quality signal leading the return
rank, but add only a small tilt to a directional book and nothing to the
market-neutral core, in line with the practitioner view that backward-looking
accounting ratios carry little tradeable signal. The tradeable book therefore
uses market data alone, and
that is the sense in which the paper invests with three matrices and nothing
else.

Two design choices make the framework uniform and estimable. The first treats
the target and the covariates on the same footing: each characteristic is
ranked across the firms present on a date and bucketed into ten equal groups,
so a continuous signal of arbitrary scale becomes an ordinal in $\{1,\dots,10\}$,
insensitive to outliers and to any monotone reparameterization, and a
heterogeneous library of firm characteristics reduces to a common discrete
representation. The second parameterizes the conditioning by a log-linear
conditional-logit model that nests the unconditional chain at zero coefficients,
so adding covariates is a strict refinement measured against the OMD-Stocks
baseline.

The discreteness pays off again in the diagnostics. We show that
conditioning the chain on a covariate can only increase its entropy
production, a consequence of the convexity of the entropy-production
functional, so the gap between the conditioned and the pooled entropy
production measures the directional structure, the arrow of time, that
the covariate resolves and that aggregation over firms had cancelled.
We also form the transfer entropy between a characteristic's bucket
process and the rank process. Because both series are already discrete,
this is an exact sum, and its asymmetry gives a directional reading of
whether a characteristic drives the rank or merely follows it. The
conditioned chains then supply a selection rule. Ranking securities by
their characteristic-conditioned forward-rank forecast, in the return
chain for expected performance and in the volatility chain for expected
risk, is a rank-based strategy in the spirit of stochastic portfolio
theory \cite{fernholz2002,banner2005}, now informed by the full
covariate state.

The empirical results follow this logic and separate the two chains. Among the
predictors, conditioning on size, beta, illiquidity, or volatility resolves
several times the pooled entropy production, and the transfer entropy marks the
risk descriptors as leaders of the ranking while momentum only follows it.
Under forecast the chains diverge: the volatility ranking is predictable out of
sample and carries memory beyond a single step, while the return ranking stays
close to unforecastable. Turned into a portfolio, a market-neutral momentum long-short blended
adaptively with a long-only sleeve beats the market out of sample, and beats the
classical minimum-variance and maximum-diversification portfolios. A forward
test over a clean, non-overlapping window from January 2025 to July 2026, which
the study never saw, confirms it at a daily-marked Sharpe of $1.32$ against the
market's $1.14$ and roughly double its return. Diversifying the long sleeve by residual distance sharpens
the book further. A convex information-leader overlay adds crash insurance, and
neither conviction nor fragility sizing improves the risk-adjusted result. A
second ablation finds that the spectral early-warning signals of OMD-Stocks and
OMD do not time the regime weight better than the sign of the recent market
return. Trading volume, read through the same chains, has an arrow of time far
stronger than the price ranks, and as share turnover it forecasts the next day's
return ranking out of sample, though at a horizon too fast to trade net of cost.

The paper is organized as follows. Section~\ref{sec:chains} builds the three
matrices and the covariate-conditioned ranking chains, and
Section~\ref{sec:analysis} estimates and diagnoses them. Section~\ref{sec:portfolio}
constructs the portfolio, calibrates it, and reports its results in sample over
2018--2021 and out of sample over 2022--2024, and Section~\ref{sec:freshoos}
carries the analysis to a clean, non-overlapping test set spanning January 2025
to July 2026. Section~\ref{sec:summary} concludes. The appendices collect the
ablations and the signal studies: Appendix~\ref{sec:fundsection} the fundamentals
over half a century, Appendix~\ref{sec:spectralablation} the spectral
early-warning signals, and Appendix~\ref{sec:signalstudies} the
transfer-entropy market leaders, the alternative early-warning predictors, the
regime-aware signals, and the information in trading volume.

\section{OMD-Portfolio model}
\label{sec:chains}

\subsection{Three matrices of OMD-Portfolio}
\label{sec:threematrices}

The OMD-Stocks framework represents the whole dynamics of a market of $N$
securities by three fixed-size matrices read from the price record alone
\cite{halperin2026omdstocks}. The first is the arccos geodesic distance matrix
of the return correlations, a snapshot of the cross-sectional geometry at each
date. The other two are the transition matrices of two Markov chains that rank
the names by trailing return and by trailing volatility, the dynamics of the
performance order and the risk order. We recall the construction of all three,
since the portfolio is built on them, and then extend the two chains to depend
on covariates, which is where this paper departs from OMD-Stocks.

Let there be $N$ securities indexed by $i$, observed on daily dates $t$. Over a
trailing window of $L$ observations, about a year, the correlation matrix
$C(t)$ is the Pearson correlation of the daily returns, which is the
cosine-similarity Gram of the demeaned, unit-normalized return vectors,
$C_{ij}(t)=\cos\theta_{ij}(t)$ with unit diagonal. The first matrix is the
element-wise arccos of that Gram,
\begin{equation}
\label{eq:mmatrix}
M_{ij}(t) = \arccos G_{ij}(t), \qquad
G_{ij}(t) = \frac{C_{ij}(t)}{\sqrt{C_{ii}(t)\,C_{jj}(t)}},
\end{equation}
the geodesic angle between the return directions on the unit sphere, with
$M_{ij}\in[0,\pi]$. The diagonal vanishes automatically, since $G_{ii}=1$ gives
$M_{ii}=\arccos 1 = 0$. For a correlation input the Gram is already $G=C$, so
$M=\arccos C$, and the renormalization matters only when the input is not itself
a correlation, as for the market-removed residual of Section~\ref{sec:portfolio}.
Because $\arccos$ decreases monotonically on $[-1,1]$, the distance is a
decreasing function of the correlation, equal to $0$ at perfect correlation,
$\pi/2$ at zero correlation, and $\pi$ at perfect anticorrelation. The more
strongly two names co-move, the closer they sit; a strong negative correlation,
by contrast, places them far apart near $\pi$. Recasting the correlations as a
geodesic distance matrix, rather than keeping the correlation matrix $C(t)$, is
more than a change of variable. As stressed in the OMD-Stocks framework
\cite{halperin2026omdstocks}, this reframing places the cross section within the
mathematical theory of random distance matrices
\cite{halperin2026ibbs,bogomolny2003,bogomolny2007}, and lets the spectral
analysis of $M(t)$ serve as a descriptive statistic of learning in
machine-learning models \cite{halperin2026omd} and of relaxation in physical
systems \cite{halperin2026fdm}. Because the S\&P 500 constituent list changes over a long historical period, the
OMD-Stocks method works with a variable number $N$ of stocks, held fixed for a
given lookback window of length $L$. As found in \cite{halperin2026omdstocks},
for different aspects of the analysis it makes sense to explore $L$ ranging from
six months to two years, and for each such choice we take values of $N < 500$
from the condition that the ratio $N/L$ is fixed and slightly above one, while
keeping both $N$ and $L$ above one hundred. The fixed ratio ensures that all our
measurements have similar levels of statistical noise, whose impact can be
estimated by tools from random matrix theory. The distance matrix enters
the portfolio in Section~\ref{sec:mmatrix}, through its leading eigenvector, its
row-mean centrality, and the directed transfer-entropy network read from it.

The other two matrices are the ranking chains, one for relative performance and
one for relative risk. Two statistics order each security against its peers. The
return statistic is the rolling mean return over a window $\tau_R$,
\begin{equation}
\label{eq:rbar}
\bar r_i(t) = \frac{1}{\tau_R}\sum_{s=t-\tau_R+1}^{t} r_i(s),
\end{equation}
a performance ordering, and the volatility statistic $\bar v_i(t)$ is the
rolling realized volatility over a window $\tau_V$, a risk ordering. We orient
the two rankings so that rank one is the most desirable name in each, the
highest mean return in the return chain and the lowest realized volatility in
the volatility chain. The two therefore run in opposite directions along their
statistics, and rank $N$ is the least desirable in each, the worst performer or
the most volatile. Both orderings are market-neutral by construction, since a common move
added to every return, or a common scaling of every volatility, leaves the order
unchanged, so each ranking discards the market factor and keeps the relative,
cross-sectional dynamics. For either statistic let
$\mathrm{rk}_i(t)\in\{1,\dots,N\}$ be the rank of security $i$ on date $t$,
coarse-grained into $K$ equal deciles,
\begin{equation}
\label{eq:class}
c_i(t) = \Big\lceil K\,\frac{\mathrm{rk}_i(t)}{N} \Big\rceil
\in \{1,\dots,K\}, \qquad K=10 ,
\end{equation}
which applied to the two statistics gives the class processes $a^R_i(t)$ and
$a^V_i(t)$, the decile of a security in the performance ranking and in the risk
ranking.

Each ordering's one-step dynamics over the horizon $\Delta$ are carried by a
time-dependent transition matrix,
\begin{equation}
\label{eq:PX}
P^X_{ab}(t) = \mathrm{Prob}\big(a^X_i(t+\Delta)=b \,\big|\, a^X_i(t)=a\big),
\qquad X\in\{R,V\},
\end{equation}
the return chain $P^R$ for relative performance and the volatility chain $P^V$
for relative risk, each a fixed-size $K\times K$ observable. We estimate it by
pooling the decile transitions of all names over a window, writing the pooled
joint distribution of consecutive classes as $\mu^X$,
\begin{equation}
\label{eq:P}
\mu^X_{ab} = \frac{1}{Z}\sum_{i,t}
\mathbf 1\{a^X_i(t)=a\}\,\mathbf 1\{a^X_i(t+\Delta)=b\}, \qquad
P^X_{ab} = \frac{\mu^X_{ab}}{\sum_{b'}\mu^X_{ab'}},
\end{equation}
with $Z$ the number of pooled transitions and a small Laplace pseudocount added
so no reverse cell vanishes. Because each rank is occupied by exactly one name on
every date, the class marginal is uniform, exactly $N/K$ securities in each
decile, so it is identical before and after a step and the content lies entirely
in the transition structure. This uniform marginal makes the entropy production
of each chain the pure Kullback--Leibler divergence between the forward joint and
its time reverse,
\begin{equation}
\label{eq:ep}
\sigma^X = \sum_{a,b} \mu^X_{ab}\,\log\frac{\mu^X_{ab}}{\mu^X_{ba}} \; \ge 0,
\qquad X\in\{R,V\},
\end{equation}
which vanishes if and only if $\mu^X$ is symmetric, that is when the rank
dynamics satisfy detailed balance. Where a statement holds for either chain we
drop the superscript and write $P_{ab}$, $\mu_{ab}$, and $\sigma$. The
unconditional chains capture the persistence of leaders and laggards and the
irreversibility of the relative order, but they cannot express that a security's
odds of climbing depend on what else is known about it. We now add that
dependence.

\subsection{Market-neutral and regime-aware chains}
\label{sec:regimeaware}

The rankings of Section~\ref{sec:threematrices} are market-neutral by design, and
that is right for security selection, where the question is which names will
outperform their peers and the common market move is a nuisance to be removed. For
other questions, the state of the market itself, the timing of risk, and the
regime a portfolio faces, the market move is the signal rather than the nuisance,
and a market-neutral chain cannot see it. Its cross section fills the ten deciles
evenly on every date, so a market-wide rally or crash leaves the class
distribution unchanged. We therefore build a second family of chains that keeps
the market move.

\begin{figure}[H]
\centering
\includegraphics[width=0.9\textwidth]{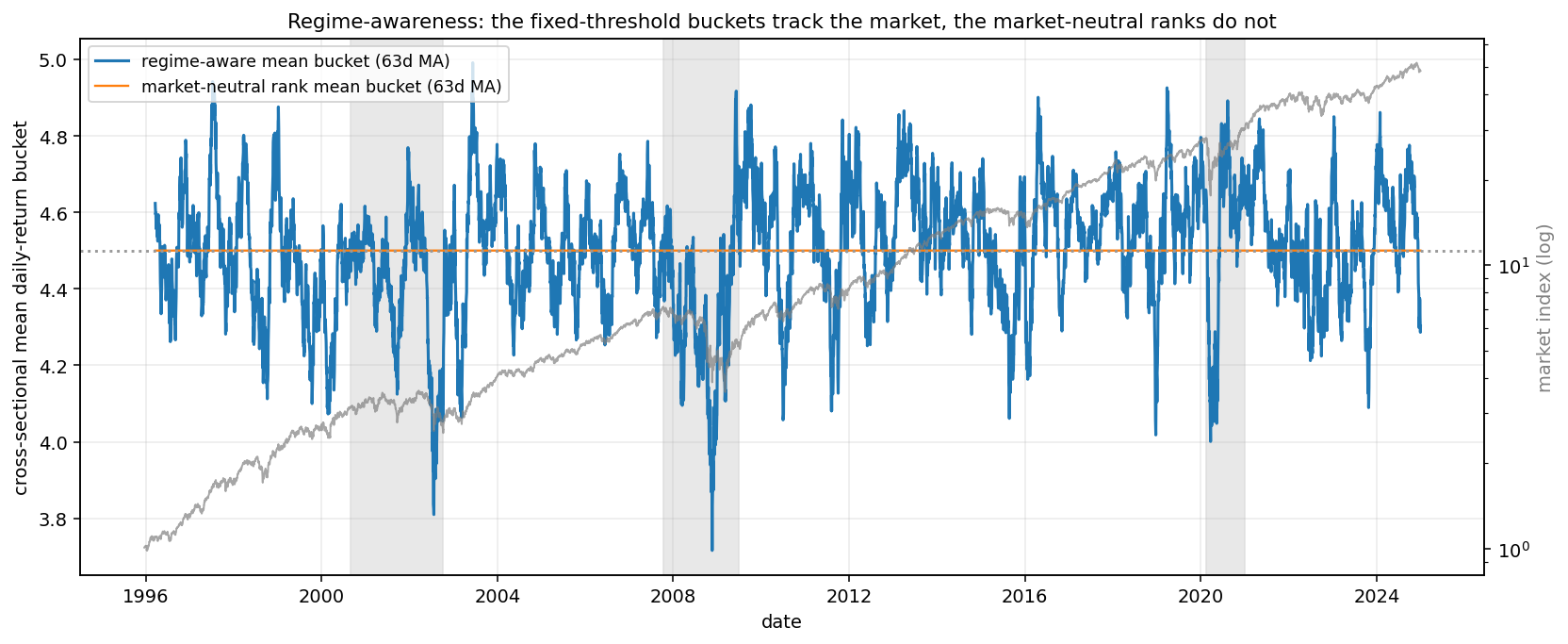}
\caption{The regime-aware buckets carry the market regime. The cross-sectional
mean daily-return bucket of the regime-aware chain (blue) swings with the market,
falling in the 2001, 2008, and 2020 crises (shaded) against a rising market index
(grey), while the same mean for the market-neutral rank chain (orange) is pinned
at the midpoint on every date. Its standard deviation is $1.6$ deciles for the
regime-aware chain and zero for the ranks.}
\label{fig:regimecheck}
\end{figure}

The construction replaces the per-period cross-sectional rank by a fixed absolute
threshold. On the training sample, which we close at the end of 2022, we measure
the mean market daily and monthly return and the mean market daily and monthly
volatility, the market being the equal-weight index of the universe. For each name
we form the excess of its return and of its volatility over these market means, at
both frequencies. For each of the four combinations, daily and monthly return and
daily and monthly volatility, we set the nine cut points of the ten buckets as the
deciles of the pooled excess distribution over the training sample, so the buckets
are equally populated on average. The bucket a name occupies on a given date is
then read off these fixed thresholds, not recomputed from the cross section, so it
records how the name did against the typical market rather than against its peers
that day. On a rally most names sit in high buckets and on a crash in low buckets,
and the cross section of buckets carries the regime, while the cross-sectional
mean bucket of the market-neutral ranks is constant by construction
(Figure~\ref{fig:regimecheck}). We keep the daily and the monthly chains as
separate objects estimated directly from their own transitions, not the monthly
chain as a power of the daily one, since the two horizons carry different memory
as Figure~\ref{fig:markov} shows.

The two families differ in one structural way that matters for the diagnostics.
The market-neutral chain has a uniform class marginal, exactly $N/K$ names per
decile on every date, so its marginal is stationary and the entropy production of
Eq.~\eqref{eq:ep} reduces to its transition, or flux, part. The regime-aware chain
has a non-uniform marginal that drifts within a window, low-heavy in a crash and
high-heavy in a rally, so its entropy production carries a second, boundary term
from the marginal drift,
\begin{equation}
\label{eq:epdecomp}
\sigma = \underbrace{\sum_{ab}\mu_{ab}\log\frac{P_{ab}}{P_{ba}}}_{\text{flux}}
       + \underbrace{\sum_{ab}\mu_{ab}\log\frac{p_a}{p_b}}_{\text{boundary}},
\end{equation}
with $p_a$ the class marginal and $P_{ab}=\mu_{ab}/p_a$ the transition matrix. The
full entropy production is the Kullback--Leibler divergence
$\sum_{ab}\mu_{ab}\log(\mu_{ab}/\mu_{ba})$ of Eq.~\eqref{eq:ep} in both cases,
which carries both terms automatically. The boundary term vanishes identically for
the market-neutral chain and is the part of the arrow of time that only the
regime-aware chain can express. The market-neutral chains remain the ones the
tradeable selection book of Section~\ref{sec:portfolio} uses, and the regime-aware
chains supply the market-state diagnostics and the new signals we turn to next.

\subsection{The model and its hyperparameters}
\label{sec:model}

The three matrices are not read in isolation; they feed a single pipeline, and
each informs the next. The distance matrix supplies three name-level covariates,
all read from the full matrix over a trailing window. The market loading is the
magnitude of a name's entry in the leading eigenvector, its exposure to the
single dominant market mode, and the distance centrality is the row mean of the
arccos distance, a name's average angular proximity to the whole cross section.
The two are related, since a name that loads strongly on the market mode sits
close to the others, but they are not identical, the loading reading only the
top eigenvector while the centrality summarizes all the correlation modes at
once. The third covariate is the directed lead-lag transfer entropy of
Section~\ref{sec:mmatrix}, which reads the direction of influence between a name
and the market. The market-removed residual distance is a separate object, used
later to diversify the long sleeve rather than to condition the chains. These
covariates condition the two ranking chains, so the correlation geometry enters
the dynamics of the ranks. Each conditioned chain then produces a one-step
forward-rank forecast, the probability $p^R_i$ that a name lands in the top
return decile next month and the probability $p^V_i$ that it lands in the
calmest volatility decile. Those forecasts drive selection: a long-only sleeve
scores each name by the convex combination $s_i=(1-\lambda)\,p^R_i+\lambda\,
p^V_i$ and holds the $2K$ highest, and a market-neutral long-short holds the top
$K$ and shorts the bottom $K$ by $p^R$ alone. The distance matrix enters a
second time on the portfolio, through the residual-distance diversification of
the long sleeve and the transfer-entropy hub overlay that insures the protective
sleeve, and a state-dependent weight $\theta$ blends the two sleeves into the
tradeable book.

Two objective functions are optimized, at two levels. The chain coefficients are
fit by maximum likelihood, the multinomial conditional-logit of the next class
on the covariate buckets of Section~\ref{sec:mle}, refit walk-forward so that no
future observation enters a forecast. The portfolio hyperparameters
$\Phi=(K,\mathrm{tol},\lambda,\theta,\dots)$ of Table~\ref{tab:hyper} are then
set at a second level, by maximizing the annualized Sharpe ratio of the combined
book's net monthly return $r_t(\Phi)$ over the validation period $\mathcal V$
(through 2021),
\begin{equation}
\label{eq:sharpeobj}
\hat\Phi = \operatorname*{arg\,max}_{\Phi}\ \mathrm{Sh}_{\mathcal V}(\Phi),
\qquad
\mathrm{Sh}_{\mathcal V}(\Phi) = \sqrt{12}\;
\frac{\operatorname{mean}_{t\in\mathcal V}\, r_t(\Phi)}
     {\operatorname{std}_{t\in\mathcal V}\, r_t(\Phi)},
\end{equation}
and read out of sample. The one exception is the diversification tilt $\gamma$,
set by the realized diversification it achieves rather than by the Sharpe, since
aiming at the return over-tilts and misses the out-of-sample peak
(Section~\ref{sec:portfolio}).

The final model carries the small set of hyperparameters of
Table~\ref{tab:hyper}. Each is set once on the validation period through 2021
and held fixed out of sample, and the experimental sections give the selection
and its sensitivity in detail (Sections~\ref{sec:portfolio}
and~\ref{sec:freshoos}).

\begin{table}[htbp]
\centering
\small
\caption{Hyperparameters of the final model, each set on the validation period
(through 2021) and read out of sample.}
\label{tab:hyper}
\begin{tabular}{lll}
\toprule
symbol & value & meaning \\
\midrule
$K$ & $15$ & book size: $2K$ long-only names, $K$ long and $K$ short in the long-short \\
tol & $0.08$ & no-trade band: swap a held name only if a challenger's score beats it by tol \\
$\lambda$ & $0$ / $0.75$ & long-only return-to-volatility weight, $0$ in rising markets, $0.75$ in falling \\
$\theta$ & $1$ / $0.4$ & blend weight, fully long-only in rising markets, $0.4$ toward the protective sleeve in falling \\
lookback & $3$ mo & trailing market-return window setting the $\lambda$ and $\theta$ regime \\
$\gamma$ & $0.5$ & residual-distance diversification tilt of the long sleeve \\
hub weight & $0.25$ & weight of the transfer-entropy hub long-short in the insurance overlay \\
refit & $12$ mo & walk-forward refit cadence of the chain coefficients \\
\bottomrule
\end{tabular}
\end{table}

\subsection{Cross-sectional rank-bucketing of features}
\label{sec:bucket}

A covariate is any firm characteristic measured on date $t$, for
instance a book-to-market ratio, a trailing twelve-month momentum, or a
short-window idiosyncratic volatility. Raw characteristics live on
incomparable scales, carry heavy tails, and are defined only up to the
monotone transforms that a practitioner might apply. To use them
uniformly we subject every characteristic to the same treatment we
applied to the ranking statistics in Eq.~\eqref{eq:class}. For a raw
feature $f^{(m)}$, $m=1,\dots,M$, we rank its values across the
securities present on date $t$ and bucket the rank into $K_f$ equal
groups,
\begin{equation}
\label{eq:xbucket}
x^{(m)}_i(t) = \Big\lceil K_f\,
\frac{\mathrm{rk}_i(f^{(m)},t)}{N} \Big\rceil
\in \{1,\dots,K_f\}, \qquad K_f = 10 .
\end{equation}
The bucketed covariate $x^{(m)}_i(t)$ is an ordinal decile of the
characteristic, and the vector $x_i(t) = (x^{(1)}_i(t),\dots,
x^{(M)}_i(t))$ is the discrete covariate state of security $i$.

This preprocessing has three properties that the rest of the
construction relies on. It is invariant under any strictly monotone
transform of the raw feature, so a characteristic and its logarithm, or
its winsorized version, produce the same buckets, which removes a
source of arbitrariness from the signal definition. It is robust to
outliers and to the cross-sectional dispersion of the feature, since
only the order matters, so a single extreme firm cannot dominate. And
it renders every feature on the identical ten-level scale, which makes a
large and heterogeneous characteristic library directly comparable and
gives each covariate a uniform marginal by construction. The ranking
statistics of Section~\ref{sec:chains} are themselves the $K_f=K$ case
of Eq.~\eqref{eq:xbucket} applied to the trailing return and the
trailing volatility, so the target ranks and the covariates are
instances of one transform. This uniformity is what lets the same
information-theoretic tools act on the ranks, on the covariates, and on
their interaction.

\subsection{Markov chains with covariates}
\label{sec:covariates}

We model each ranking chain in continuous time and let its transition
rates depend on the covariate state, specialized here to discrete
rank-based predictors. The class process
$c_i(t)$ of a security is a continuous-time Markov chain on the $K$
rank classes, described by a generator matrix $Q$ whose off-diagonal
entries are the transition intensities and whose rows sum to zero.
Our Markov chain (MC) with covariates framework for the ranking dynamics of both
chains is mathematically identical to the MC modeling frameworks for rating
migrations in credit portfolios, where the covariates can be either common
macroeconomic factors or firm-specific factors \cite{lando2002,kavvathas2001}.
Every row of $Q$ is a proper generator row, and
the chain is ergodic with the uniform stationary marginal that the
decile construction fixes. The finite-horizon transition probabilities
are the matrix exponential
\begin{equation}
\label{eq:matexp}
P_{ab}(\Delta) = \big[\exp(\Delta\,Q)\big]_{ab}
= \delta_{ab} + \Delta\,Q_{ab}
+ \tfrac12 \Delta^2 [Q^2]_{ab} + \cdots .
\end{equation}

We couple the intensities to the covariate state through a log-linear
form,
\begin{equation}
\label{eq:qcov}
q_{ab}(x) = \begin{cases}
\exp\!\big\{\eta_{ab} + \gamma_{ab}\cdot\phi(x)\big\}, & a\neq b,\\[3pt]
-\sum_{b'\neq a} q_{ab'}(x), & a=b,
\end{cases}
\end{equation}
where $\phi(x)$ scores the discrete covariate state, either as the
centered bucket levels $\phi(x^{(m)}) = (x^{(m)}-\tfrac{K_f+1}{2})/K_f$
or, for full flexibility, as the vector of bucket indicators. The
baseline intensity $\exp(\eta_{ab})$ is the unconditional rate,
recovered when $\gamma_{ab}=0$, so the covariate model refines the
unconditional chain of Section~\ref{sec:chains}, and $\gamma_{ab}$
measures how the covariate accelerates or damps the $a\to b$ move.

Because the covariates are bucketed, the predictor takes only finitely
many values, at most $K_f^{M}$ distinct states $x$ for $M$ covariates.
This is the simplification that discreteness buys. The scenario
integral of the continuous-factor model becomes a finite sum over
covariate cells, and every intensity $q_{ab}(x)$ is constant within a
cell.

\subsection{Maximum-likelihood estimation}
\label{sec:mle}

The generator is calibrated by maximum likelihood on the observed class
paths, pooling securities under the assumption that their transitions
are independent given the covariate state. The sufficient statistics of
a continuous-time chain are transition counts and exposure times.
Grouping by covariate cell $x$, let $\Delta N_{ab}(x)$ be the number of
$a\to b$ transitions among the security-days whose origin has covariate
state $x$, and $R_a(x)$ the total time spent in class $a$ with that
covariate state. For a continuous-time chain with transition intensities
$q_{ab}(x)=e^{\,\eta_{ab}+\gamma_{ab}\cdot\phi(x)}$, the negative log-likelihood
in these statistics is, following Lando and Sk\o{}deberg \cite{lando2002},
\begin{equation}
\label{eq:nll0}
-\log L = \sum_{x}\sum_a\sum_{b\neq a}\Big[R_a(x)\,q_{ab}(x)
- \Delta N_{ab}(x)\,\log q_{ab}(x)\Big].
\end{equation}
Substituting the intensity parameterization and dropping terms independent of
the parameters expands this into
\begin{equation}
\label{eq:nll}
-\log L = \sum_{x}\sum_a R_a(x)\!\sum_{b\neq a}
e^{\,\eta_{ab}+\gamma_{ab}\cdot\phi(x)}
\;-\; \sum_{a}\sum_{b\neq a}\Big[\Delta N_{ab}\,\eta_{ab}
+ \gamma_{ab}\!\sum_x \Delta N_{ab}(x)\,\phi(x)\Big],
\end{equation}
with $\Delta N_{ab}=\sum_x \Delta N_{ab}(x)$ the total count.
Minimizing over the baseline $\eta_{ab}$ has the closed form
\begin{equation}
\label{eq:eta}
e^{\eta_{ab}} = \frac{\Delta N_{ab}}
{\sum_x R_a(x)\,e^{\gamma_{ab}\cdot\phi(x)}},
\end{equation}
which, substituted back, decomposes the estimation into one independent
objective for each ordered class pair,
\begin{equation}
\label{eq:pairnll}
-\log L_{ab} = \Delta N_{ab}\,
\log\!\Big(\sum_x R_a(x)\,e^{\gamma_{ab}\cdot\phi(x)}\Big)
\;-\; \gamma_{ab}\sum_x \Delta N_{ab}(x)\,\phi(x).
\end{equation}
Each $-\log L_{ab}$ is convex in $\gamma_{ab}$, so the estimate is
unique and cheap, and a small ridge penalty
$\varepsilon\lVert\gamma_{ab}\rVert^2$ with $\varepsilon\approx0.1$
stabilizes pairs with few observations while preserving convexity. The
calibrated intensity is
\begin{equation}
\label{eq:qcalib}
q_{ab}(x) = \Delta N_{ab}\,
\frac{e^{\gamma_{ab}\cdot\phi(x)}}
{\sum_{x'} R_a(x')\,e^{\gamma_{ab}\cdot\phi(x')}},
\end{equation}
which at $\gamma_{ab}=0$ reduces to the exposure-and-count estimator
$q_{ab}=\Delta N_{ab}/\sum_x R_a(x)$ of covariate-free continuous-time
rating estimation \cite{lando2002,kavvathas2001}.

Two consequences of the discrete, rank-based predictors are worth
drawing out. First, in the fully flexible parameterization where
$\phi(x)$ is the set of bucket indicators, the per-pair estimate has
the explicit solution $q_{ab}(x)=\Delta N_{ab}(x)/R_a(x)$, the
empirical transition rate within each covariate cell. The covariate
model then reduces to estimating a separate generator inside every
bucket, the nonparametric conditional-transition tensor, and the scored
form is its parsimonious smoothing when the cells are too thin to fill.
Second, the exposure $R_a(x)$ is read directly off the daily panel. On
a daily grid a security spends one step $\Delta t$ in its recorded
class between observations, so $R_a(x)$ is the count of security-days
in class $a$ with covariate state $x$ times $\Delta t$, and the
one-step transition matrix that the OMD-Stocks framework estimates
\cite{halperin2026omdstocks} is
$P(x)=\exp(\Delta t\,Q(x))\approx I+\Delta t\,Q(x)$. The continuous-time
generator and the discrete-time transition matrix are two readings of
the same calibrated object, and the covariate-conditioned transition
distribution $P_{ab}(x)$ feeds the diagnostics of the next sections.

\subsection{Conditional entropy production}
\label{sec:condep}

The entropy production of Eq.~\eqref{eq:ep} measures the
irreversibility of the pooled chain. With covariates in hand we can ask
a sharper question. How much of that irreversibility, the arrow of time
of the relative order, is carried by the covariate, and how much
survives once we condition on it. Let $\mu_{ab\mid x}$ be the joint
transition distribution restricted to origins with covariate state $x$,
and $p(x)$ the fraction of transitions with that state. The pooled
joint is the covariate average $\mu_{ab} = \sum_x p(x)\,\mu_{ab\mid
x}$, and the conditional entropy production is the covariate-weighted
mean of the per-state entropy productions,
\begin{equation}
\label{eq:condep}
\sigma(x) = \sum_{a,b}\mu_{ab\mid x}\,
\log\frac{\mu_{ab\mid x}}{\mu_{ba\mid x}}, \qquad
\sigma_{\mathrm{cond}} = \sum_x p(x)\,\sigma(x).
\end{equation}

Conditioning cannot decrease the measured entropy production. The
functional $\sigma(\mu) = \sum_{ab}\mu_{ab}\log(\mu_{ab}/\mu_{ba})$ is
the Kullback--Leibler divergence $\mathrm{KL}(\mu\,\|\,\mu^{\!\top})$
between the joint and its transpose. The transpose is linear in $\mu$,
and the divergence is jointly convex in its two arguments, so $\sigma$
is convex in $\mu$. Jensen's inequality applied to the covariate
average $\mu = \sum_x p(x)\mu_{\cdot\mid x}$ then gives
\begin{equation}
\label{eq:jensen}
\sigma \;=\; \sigma\Big(\textstyle\sum_x p(x)\,\mu_{\cdot\mid x}\Big)
\;\le\; \sum_x p(x)\,\sigma(\mu_{\cdot\mid x})
\;=\; \sigma_{\mathrm{cond}}.
\end{equation}
The gap $\Delta\sigma = \sigma_{\mathrm{cond}} - \sigma \ge 0$ is the
irreversibility that pooling over the covariate had hidden. When a
characteristic is uninformative about the direction of the flow the two
sides of Eq.~\eqref{eq:jensen} coincide and $\Delta\sigma=0$, whereas a
characteristic that lines up with the systematic circulation of the
ranking resolves a positive $\Delta\sigma$. The quantity $\Delta\sigma$
is thus a covariate-attribution of the arrow of time, a measure of how
much of the market's directional structure a given signal explains,
computed entirely from plug-in counts on the discrete chain.

A direct time-reversal check confirms that this reading is the genuine
irreversibility of the data rather than an artifact of how the covariate is
built. Because $\sigma$ compares the forward joint to its transpose, reversing
the return history in time and recounting the transitions reproduces the
transpose to machine precision and leaves the pooled $\sigma$ unchanged at every
lag, as the detailed-balance condition $\mu = \mu^{\!\top}$ requires. The
covariate part behaves as it must under the same reversal. A characteristic that
is a function of the past, such as the trailing return that carries the leverage
effect, is not time-symmetric: recomputing it on the reversed history and
re-estimating the transition gives a different, here sign-reversed, dependence,
while a contemporaneous or slowly-varying characteristic such as the current
volatility keeps its coefficient. That coefficient change under reversal is the
same time-asymmetry that $\Delta\sigma$ measures. The covariate resolves a
positive arrow of time exactly when its predictive relation to the future is
directional, so the conditional entropy production inherits its irreversibility
from the data and not from the choice of estimator.

\subsection{Transfer entropy and directional causality}
\label{sec:te}

The entropy production reads the irreversibility of the rank process.
A separate question is directional. Does a characteristic carry
information about where a security's rank will go next, beyond what the
rank's own history already implies, and does the influence run from the
characteristic to the rank or the other way. The transfer entropy
answers this \cite{schreiber2000}. For a security's rank-class path
$\{a_i(t)\}$ and covariate-bucket path $\{x_i(t)\}$, the transfer
entropy from the covariate to the rank is
\begin{equation}
\label{eq:te}
\TE(X\!\to\!A) = \sum
p\big(a_{t+1},a_t,x_t\big)\,
\log\frac{p\big(a_{t+1}\mid a_t,x_t\big)}
{p\big(a_{t+1}\mid a_t\big)},
\end{equation}
pooled over securities and dates, the reduction in uncertainty about the
next rank gained from the present covariate bucket, over and above the
rank's own present. The reverse quantity $\TE(A\!\to\!X)$ exchanges the
roles of the two series and measures whether the rank predicts the
characteristic. Because both paths are already bucketed into ten
levels, Eq.~\eqref{eq:te} is an exact plug-in sum over a discrete joint
distribution, with none of the nearest-neighbour or kernel estimation
that continuous characteristics would demand, the same advantage the
discrete chain gives the entropy production.

The asymmetry of the two transfer entropies gives a directional
reading. The net flow $\TE(X\!\to\!A) - \TE(A\!\to\!X)$ is positive
when the characteristic leads the rank and negative when the rank leads
the characteristic. This distinction has content for security
selection. A fundamental signal such as a value or quality ratio is
plausibly exogenous to short-horizon price action, so a positive net
flow into the return rank is evidence that the signal anticipates
performance rather than merely recording it. A signal that is itself a
function of past returns, such as a trailing-return momentum, is
mechanically tied to the recent return rank, and the framework should
recover a near-tautological or reversed flow for it, which serves as a
consistency check on the reading. Conditioning the transfer entropy on
a third bucketed characteristic probes causality while controlling for
a candidate confounder, in the same plug-in form. The transfer-entropy
map over a characteristic library is then a directed, discrete
screen for which signals genuinely drive the performance and risk
rankings, and which follow them.

\section{Model analysis and estimation}
\label{sec:analysis}

\subsection{From conditioned chains to portfolios}
\label{sec:selection}

The conditioned chains forecast where a security's rank will go, and
that forecast is the selection signal. By the orientation above, class
one is the best performer in the return chain and the calmest name in
the volatility chain, and class $K$ the worst performer or the most
volatile. From the covariate-conditioned
forward distribution $P_{ab}(x)$ of Eq.~\eqref{eq:matexp} we read for
each security the two calibrated probabilities that matter for
selection by the next rebalancing date, the chance it lands in the top
two return deciles and the chance it lands in the two calmest
volatility deciles,
\begin{equation}
\label{eq:score}
\pi^R_i(t) = \!\!\sum_{b\le 2}\! P^R_{c^R_i(t)\,b}\big(x_i(t)\big),
\qquad
\pi^V_i(t) = \!\!\sum_{b\le 2}\! P^V_{c^V_i(t)\,b}\big(x_i(t)\big),
\end{equation}
the probability of being among the best performers and among the
calmest names. Both are desirable, so the selection score is their
convex blend with weight $\lambda\in[0,1]$,
\begin{equation}
\label{eq:combined}
S_i(t) = (1-\lambda)\,\pi^R_i(t) + \lambda\,\pi^V_i(t),
\end{equation}
tilting from pure performance at $\lambda=0$ toward pure calm at
$\lambda=1$. The score ranks the universe into a candidate set, and we
long the $K_{\mathrm{sel}}$ names with the highest score and short the
$K_{\mathrm{sel}}$ lowest, equal weight and dollar neutral. Both
$\lambda$ and the book size $K_{\mathrm{sel}}$ are hyperparameters,
chosen on a validation period, alongside the no-trade tolerance of
Section~\ref{sec:portfolio} that turns the candidate set into an actual
rebalancing.
This is a rank-based strategy of the kind studied in stochastic
portfolio theory \cite{fernholz2002,banner2005}, with the difference
that the transition probabilities are now conditioned on the full
covariate state rather than on rank alone. The transfer-entropy screen of
Section~\ref{sec:te} tells which covariates to admit into the
conditioning, keeping the characteristics that lead the rank and
dropping those that only follow it, and the entropy-production
attribution of Section~\ref{sec:condep} tells how much directional
structure each admitted covariate resolves. The construction is thus a
single discrete pipeline, from raw characteristics to bucketed
covariates, to conditioned transition matrices, to a forward-rank
selection score, in which every step is estimated by plug-in counts and
every intermediate object is interpretable.

\subsection{Predictors and diagnostics}
\label{sec:empirical}

The empirical part proceeds in the order the construction suggests. We
first read the two diagnostics of the conditioned chains, the entropy
production and the transfer entropy, on the volatility chain, and then
turn to two further families of predictors, those read from the
distance matrix itself (Section~\ref{sec:mmatrix}) and the accounting
fundamentals for the return chain (Appendix~\ref{sec:fundamentals}).
Section~\ref{sec:calibration} then tests how well the calibrated chains
forecast, one step and over multi-step trajectories, and
Section~\ref{sec:portfolio} builds the portfolios and reports the
results.

The experiments draw on three market datasets. Daily prices come
from CRSP over 2005--2024. The tradeable study uses the $N=331$ names that are
S\&P 500 members throughout 2015--2024, ranked each month into deciles by
trailing return and by trailing volatility, so the two chains are read on a
monthly grid unless a shorter horizon is stated. Company fundamentals, which
enter only the ablation of Appendix~\ref{sec:fundsection}, are Compustat
financial ratios over 1970--2025, carried point-in-time by their public date.
The second out-of-sample test of Section~\ref{sec:freshoos} rebuilds the panel
from Yahoo Finance, an independent daily price and volume source running from
2018 to July 2026, so its recent window lies outside the CRSP record and outside
the rest of the study.

Every portfolio result below is net of transaction costs, charged on the
notional traded each time a name's weight changes. Our base case is five basis
points, realistic for the liquid large-capitalization names of the S\&P 500,
where the spread, commissions, and a modest market impact run to about three to
five, and we carry a conservative ten for comparison so that every headline
result can be read at both, the main numbers and figure lines being the
five-basis-point case. Costs at either level are large enough to matter, and
they motivate the no-trade band of Section~\ref{sec:portfolio}. The combined
book is charged on its net per-name weight, the two sleeves held in one account,
so a name carried in both is traded only on its net position; this netting trims
about a seventh of the traded notional over the recent window and is already
inside every number below.

We begin with the covariate-conditioned volatility chain, using only
characteristics that the daily market record supports without external
accounting data. The target is the decile of
the trailing 21-day realized volatility, whose one-step transitions are
the volatility chain of Section~\ref{sec:chains}. Five covariates
enter, each ranked cross-sectionally and bucketed into deciles by
Eq.~\eqref{eq:xbucket}: the log market capitalization, the trailing
252-day market beta, the twelve-minus-one momentum, the Amihud
illiquidity,\footnote{The Amihud illiquidity is the average over the
trailing month of a name's absolute daily return divided by its daily dollar
volume, a price-impact-per-dollar measure that is large for a name whose price
moves far on little turnover.} and the trailing 63-day realized volatility, all
read from prices, volume, and shares outstanding alone.

\begin{figure}[htbp]
\centering
\includegraphics[width=0.98\textwidth]{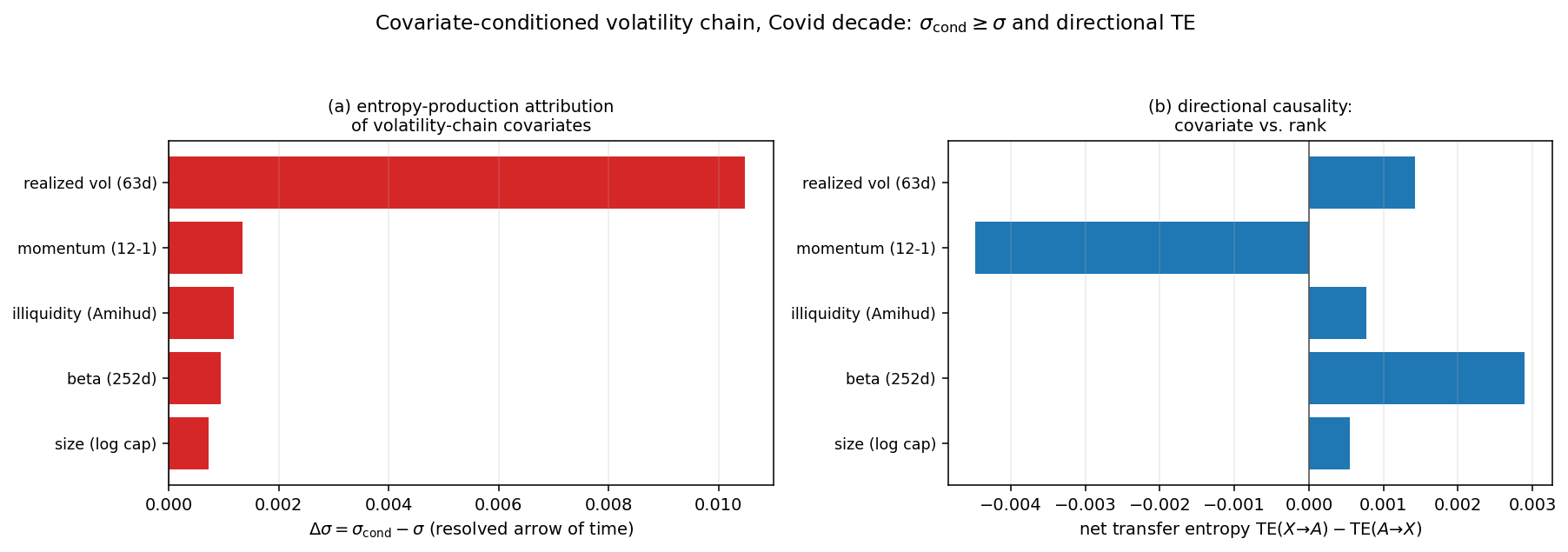}
\caption{Covariate-conditioned volatility chain, S\&P 500 over
2015--2024. (a) The entropy-production attribution
$\Delta\sigma=\sigma_{\mathrm{cond}}-\sigma$ of five price-based
covariates, the irreversibility each resolves beyond the firm-pooled
chain, all positive as the convexity bound requires. (b) The net
transfer entropy $\TE(X\!\to\!A)-\TE(A\!\to\!X)$ between each covariate
and the volatility rank. Size, beta, and illiquidity lead the rank
(positive), while momentum, a function of past returns, follows it
(negative).}
\label{fig:covariate}
\end{figure}

Table~\ref{tab:covariates} and Figure~\ref{fig:covariate} report the
conditional entropy production and the transfer entropy of each
covariate. The pooled volatility chain has $\sigma=2.0\times10^{-4}$
nats per step. Conditioning on any single covariate raises the entropy
production, to between $9\times10^{-4}$ for size and $1.1\times10^{-2}$
for the 63-day volatility, so $\sigma_{\mathrm{cond}}$ runs from five to
fifty times the pooled value. The convexity bound
$\sigma_{\mathrm{cond}}\ge\sigma$ of Eq.~\eqref{eq:jensen} holds in
every case, and the gap $\Delta\sigma$ is large. Much of the arrow of
time in the relative-risk order is therefore invisible to the
firm-pooled chain and is resolved only once the chain is conditioned on
where a name sits in the cross section of a characteristic. The 63-day
volatility resolves the most, which is the persistence of volatility
itself, a longer-horizon volatility predicting the fast-up, slow-down
circulation of the shorter-horizon rank.

The transfer entropy separates the covariates by direction. For size,
beta, and illiquidity the net flow
$\TE(X\!\to\!A)-\TE(A\!\to\!X)$ is positive, so the characteristic
leads the volatility rank, the reading expected of an exogenous risk
descriptor. For momentum the net flow is negative,
$-4.5\times10^{-3}$, so the rank leads the characteristic rather than
the reverse. This is the consistency check anticipated in
Section~\ref{sec:te}. The twelve-minus-one momentum is a function of
past returns, so it echoes the recent rank rather than anticipating it,
and the directional transfer entropy recovers exactly that. The two
diagnostics thus play complementary roles, the entropy-production
attribution $\Delta\sigma$ measuring how much directional structure a
covariate resolves, and the net transfer entropy telling whether the
covariate drives the ranking or trails it.

\begin{table}[htbp]
\centering
\small
\caption{Covariate-conditioned volatility chain, S\&P 500, 2015--2024.
Conditional entropy production $\sigma_{\mathrm{cond}}$, its excess
$\Delta\sigma=\sigma_{\mathrm{cond}}-\sigma$ over the pooled
$\sigma=0.20\times10^{-3}$, and the net transfer entropy
$\TE(X\!\to\!A)-\TE(A\!\to\!X)$, all in units of $10^{-3}$ nats per
step, ordered by $\Delta\sigma$. A positive net transfer entropy marks
a covariate that leads the rank, a negative one a covariate that
follows it.}
\label{tab:covariates}
\begin{tabular}{lccc}
\toprule
covariate & $\sigma_{\mathrm{cond}}$ & $\Delta\sigma$
& net $\TE$ \\
\midrule
realized vol (63d)   & $10.7$ & $10.5$ & $+1.42$ \\
momentum (12-1)      & $1.54$ & $1.34$ & $-4.47$ \\
illiquidity (Amihud) & $1.39$ & $1.19$ & $+0.77$ \\
beta (252d)          & $1.15$ & $0.95$ & $+2.90$ \\
size (log cap)       & $0.92$ & $0.72$ & $+0.55$ \\
\bottomrule
\end{tabular}
\end{table}

\subsection{Distance-matrix predictors}
\label{sec:mmatrix}

The return chain is hard to forecast from the market characteristics
above, so we add a third family of predictors read from the distance
matrix $M(t)$ itself, the object the ranking chains sit inside
\cite{halperin2026omdstocks}. Over a trailing year we compute three
quantities for each name and bucket each into deciles by
Eq.~\eqref{eq:xbucket}. The market loading is the name's weight in the
leading eigenvector of the correlation matrix, its exposure to the
market mode. The distance centrality is its mean arccos distance to the
rest of the cross section, high for an idiosyncratic, peripheral name
and low for one that moves with the crowd. The lead-lag score is the
transfer entropy the name sends to the market minus the transfer
entropy it receives, positive for a bellwether the market follows and
negative for a follower, the leader-and-follower reading of the network
carried down to the name.

Table~\ref{tab:mmatrix} reports the out-of-sample predictive
log-likelihood gain over the covariate-free baseline for the monthly
chains, with the market covariates alone and with the market covariates
augmented by the three distance-matrix predictors. The two chains part here as
they will throughout. The market covariates alone raise the
volatility-chain log-likelihood but push the return-chain one below its
baseline out of sample, an overfit. Adding the distance-matrix
predictors improves the volatility chain further, to $+0.108$, and for
the return chain mainly corrects the overfit, lifting it just above the
baseline to a marginal $+0.007$. The distance-matrix structure, being
slower and steadier than the fast market characteristics, survives out
of sample where they do not, but the return-chain gain is small, and the
one-step tests of Section~\ref{sec:calibration} show it does not become a
forecast the book could use. The volatility chain is the one the
predictors genuinely help.

One caveat on the second covariate is worth stating. The distance centrality and
the market loading are two readings of the same market-centrality, since a name
central in the full geometry is one that moves with the market, so the two are
correlated across the cross section at a rank correlation near $-0.6$. A cleaner
alternative computes the centrality in the geometry orthogonal to the market.
Regressing the equal-weight market out of the return window and taking the
leading eigenvector of the residual co-movement network, the same market removal
the residual-distance diversification of Section~\ref{sec:portfolio} uses, gives
a residual eigenvector centrality far less redundant with the market loading, at
a rank correlation near $-0.4$. It leaves the predictive log-likelihood of both
chains essentially unchanged, and across both test datasets the tradeable book
is comparable, the development out-of-sample Sharpe a fraction below the
full-matrix version and the recent-window diversified book a fraction above it.
We therefore keep the simpler full-matrix centrality in the tradeable book, and
note the residual one as a market-orthogonal alternative that performs as well
on both test periods.

\begin{table}[htbp]
\centering
\small
\caption{Monthly chains, S\&P 500 2015--2024. Out-of-sample predictive
log-likelihood gain over the covariate-free baseline, in nats per step,
for the market covariates and for the market covariates augmented by
the three distance-matrix predictors. The distance-matrix predictors
turn the return chain from an out-of-sample overfit into a marginal
positive, and improve the volatility chain.}
\label{tab:mmatrix}
\begin{tabular}{lcc}
\toprule
chain & market covariates & market $+$ $M$-matrix \\
\midrule
return     & $-0.008$ & $+0.007$ \\
volatility & $+0.091$ & $+0.108$ \\
\bottomrule
\end{tabular}
\end{table}

The lead-lag score is a transfer-entropy score, and it earns its place
as a hedge rather than a forecast. As a predictor it is neutral. Dropping
it from the set leaves the out-of-sample log-likelihood unchanged, even a
shade higher, at $+0.008$ for the return chain and $+0.110$ for the
volatility chain, and the full directed stock-to-stock transfer-entropy
network, its out-strength, in-strength, and centrality all tested as
predictors, adds nothing to either chain. Its value is in the portfolio.
The long-short of Section~\ref{sec:portfolio} is market-neutral, beta
$-0.02$, and gains $1.13\%$ a month across the down markets, but without
the lead-lag score the beta rises to $+0.06$, that down-market gain
halves to $0.57\%$, and the combined book's out-of-sample Sharpe slips
from $1.06$ to $0.99$. The transfer-entropy structure hedges the book
even where it does not forecast the rank, so we keep it.

\subsection{Calibration and multi-step forecasts}
\label{sec:calibration}

\begin{figure}[htbp]
\centering
\includegraphics[width=0.98\textwidth]{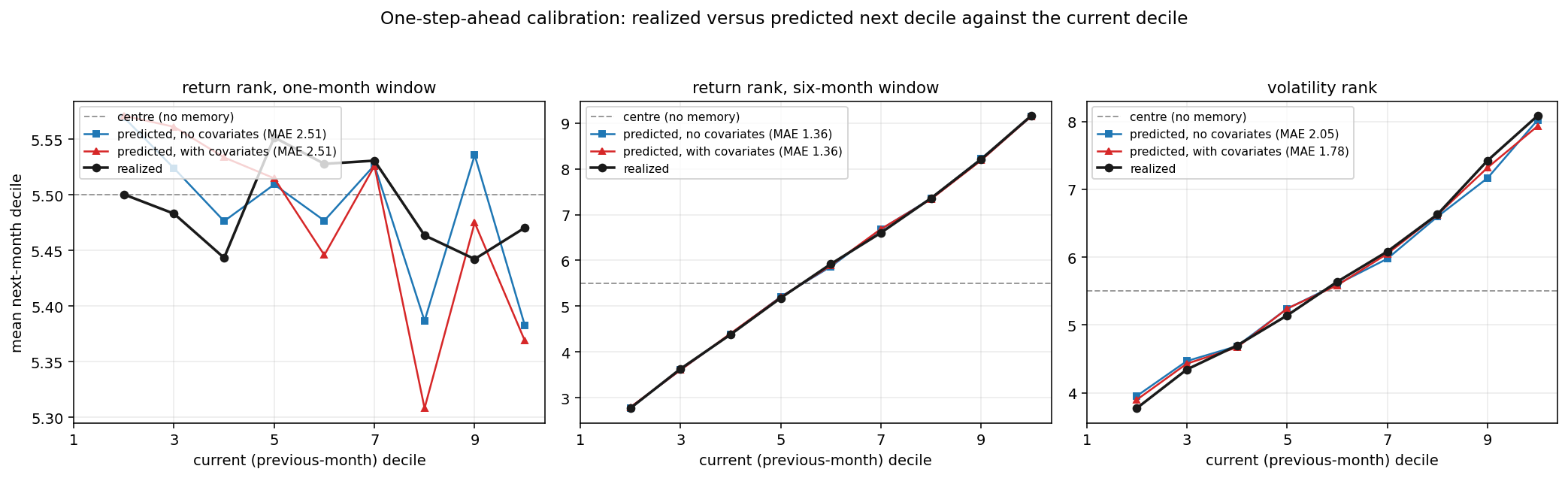}
\caption{One-step-ahead calibration, the mean next-month decile against
the current decile, realized (black) and predicted from the previous
decile, covariate-free (blue) and with covariates (red), accumulated
walk-forward. The dashed line is the centre a memoryless chain predicts.
The one-month return rank (left) is flat, near-random. The six-month
return rank (middle), the window the portfolio ranks on, is strongly
persistent, tracking the diagonal, but the persistence is the mechanical
overlap of its windows and the covariates add nothing. The volatility
rank (right) is persistent for real, and the covariates sharpen it,
lowering the mean absolute error from $2.05$ to $1.78$.}
\label{fig:calibration}
\end{figure}

A conditioned chain that resolves entropy production need not forecast
individual transitions, so we test the forecast directly. The strategy
rebalances monthly, so the transition matrices are estimated on a
monthly grid, and the test that matters is one step ahead on a rolling
basis. Each month we predict this month's decile for every name from
last month's decile, once from the covariate-free transition matrix and
once from the covariate-conditioned model $P_{ab}(x)$ of
Eq.~\eqref{eq:matexp}, both fit on the history up to that month, and
compare the two against the realized decile. Figure~\ref{fig:calibration}
plots, for each current decile, the mean next decile that is realized and
that each model predicts. The three panels separate two things a
persistent transition can mean. The one-month return rank is near-random,
its realized curve flat at the centre whatever the current decile, and
the covariates do not move it, a mean absolute error of $2.5$ deciles
either way. The six-month return rank, the window the portfolio ranks on,
is strongly persistent instead, its next decile tracking the current one
almost along the diagonal at an error of $1.4$. That persistence is
mechanical, the overlapping six-month windows sharing five months and
predicting their own next value, so the covariates add nothing to it and
it earns no forward-return edge. The volatility rank is persistent for a
different reason. Volatility clusters, so its rank genuinely carries, and
there the covariates sharpen the forecast, lowering the error from $2.05$
to $1.78$, a gain the predictive log-likelihood confirms, $0.116$ against
$0.007$ or less for the return chain. Relative risk is forecastable with
skill one step ahead. Relative return is not, whether it reads as random
at one month or as trivially self-predicting at six.

\begin{figure}[H]
\centering
\includegraphics[width=0.98\textwidth]{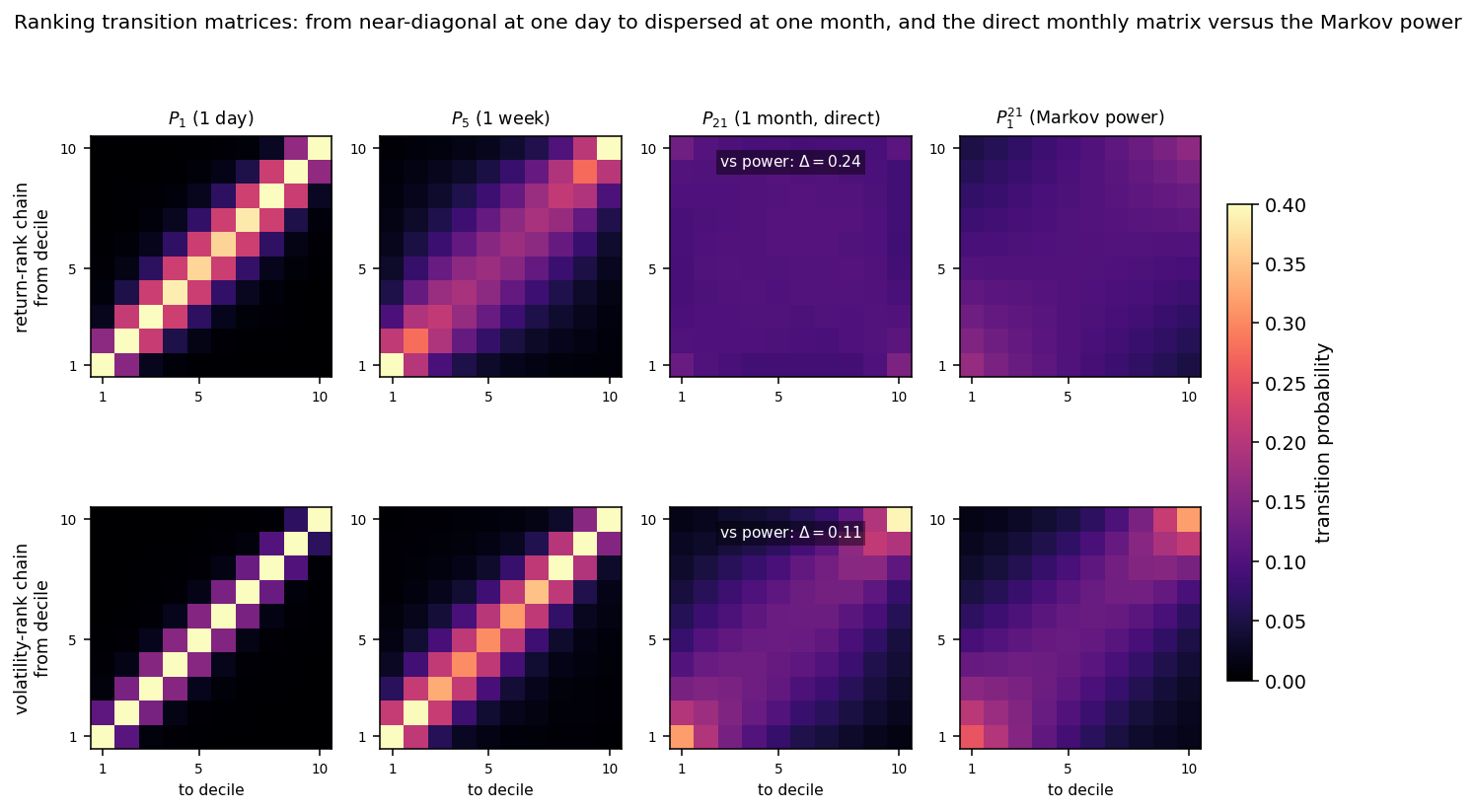}
\caption{Ranking transition matrices as heatmaps, pooled over 2015--2024.
The one-day matrix $P_1$ is sharply diagonal, both chains persistent from
one day to the next, and it disperses with the step. By one month the
return matrix $P_{21}$ (top) has nearly uniform rows, its rank forgotten,
while the volatility matrix keeps bright corners where the calmest and
most volatile deciles persist. The last column is the twenty-one-day
power $P_1^{21}$, the monthly matrix a first-order chain would predict. It
differs from the direct $P_{21}$, and the normalized distance $\Delta$
between them is the non-Markovianity, larger for the return chain.}
\label{fig:heatmaps}
\end{figure}

A secondary check over multiple steps explains why the volatility chain
is the forecastable one, and points to a sharper predictor. Raising the one-step monthly transition matrix $P$
to the power $n$ gives the Markov forecast $n$ steps ahead. For the
one-month return chain it matches the realized dynamics to within $0.04$
deciles, its top-decile mean falling to the centre in a single step, so
that chain is first-order and mixes at once, the near-random matrix of
Fig.~\ref{fig:heatmaps}. The volatility
chain is different. Its realized ranks stay separated, the top decile
still near eight and the bottom near three and a half after ten months,
while the covariate-free $P^n$ predicts convergence to the centre by
month six, a gap of $1.8$ deciles. The volatility ranking has memory
beyond one step, the persistence of volatility clustering, which a
first-order chain underestimates.

That memory can be fed back in as a predictor. Adding a longer-window
realized volatility, ranked cross-sectionally and bucketed like any
other feature, injects the persistence the one-step chain misses. Our
final single-name predictor set for the volatility chain is size, beta,
Amihud illiquidity, the sixty-three-day realized volatility, the
one-hundred-and-twenty-six and two-hundred-and-fifty-two-day realized
volatilities, and the three distance-matrix predictors of
Section~\ref{sec:mmatrix}. Coefficients are calibrated per origin decile
on the training and validation periods, and each name's own covariate
values turn the shared coefficients into that name's one-step
probabilities. With this set the out-of-sample predictive log-likelihood
gain on the monthly volatility chain rises from $0.091$ with the
sixty-three-day volatility alone to $0.116$ once the long-memory
volatilities and the distance-matrix predictors are added, the single
largest gain from any predictor group.

\begin{figure}[htbp]
\centering
\includegraphics[width=0.98\textwidth]{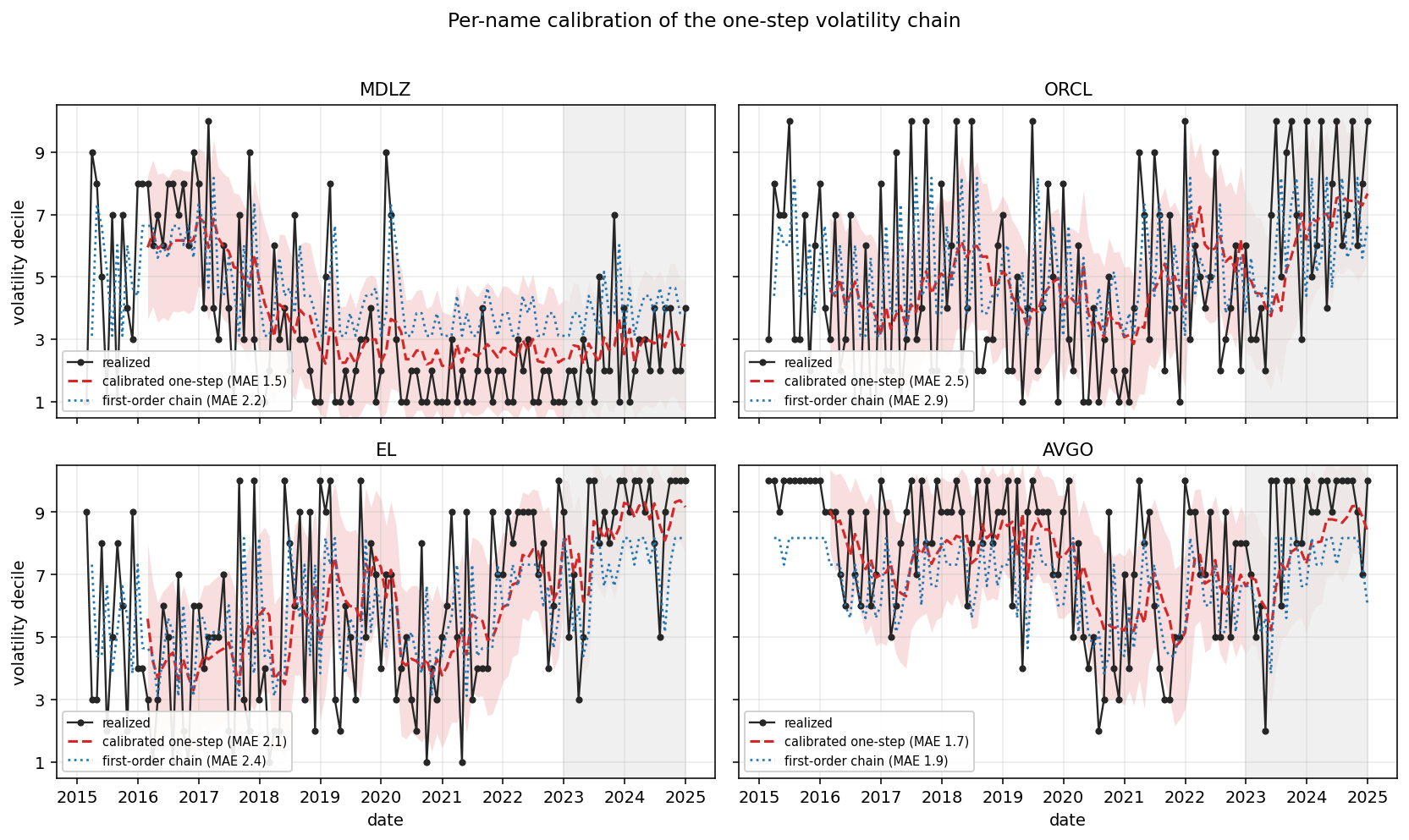}
\caption{Per-name calibration of the one-step volatility chain, four
stocks spanning the volatility spectrum. The black path is the realized
volatility decile month by month. The red dashed path is the most
plausible next decile, the mean of the calibrated one-step distribution
evaluated at the name's own decile and covariates, with the shaded band
one predictive standard deviation. The blue dotted path is the
covariate-free first-order chain, which mean-reverts to the centre too
fast. The calibrated forecast tracks each name's realized level and
drift, in sample and out (grey), with a smaller error on every name. The
first-order chain misses the persistence. The out-of-sample period
(2023--2024) is shaded.}
\label{fig:trajectories}
\end{figure}

Figure~\ref{fig:trajectories} makes the calibration visible name by
name. For four stocks spanning the volatility spectrum, from the placid
MDLZ to the volatile AVGO, we read off at each month the most plausible
next decile, the mean of the calibrated one-step distribution evaluated
at the name's current decile and covariates, and lay this predicted path
over the realized decile path. The calibrated forecast follows each
name's realized level and its slow drift, MDLZ decaying from the middle
to the floor after 2018 and ORCL and EL climbing back through
2020--2024, and tracks it out of sample with a mean absolute error of
one and a half to two and a half deciles. The covariate-free first-order
chain, predicting from the current decile alone, mean-reverts to the
centre and misses the persistence, with a larger error on every name.
The gap between the two lines is the memory the long-memory predictors
restore.

The return chain resists the fast characteristics, but one predictor
does reach it, and classical portfolio theory says which. Risk and
return are balanced in equilibrium, so a name carrying more volatility
should be priced to earn more, and the cross-sectional rank of trailing
volatility should then carry information about the next return rank.
Volatility is the one obvious characteristic absent from the
return-chain predictor set. Ranking the trailing realized volatility
over a lookback chosen on the validation period, sixty-three days, and
bucketing it into deciles like any other feature, we condition the
return chain on it alone. Out of sample it raises the predictive
log-likelihood by $0.018$ nats per step over the vanilla chain, more
than double the $0.007$ of the full market-plus-distance-matrix set of
Table~\ref{tab:mmatrix} and the largest gain any predictor brings to
the return chain. Stacking volatility with the other characteristics
erases the gain, down to $0.008$, an overfit the single feature avoids.

\begin{figure}[htbp]
\centering
\includegraphics[width=0.62\textwidth]{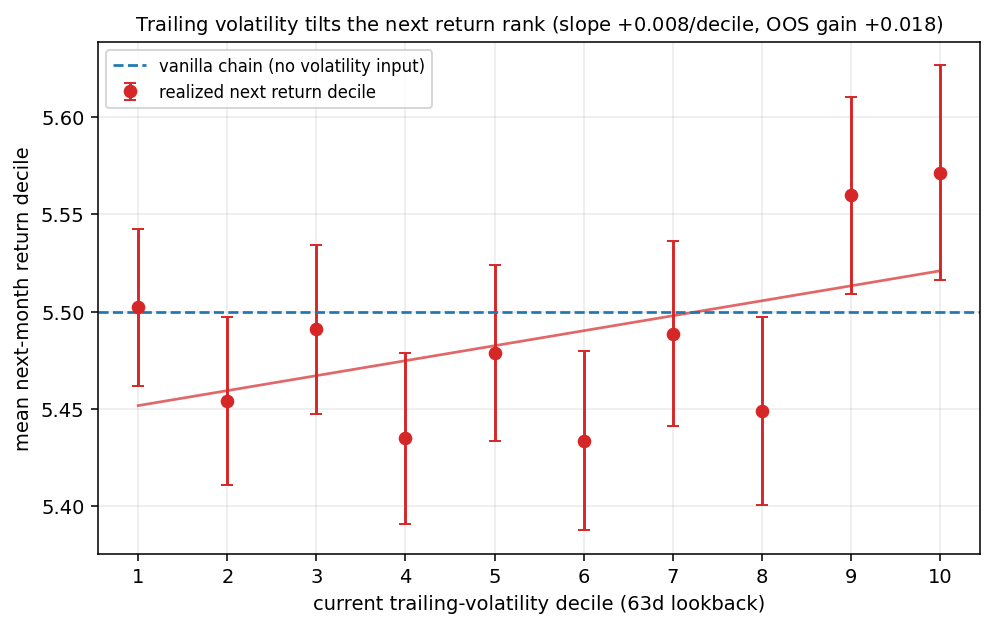}
\caption{Trailing volatility as a predictor of the return rank, the mean
realized next-month return decile against the current trailing-volatility
decile (sixty-three-day lookback), with one standard error. The relation
rises with volatility, while the vanilla chain, ignoring volatility,
predicts the centre for all. The out-of-sample predictive log-likelihood
gain over the vanilla chain is $0.018$ nats per step.}
\label{fig:returnvol}
\end{figure}

The signal is a risk-return tilt, modest in the mean and clearest at
the extremes (Fig.~\ref{fig:returnvol}). Sorted by trailing volatility,
the mean next-month return decile rises with the volatility decile,
where the vanilla chain, blind to volatility, predicts the centre for
all. The effect is small, a fifth of a decile across the range,
consistent with a return rank that stays close to unforecastable. That
trailing volatility is nonetheless its single most informative predictor
is the risk-return balance of classical portfolio theory showing through
the ranking.

The mirror question, whether trailing mean return predicts the
volatility rank, has a different answer, and the asymmetry is
instructive. The coupling itself is present, and strongly. Sorted by
trailing return, the next-month volatility decile is highest for the
largest losers, lowest in the middle, and elevated again for the largest
winners, so volatility follows the magnitude of returns with the
leverage asymmetry that puts the losers above the winners. The
one-month gap between the extreme return deciles, three quarters of a
volatility decile, is larger than any tilt the reverse direction
produces.

Yet as a predictor of the volatility rank trailing return adds nothing.
On its own it does not beat the vanilla chain out of sample, a gain of
$0.001$, and tuned on the validation period it is negative at every
lookback. Added to the established volatility set, it changes the
out-of-sample log-likelihood by less than the validation noise. The
reason is that the volatility chain is already well forecast, at $0.118$
nats per step, by its own persistence and the long-memory volatility
predictors, which encode the very return-magnitude state that trailing
return would report. The same return-volatility coupling that is the
single most valuable handle on the near-unforecastable return chain is
redundant for the volatility chain, which forecasts itself. Information
couples both ways between the two chains, but its predictive value flows
only one way.

\begin{figure}[H]
\centering
\includegraphics[width=0.9\textwidth]{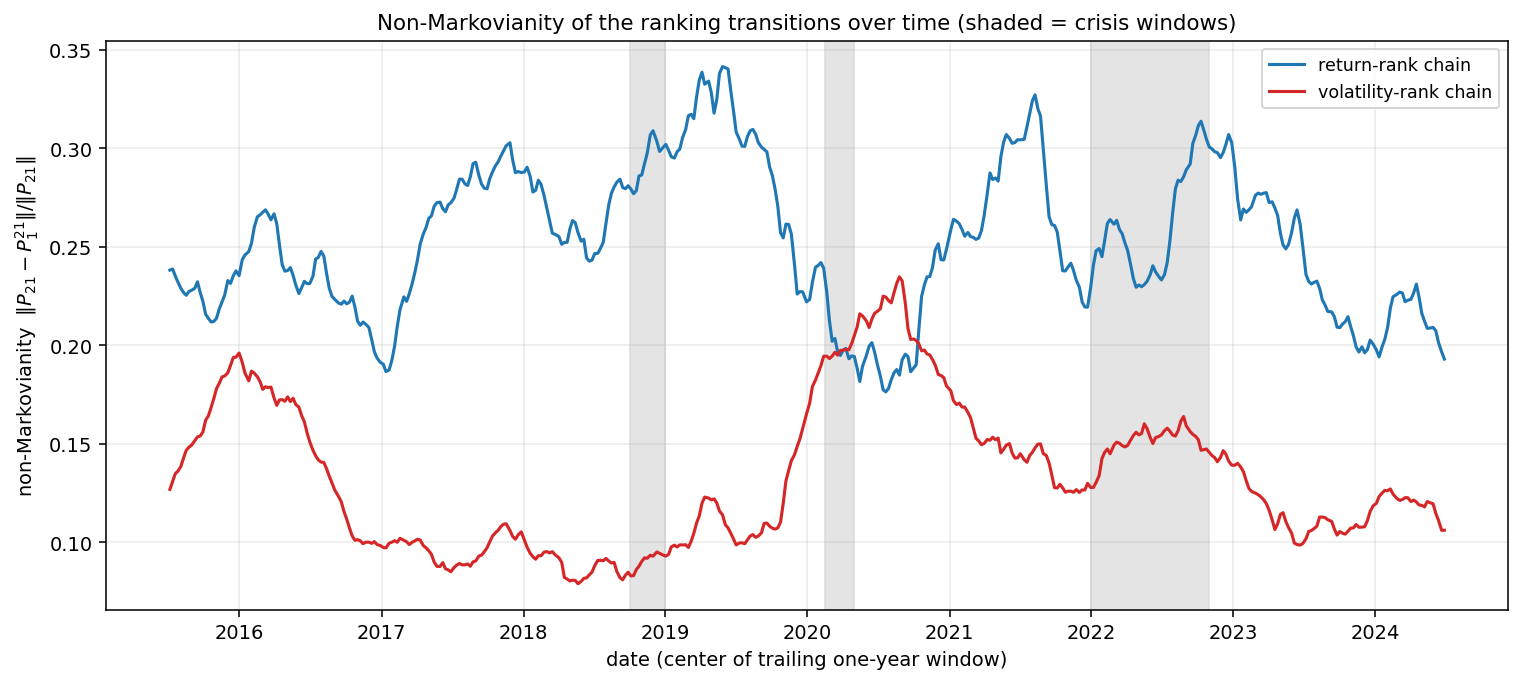}
\caption{Non-Markovianity of the ranking transitions on a one-year
rolling window, the normalized Frobenius distance
$\|P_{21}-P_1^{21}\|/\|P_{21}\|$ between the monthly transition matrix
and the twenty-one-day power of the one-day transition matrix. The
volatility chain (red) sits near $0.10$ in calm years and doubles around
the Covid crash and the 2022 bear (shaded), where its memory is
strongest. The return chain (blue) sits higher and noisier, its ranking
window resonating with the monthly horizon. A first-order Markov chain
is a good description in ordinary times and a weaker one at crisis
peaks.}
\label{fig:markov}
\end{figure}

The one-step test treats each transition on its own, but the monthly
step rests on an assumption worth checking, that the ranking is Markov. A
first-order daily chain would have its monthly transition matrix equal
the one-day matrix raised to the twenty-one-day power. The heatmaps show
what happens instead (Fig.~\ref{fig:heatmaps}). The one-day matrix is
sharply diagonal for both chains. By one month the return matrix has
dispersed to nearly uniform rows, its rank forgotten, while the
volatility matrix keeps bright corners where the extreme deciles persist.
In each, the direct monthly matrix differs from the daily power, and that
difference, the normalized Frobenius distance, is the non-Markovianity.
On a one-year rolling window it is far from constant
(Fig.~\ref{fig:markov}): for the volatility chain it sits near $0.10$ in
calm years and doubles, to $0.23$, around the Covid crash, rising again
through the 2022 bear, so the ranking's memory is strongest when
volatility clusters hardest.

The return chain sits higher, near $0.24$, which resolves an apparent
tension with the OMD study \cite{halperin2026omdstocks} and its
above-random return-rank stay probabilities. Those are daily. The
twenty-one-day return window barely shifts from one day to the next, so
the daily rank is strongly persistent, its stay probabilities running
from $0.4$ to $0.8$ (Fig.~\ref{fig:heatmaps}, left). Over a month the
window turns over completely, the stay probabilities collapse to near
$0.10$, and the daily persistence does not compound to the monthly
transition. The return rank that is strongly persistent one day ahead is
near-random one month ahead, the rebalancing horizon. The volatility
chain is the opposite: its monthly stay probabilities stay from $0.33$ to
$0.40$, so its persistence survives the horizon, and there the
non-Markovianity, and the memory the covariates exploit, are real and
peak with the crises.

One might ask whether the one-month ranking window is what makes the
return chain look random, and whether a longer window would restore
forecastability. A longer window restores persistence but not
forecastability. Ranking on the trailing return over one, three, six,
ten, and twelve months, the extreme-decile stay probability climbs from
$0.12$ to $0.72$ and the one-step rank error falls from $2.5$ to $1.0$
deciles, so the transition matrix does grow persistent. That persistence
is mechanical. A ten-month window shifts by only a tenth each month, so
the rank barely moves and predicts its own next value trivially, the same
overlapping-window effect the daily twenty-one-day rank shows. The
covariates add nothing to the point forecast at any window, the rank
error with and without them agreeing to a hundredth of a decile
throughout. And the quantity a monthly book earns, the next-month return
of the top decile minus the bottom, stays small and negative at every
window, weakest at the six-to-ten-month momentum horizons and larger at
the one-month reversal, but never a positive forward edge. Coarsening
the buckets helps as little as lengthening the window. Pairing the ten deciles
into five shrinks the error only because it shrinks the scale; against the
no-skill level, which falls in step, the accuracy is unchanged for the return
chain and slightly lower for the volatility chain, whose finer distinctions the
coarsening discards. Relative return is unforecastable
at the monthly horizon whatever the ranking window or the number of
buckets, and the near-random one-month decile chain shows it plainly.

\subsection{Diagnostics under the regime-aware chains}
\label{sec:regimediag}

\begin{figure}[H]
\centering
\includegraphics[width=0.95\textwidth]{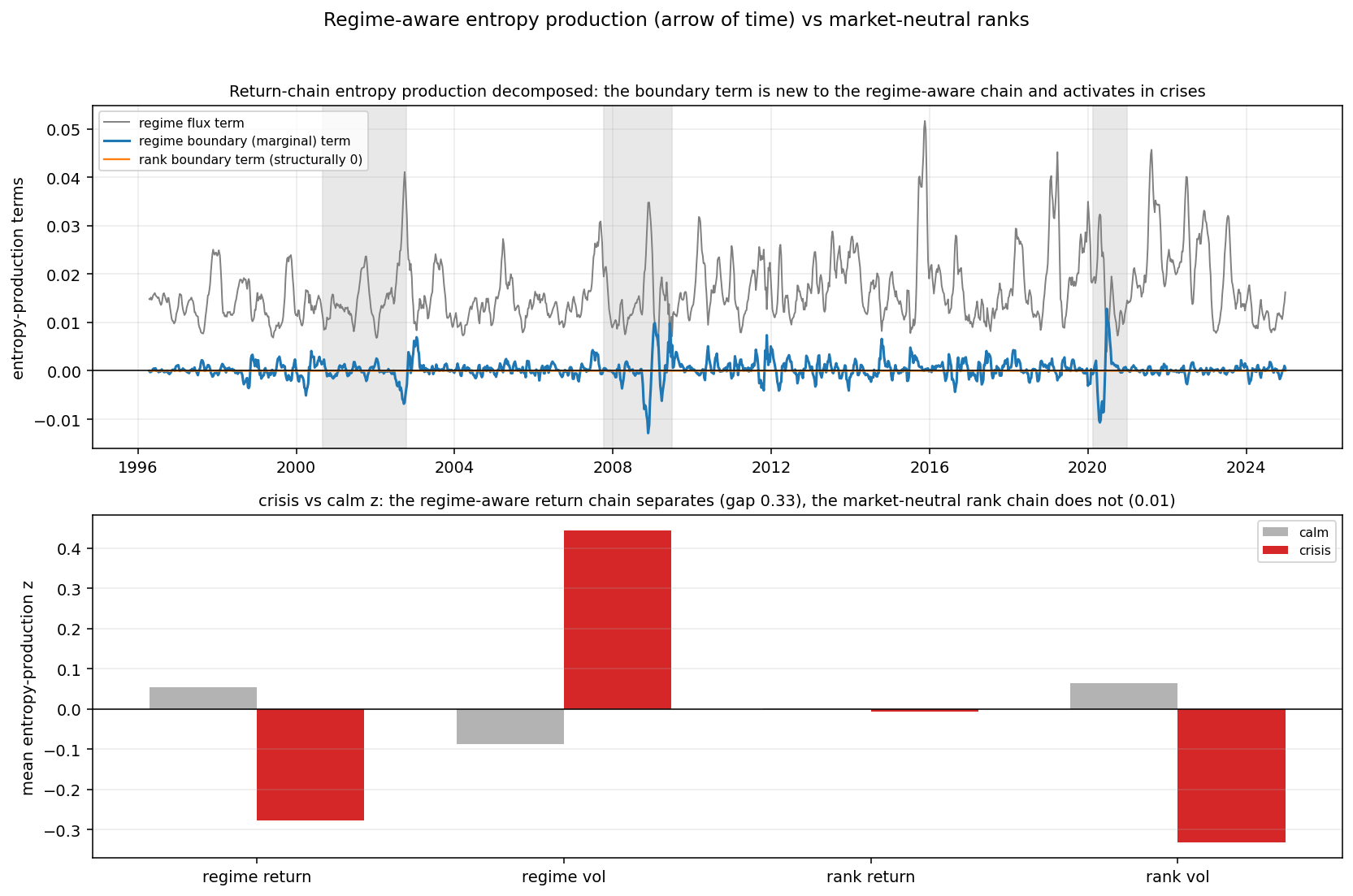}
\caption{Entropy production of the ranking chains, regime-aware versus
market-neutral. Top: the return-chain entropy production split into its flux and
boundary terms. The boundary term is new to the regime-aware chain (blue) and
activates in the 2008 and 2020 crises, while the market-neutral boundary term
(orange) is structurally zero. Bottom: the crisis-versus-calm standardized
entropy production. The regime-aware return chain separates the two regimes and
the market-neutral rank return chain does not.}
\label{fig:epregime}
\end{figure}

The entropy production of Section~\ref{sec:condep} reads the same way on the
regime-aware chains, but now it sees the market regime the ranks had removed. On
the market-neutral return chain it carries no crisis signal, its standardized
value differing between crisis and calm periods by $0.01$, because the boundary
term of Eq.~\eqref{eq:epdecomp} is identically zero and the reshuffling of
relative order is about as irreversible in a crash as in a calm year. On the
regime-aware return chain the same measure separates crisis from calm by $0.33$ of
a standard deviation, and the regime-aware volatility chain's arrow of time rises
in crises by $0.53$. The boundary term, structurally absent from the ranks, is
small in ordinary times and lights up in the 2008 and 2020 crashes, when the whole
cross section drifts down together and the class marginal moves hardest
(Figure~\ref{fig:epregime}).

Keeping the market mode also binds the arrow of time to the correlation geometry
the distance matrix measures. Read from the regime-aware chain, the
entropy-production score moves with the spectral concentration of the
return-correlation matrix, its Spearman correlation with the rank-decay exponent
of the distance-matrix spectrum rising to $0.31$ and with the month-on-month
eigenvector rotation of the correlation matrix to $0.42$, from near zero under the
market-neutral ranks. The irreversibility of the ranking dynamics and the
concentration of the cross-sectional geometry are, under the regime-aware reading,
one coupled signature of market stress rather than two independent ones. The
transfer entropy of Section~\ref{sec:te} carries over in the same spirit, and we
return to it in Appendix~\ref{sec:teleaders}, where the leaders read from the
regime-aware chains gain out-of-sample forecast power the market-neutral version
lacked.

\subsection{Information dissipation length}
\label{sec:idl}

A further market-state diagnostic comes from the information dissipation length
introduced by Quax, Kandhai, and Sloot \cite{quax2013} as an early-warning signal
for the 2008 collapse. Their construction orders the units of a system along a
logical distance, measures the mutual information between pairs as a function of
that distance, fits the decay $\mathrm{MI}(d)=a+b\,f^{d}$, and reports the length
as the half-life $\log(1/2)/\log f$ of the decay. A larger length means that
information about one name reaches farther across the market before it dissipates,
the mark of the long-range correlation that precedes a systemic transition.

\begin{figure}[H]
\centering
\includegraphics[width=0.95\textwidth]{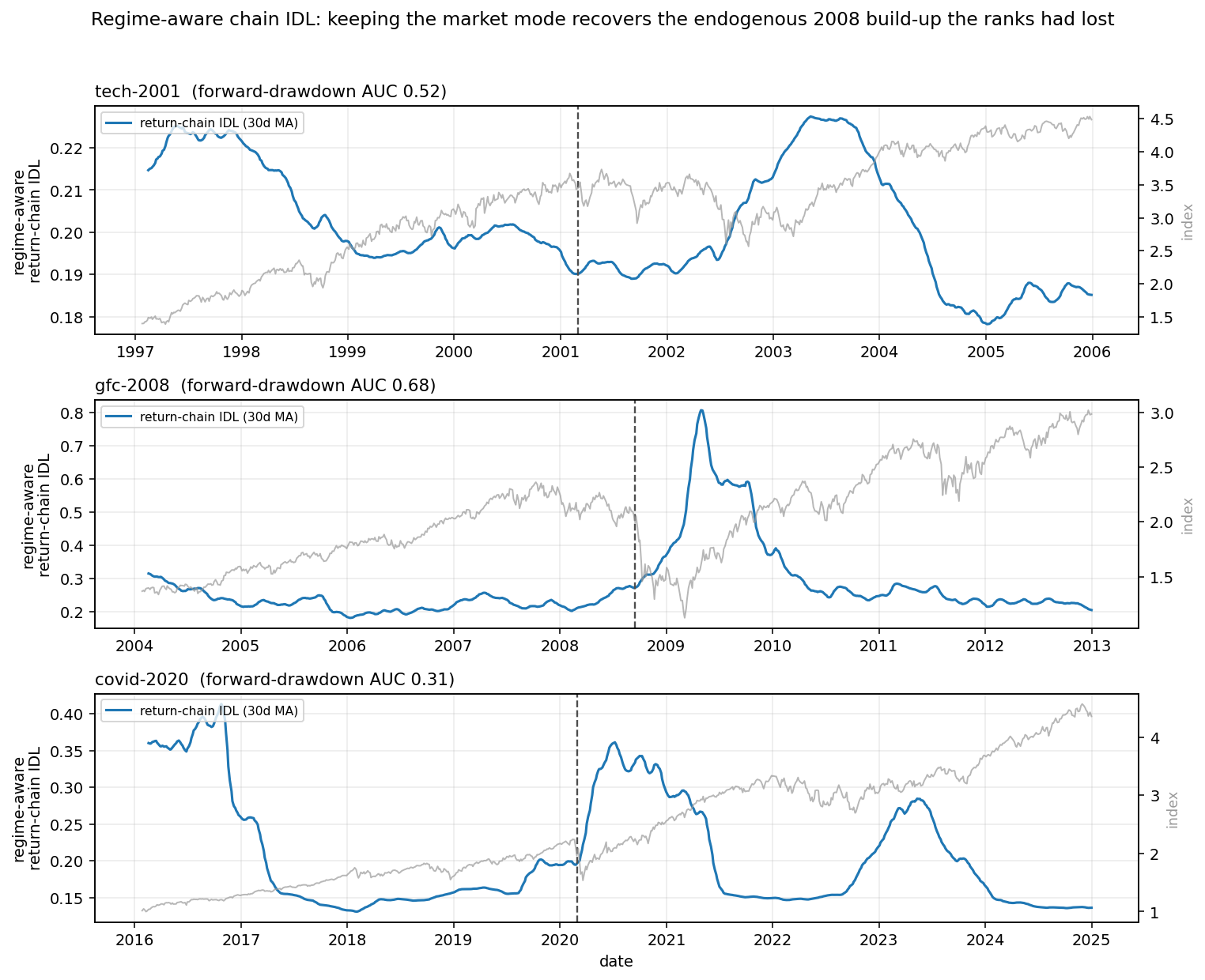}
\caption{Information dissipation length from the regime-aware return chain through
the three crises, with the equal-weight index in grey and the crisis onset dashed.
The length builds up into the endogenous 2008 crisis (area under the ROC curve
$0.68$ against forward drawdowns), does not build up before the exogenous 2020
shock ($0.31$), and sits at chance before the dispersed 2001 unwind ($0.52$).
Keeping the market mode recovers the endogenous build-up the market-neutral ranks
had lost.}
\label{fig:idlregime}
\end{figure}

A market of stocks carries no natural distance axis, but the distance matrix
$M(t)=\arccos G(t)$ supplies one, the geodesic distance between names on the
correlation manifold. We take the mutual information from the ranking chains
themselves, an exact discrete quantity on the decile states, with the axis fixed
from a baseline calm period so the measure does not collapse into an identity. On
the market-neutral chains the length is weak, near chance as a forward-drawdown
alarm, because the ranks remove the common market mode where the long-range
correlation lives. On the regime-aware chains it recovers the signal. Its
forward-drawdown skill on the endogenous 2008 crisis rises to an area under the
ROC curve of $0.68$, above the market-neutral $0.56$, and the length climbs into
and through the 2008 drawdown. The exogenous 2020 shock leaves no such build-up,
an area under the curve of $0.31$, and the dispersed 2001 unwind sits at chance,
so the length separates the crisis that builds inside the market from the one that
arrives from outside (Figure~\ref{fig:idlregime}).

The entropy production, the arrow of time, and the information dissipation length,
read from the regime-aware chains, are all market-state signals that the
market-neutral construction could not carry. They set the stage for the portfolio
signals of the next sections, where the question turns from reading the regime to
acting on it.

\section{Portfolio construction and results}
\label{sec:portfolio}

The forecasts drive a monthly long-short. The performance leg is the
return chain ranked on a six-month window, which follows momentum rather
than the one-month reversal, and each name's score is the blend $S$ of
Eq.~\eqref{eq:combined}, its calibrated probability of the top return
decile weighed against its probability of the calmest volatility decile,
read from the covariate-conditioned transitions with the market and
distance-matrix predictors. The
truncation rule is a top-and-bottom cut, long the $K$ names with the
highest score and short the $K$ lowest, equal weight and dollar neutral.
A long-only sleeve holds instead the $2K$ highest-scoring names, fully
invested. Both are rebalanced monthly, net of five basis points a name
traded, ten for comparison.
This score is already a soft group-relative rule: it is high only near the
top, since the chain rarely reaches the top two deciles from the middle, and it
drifts gradually, so the no-trade band holds turnover down. Imposing the decile
boundary explicitly, ranking only within the current top and bottom groups,
raises turnover from $1.0$ to $1.6$ without adding forward-return signal and
lowers the out-of-sample Sharpe, so the smooth score is kept.

Three hyperparameters are chosen on the validation period. The book size
$K$ ranges over five to twenty. The volatility weight $\lambda$ of
Eq.~\eqref{eq:combined} is tuned and comes out at zero, which means the
volatility score drops out and the book is selected on the return
probability alone. Rewarding the calmest decile does not help and
slightly hurts. The momentum winners of these years were the volatile
names, so tilting the score toward calm names dilutes the return, and
because those calm names carry little relative-return signal the tilt
raises rather than lowers the realized volatility of the book. A no-trade
tolerance then controls turnover, a held name being replaced only when a
challenger's score exceeds it by more than the tolerance, so small
probability differences do not trigger a trade.

\begin{figure}[htbp]
\centering
\includegraphics[width=0.82\textwidth]{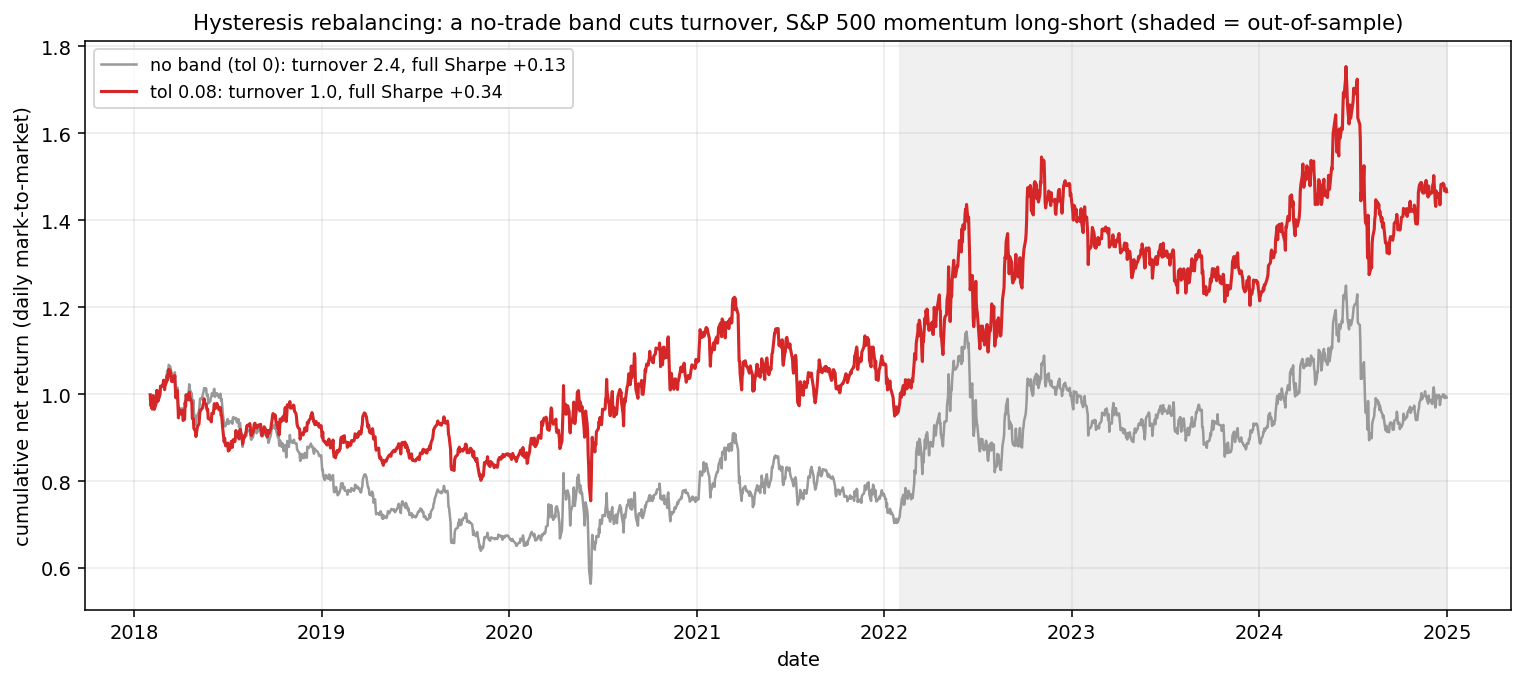}
\caption{Hysteresis rebalancing. Cumulative net return of the monthly
momentum long-short traded every month with no band (grey, turnover
$2.4$, Sharpe $0.13$) and with a no-trade band (red, turnover $0.9$,
Sharpe $0.34$), the band trading only when a challenger beats a held
name by more than a tolerance. The dotted line marks the end of the
validation period.}
\label{fig:hysteresis}
\end{figure}

The tolerance is decisive. Selecting by a predicted probability
reshuffles about sixty percent of the book each month, a turnover of
$2.4$ against a maximum of four, so at five basis points the cost drag is
twelve basis points a month, twice that at ten. The book has a real gross
edge, a Sharpe of $0.17$ and about ten percent over the span, but at the
churny extreme the drag erodes it to roughly breakeven at five basis points
and to a net loss at ten. The no-trade band fixes it. Any tolerance from
five to twenty-five percent cuts the turnover to about $0.9$ and the cost to
four or five basis points a month, nine at ten, and lifts the net
full-period Sharpe to between $0.23$ and $0.34$
(Figure~\ref{fig:hysteresis}). Within that range the exact value matters
little, and we set the band at $0.08$, as the combined book of
Section~\ref{sec:portfolio} does, at which the book returns about
forty-seven percent at a full-period Sharpe of $0.34$ and an out-of-sample
Sharpe of $0.63$.

The rebalancing frequency is a separate choice, and the monthly one is not
arbitrary. We tested rebalancing the book daily instead, updating the momentum and
volatility rankings every day and trading through the same no-trade band, with the
band re-tuned on the validation period. Daily rebalancing does not help. At every
band it earns a lower validation Sharpe than the monthly book, and for both
frequencies the Sharpe rises as the band widens, so trading less is better rather
than worse. The signals move slowly. The momentum ordering over a hundred and
twenty-six days and the volatility ordering over twenty-one barely change from one
day to the next, so a daily update captures no opportunity a monthly one misses
and only multiplies the turnover, two to four times that of the monthly book.
Monthly rebalancing already extracts what the slow signal offers, and we keep it.

The band choice is also stable across the regimes inside the validation
period. Splitting that period into its calm and its stressed part, the bull
market of 2019 and the Covid crash and recovery of 2020, the long-only sleeve
earns a Sharpe near $0.7$ in the calm year and near $0.9$ in the crisis, so it
carries signal in both. The two years pull the band in opposite directions, the
calm one toward a wider band that holds through the trend and the crisis toward
a tighter one that reacts. The compromise the full window selects, $0.08$, is
the one the stressed year prefers. Tuning the band on either year alone leaves
the out-of-sample combined Sharpe between $1.00$ and $1.06$, the range the whole
band grid spans in any case, so the headline does not turn on which regime sets
the band.

\subsection{Hyperparameter calibration by a regime-split search}
\label{sec:optuna}

We checked the hand grid of Table~\ref{tab:hyper} with the tree-structured
sampler of Optuna \cite{optuna2019}. An unconstrained search over all the
hyperparameters, maximizing the validation Sharpe, overfits: it lifts the
validation Sharpe from $0.49$ to $0.57$ and lowers the out-of-sample from
$1.06$ to $0.93$, the over-fitting the diversification tilt of
Section~\ref{sec:portfolio} already warned of. The instructive case is the book
size, and a sharper search fixes the blend and the $\lambda$ timing and
optimizes only the size and the no-trade band, separately for the long-only and
the long-short sleeve and separately on the two regimes inside the validation
window, the benign 2018--2019 and the Covid 2020--2021
(Table~\ref{tab:optuna_sleeves}).

Maximizing the Sharpe over the book size is ill-posed. The Sharpe improves
monotonically as the book grows and diversifies, so a larger book keeps looking
better until it becomes the market. The drawdown falls with diversification too,
only reinforcing the pull. Optuna's long-only optimum is therefore pinned to the
top of the search range, about $K=25$, not an interior peak, while the
long-short prefers the bottom, about $K=9$, a genuinely concentrated book the
two regimes agree on. The long-only preference is not identified by the data: on
the validation period the diversified book's Sharpe is nearly flat in the size
(Table~\ref{tab:optuna_ksweep}), and the apparent gain from a larger long-only
book shows up only on the 2025--2026 window, absent on the first test that
carries the 2022 bear. Because the net exposure is $\theta$ at every size, the
gain is market breadth, not added exposure and not selection alpha, so the
information ratio, the active return over the market, does not reward it.

No performance objective on the calibration data thus fixes the long-only size,
and the one that honors the strategy's purpose, the active return, is flat in
it. The long-short's preference for a smaller book is real but modest, and we
keep a single $K=15$ for both sleeves for parsimony, one book-size
hyperparameter consistent with the OMD-Stocks study, rather than chase a
regime-specific maximum. The lesson is the paper's discipline once more. A
modern optimizer run on the validation Sharpe overfits, and the size that
maximizes the Sharpe out of sample is diversification bought on one favorable
window, which both the held-out bear and the active objective reject.

\begin{table}[htbp]
\centering
\small
\caption{Per-sleeve optimum of the book size $K$ and the no-trade band, found by
Optuna on each validation regime separately, daily mark to market. Each row
optimizes one sleeve on one regime and reads its Sharpe on both regimes and on
the held-out 2022--2024. The long-short prefers a small, concentrated book the
two regimes agree on; the long-only optimum sits at the top of the search range
(the ill-posed maximization discussed in the text), and the band is
regime-conflicting.}
\label{tab:optuna_sleeves}
\begin{tabular}{llccccc}
\toprule
sleeve & tuned on & $K$ & band & benign & Covid & OOS \\
\midrule
long-only  & benign 2018--19 & $24$ & $0.35$ & $+0.23$ & $+0.77$ & $+0.89$ \\
long-only  & Covid 2020--21  & $25$ & $0.03$ & $+0.21$ & $+0.94$ & $+0.96$ \\
long-short & benign 2018--19 & $9$  & $0.07$ & $-0.20$ & $+0.25$ & $+0.73$ \\
long-short & Covid 2020--21  & $8$  & $0.21$ & $-0.51$ & $+0.59$ & $+0.72$ \\
\bottomrule
\end{tabular}
\end{table}

\begin{table}[htbp]
\centering
\small
\caption{The diversified book's daily-marked Sharpe against the long-only book
size $K$, the long-short fixed at $K=9$ and the blend and band unchanged, on the
validation period and the two out-of-sample tests. The size barely moves the
Sharpe on the validation period and on the first test, and matters only on the
2025--2026 window, a bull that rewards a broader book. The net exposure is
$\theta$ at every size, so the rise there is market breadth rather than exposure
or selection.}
\label{tab:optuna_ksweep}
\begin{tabular}{cccc}
\toprule
long-only $K$ & validation & first test & second test \\
\midrule
$10$ & $0.52$ & $1.10$ & $1.70$ \\
$15$ & $0.53$ & $1.09$ & $1.55$ \\
$20$ & $0.54$ & $1.07$ & $1.60$ \\
$25$ & $0.55$ & $1.08$ & $1.68$ \\
$30$ & $0.58$ & $1.06$ & $1.73$ \\
$40$ & $0.58$ & $1.07$ & $1.78$ \\
\bottomrule
\end{tabular}
\end{table}

\subsection{Regime non-stationarity and the market anchor}
\label{sec:nonstationarity}

Machine learning usually assumes the training, validation, and test sets are
drawn from one stationary distribution, so a choice that validates well
generalizes. Financial data violate this. The data-generating mechanism is the
market regime, and it differs across periods, so a hyperparameter or a model
tuned on one period need not carry to the next, as the calibration above already
showed. To see the structure behind that, we read the daily-marked Sharpe of
each component of the book across four regimes, the benign 2018--2019, the Covid
crash and recovery of 2020--2021, the 2022--2024 bear and rebound, and the
2025--2026 bull (Table~\ref{tab:regime_stability}).

The stable component is the market itself. Its Sharpe varies least across the
four, a standard deviation of $0.16$, the equity premium the most persistent
signal in the data. The style component is the opposite. The market-neutral
momentum long-short, often taken to be regime-insensitive because it carries no
market exposure, is in fact the least stable, a standard deviation of $0.70$,
swinging from $-0.54$ in the momentum-unfriendly 2018--2019 to $+1.41$ in the
2025--2026 trend, the momentum-crash pattern of Daniel and Moskowitz
\cite{danielmoskowitz2016}. Removing the market does not remove the
non-stationarity. It strips out the stable factor and leaves the volatile one.

The same logic explains why hedging the market out of the long-only sleeve makes
it worse. Subtracting the market return from the long-only sleeve, to isolate its
selection, raises its regime dependence from a standard deviation of $0.37$ to
$0.57$, because the market beta was anchoring it. The net-long sleeve inherits
the market's stability, the bare active return does not. For a strategy whose
alpha is a volatile style factor, the market exposure is a stabilizer, not a
risk. This is why the book blends the net-long sleeve with the neutral one
through the regime-timed weight rather than running a pure market-neutral book.
The market premium anchors the combination and the timing modulates the exposure.

The sound response to this non-stationarity is not to neutralize the market but
to lean on what is invariant across regimes and to select for the worst regime
rather than the average. The market premium is the invariant anchor, kept and
timed rather than hedged away. The hyperparameters and the timing are best chosen
for their weakest regime, not their mean or a single window, since a choice that
maximizes one regime overfits the next. And performance is read honestly as a
range across regimes, the weakest one the real risk, not as a single number.

This reading motivates the construction, and it also shows why the construction
is hard to beat. The strategy's poor regime is the momentum-unfriendly
2018--2019, where the long-short crashes and the long-only is flat. A price-only
defence against momentum crashes, scaling the long-short by the ratio of a target
volatility to its own trailing volatility after Barroso and Santa-Clara
\cite{barroso2015}, cuts the sleeve exactly when it is crash-prone and does lift
the 2018--2019 Sharpe, from $-0.33$ to $-0.14$. But it lowers the out-of-sample
Sharpe, from $1.06$ to $0.83$, because the same cut removes the long-short in the
2022 bear, where it is the book's protection rather than its risk. Optimizing the
mixture and the protection together for the worst regime's information ratio, the
well-posed objective, confirms the pattern, the worst regime improving and the
out-of-sample not. The two regimes want opposite things from the long-short, so
no single rule serves both, and the reactive blend that de-risks in down markets
without cutting the sleeve outright is the better all-regime compromise. What the
framework has not yet spent is the trading-volume information of
Appendix~\ref{sec:volume}, real out of sample though too fast to trade on its own,
a candidate stress signal for the timing weight. Prices, size, volume, and the
three matrices are not exhausted, but the gains left are guarded by the
non-stationarity itself.

\begin{table}[htbp]
\centering
\small
\caption{Daily-marked Sharpe of each component of the book across four market
regimes, and its standard deviation across them. The market premium is the most
stable component and the market-neutral momentum long-short the least;
subtracting the market from the long-only sleeve raises its regime dependence
rather than lowering it, since the market exposure was anchoring it. The net-long
sleeve inherits the market's stability.}
\label{tab:regime_stability}
\begin{tabular}{lccccc}
\toprule
component & 2018--19 & 2020--21 & 2022--24 & 2025--26 & std \\
\midrule
market                     & $0.70$  & $0.86$ & $0.78$ & $1.14$ & $0.16$ \\
long-only, net-long        & $0.03$  & $0.85$ & $0.92$ & $0.92$ & $0.37$ \\
combined book              & $-0.33$ & $1.04$ & $1.06$ & $1.32$ & $0.64$ \\
long-short, market-neutral & $-0.54$ & $0.25$ & $0.61$ & $1.41$ & $0.70$ \\
long-only minus market     & $-1.11$ & $0.09$ & $0.12$ & $0.32$ & $0.57$ \\
\bottomrule
\end{tabular}
\end{table}

\begin{figure}[htbp]
\centering
\includegraphics[width=0.85\textwidth]{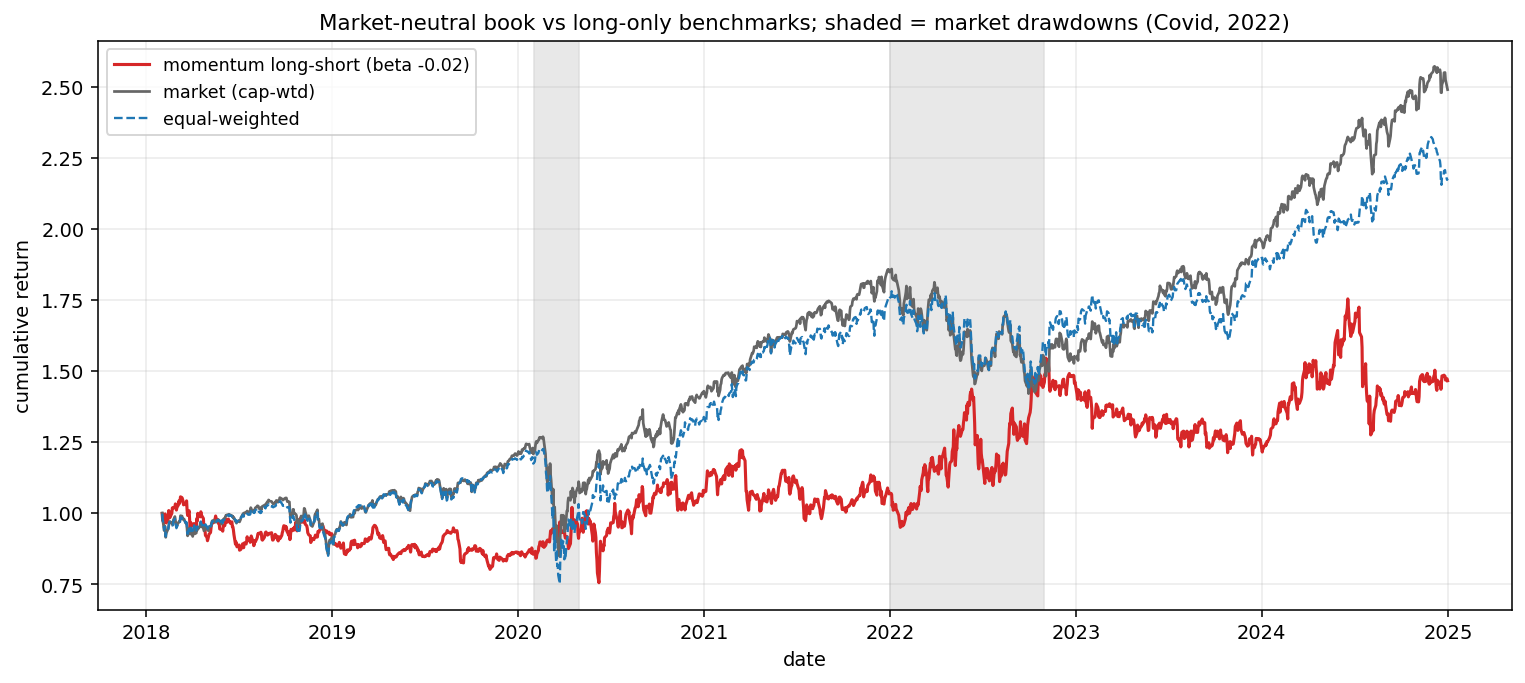}
\caption{The low-turnover momentum long-short (red) against the
capitalization-weighted market (grey) and the equal-weighted portfolio
(blue dashed), cumulative return. Beta $-0.02$ to the market, it gains
in both regimes and rises most in the shaded Covid and 2022 drawdowns,
where the long-only benchmarks fall.}
\label{fig:downturns}
\end{figure}

The low-turnover book is market-neutral, which makes it a diversifier
rather than a market substitute. Its beta to the market is $-0.02$ and
its correlation $-0.02$ (Figure~\ref{fig:downturns}), and it gains in
both regimes, about $0.35\%$ a month when the market rises and about
$1.13\%$ a month in the twenty-seven down-market months, when the market
itself loses $4.5\%$. The two drawdowns are the strongest case. Through
the Covid crash of early 2020 it gains about $3.0\%$ a month and is
positive every month, and through the 2022 bear market about $3.1\%$ a
month, in both cases while the market falls, with a Sharpe in down
months alone of $+0.83$ against the market's $-5.3$. The exception is a
sharp momentum reversal such as the late-2018 selloff, which hurts it.
Over the full span the book earns a Sharpe of $0.34$, below the
long-only equal-weighted and market Sharpes of $0.67$ and $0.77$, but it
carries no market exposure and gains most where they lose.

The volatility chain, which the selection score leaves aside at
$\lambda=0$, still has a portfolio role, on the long side. A
market-timed volatility weight has no place in the long-short, which is
already market-neutral and gains in downturns, but it belongs on the
long-only sleeve, the book that carries the market and loses when it
falls. We time $\lambda$ on the long-only score at zero when the
trailing three-month market return is positive, so the long book chases
performance, and raised to $0.75$ when it is negative, so the book leans
toward the calmest volatility decile. Chosen this way on the validation
period, where it lifts the long-only Sharpe from $0.52$ to $0.62$, the
defensive tilt cuts the book's worst drawdown from twenty-five to
eighteen percent and its average down-market month from $-4.4\%$ to
$-3.8\%$, at no cost to the full-sample risk-adjusted return. The same
tilt on the long-short is rejected on the validation period at every
setting, since a market-neutral book needs no protection from the
market.

\begin{figure}[htbp]
\centering
\includegraphics[width=0.9\textwidth]{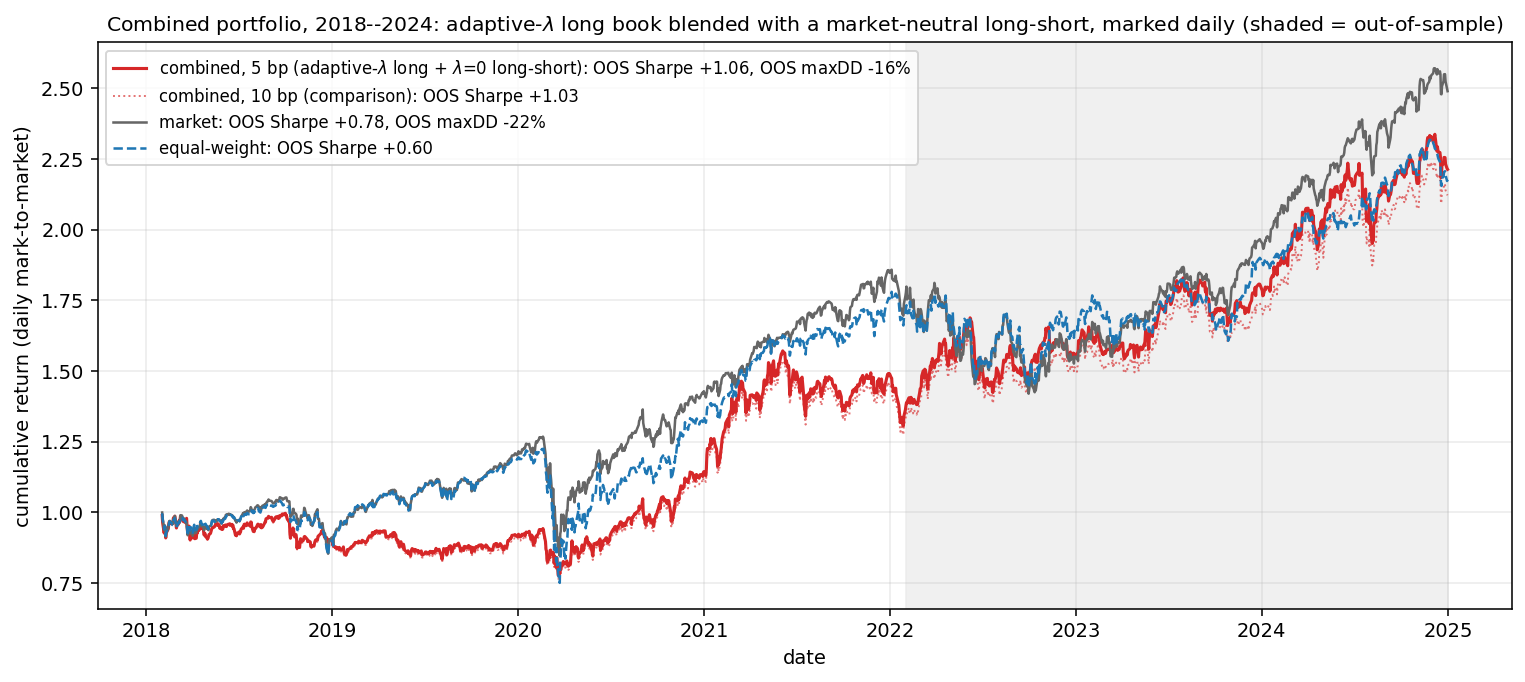}
\caption{Combined portfolio, 2018--2024, cumulative return net of five basis
points a name traded (red solid), with the ten-basis-point book for comparison
(red dotted) and the out-of-sample period (2022--2024) shaded. The book blends
the adaptive-$\lambda$ long-only sleeve with the $\lambda=0$ market-neutral
long-short at a state-dependent weight $\theta$, one when the trailing
market has risen and $0.4$ when it has fallen, chosen on the validation
period. Defensive by construction, it lags the market (grey) and the
equal-weighted portfolio (blue dashed) through the 2018--2021 bull, but
leads them out of sample, holding flat through the 2022 bear while they
fall. Metrics in Table~\ref{tab:combined}.}
\label{fig:combined}
\end{figure}

These two sleeves make the tradeable book. The long-only sleeve, with
its market-timed $\lambda$, carries the market and turns defensive in
downturns. The long-short sleeve, at $\lambda=0$, is market-neutral and
protective. We hold a state-dependent fraction $\theta$ of the book in
the long-only sleeve and $1-\theta$ in the long-short, with $\theta$ set
by the trailing three-month market return, its full weight on the long
sleeve when the market has been rising and a reduced weight when it has
been falling. The validation period selects $\theta=1$ in up markets and
$\theta=0.4$ in down, a balanced book that leans long in recoveries and
leans on the market-neutral sleeve for protection rather than switching
all or nothing. Table~\ref{tab:combined} splits the record by period. In
the 2018--2021 bull, the period on which the blend weights are chosen,
the balanced book lags the long-only benchmarks, a Sharpe of $0.49$
against the market's $0.78$, the cost of carrying a market-neutral hedge
while the mega-caps ran. Out of sample over 2022--2024 the picture
reverses. The combined book returns sixty-two percent against the market's
forty-one and the equal weight's twenty-eight, at a Sharpe of $1.06$
against $0.78$ and $0.60$, at a volatility close to the market's, and a
worst drawdown of sixteen percent against twenty-two (Figure~\ref{fig:combined}).
It holds flat through the 2022 bear while the market falls a fifth, then
keeps pace through the recovery. The strategy trades bull-market upside
for downturn protection, and the out-of-sample period, a harder market
than the training bull, is where that trade pays. The no-trade band,
decisive for the bare long-short, washes out here: at $0.08$ and at
$0.18$ the combined book's out-of-sample Sharpe is $1.06$ and $1.05$
(Table~\ref{tab:combined}), the two sleeves and the regime
timing carrying the result rather than the tolerance.

\begin{table}[htbp]
\centering
\small
\caption{Combined portfolio against the capitalization-weighted market
and the equal-weighted portfolio, in-sample (2018--2021, the period on
which the blend weights are chosen) and out-of-sample (2022--2024).
Annualized return, annualized volatility, Sharpe ratio, and worst
drawdown, net of five basis points a name traded. All performance is marked to
market daily: the monthly target weights are held and drift with prices, and the
volatility, Sharpe, and drawdown are annualized from the daily equity curve, the
industry-standard footing, while the annualized return is unchanged from the
monthly one. The combined book is shown at the two no-trade bands $0.08$ and
$0.18$; the two agree to within a point on every metric, so the out-of-sample
result does not depend on the tolerance. The
combined book's average gross exposure, its measure of leverage, is $1.12$ in
sample and $1.20$ out of sample, with a net exposure of $0.88$ and $0.79$,
against a gross and net of $1$ for the market and the equal-weighted book.}
\label{tab:combined}
\begin{tabular}{lcccc c cccc}
\toprule
& \multicolumn{4}{c}{in-sample (2018--2021)} && \multicolumn{4}{c}{out-of-sample (2022--2024)} \\
\cmidrule{2-5}\cmidrule{7-10}
book & ann.\ ret & ann.\ vol & Sharpe & max DD && ann.\ ret & ann.\ vol & Sharpe & max DD \\
\midrule
combined, band $0.18$ & $7.2\%$  & $19.7\%$ & $0.45$ & $-24\%$ && $17.9\%$ & $17.1\%$ & $1.05$ & $-15\%$ \\
combined, band $0.08$ & $8.1\%$  & $19.7\%$ & $0.49$ & $-22\%$ && $18.1\%$ & $17.0\%$ & $1.06$ & $-16\%$ \\
market       & $15.3\%$ & $21.2\%$ & $0.78$ & $-33\%$ && $12.6\%$ & $17.0\%$ & $0.78$ & $-22\%$ \\
equal-weight & $14.3\%$ & $22.4\%$ & $0.71$ & $-39\%$ && $8.7\%$  & $16.1\%$ & $0.60$ & $-19\%$ \\
\bottomrule
\end{tabular}
\end{table}

\begin{figure}[htbp]
\centering
\includegraphics[width=0.98\textwidth]{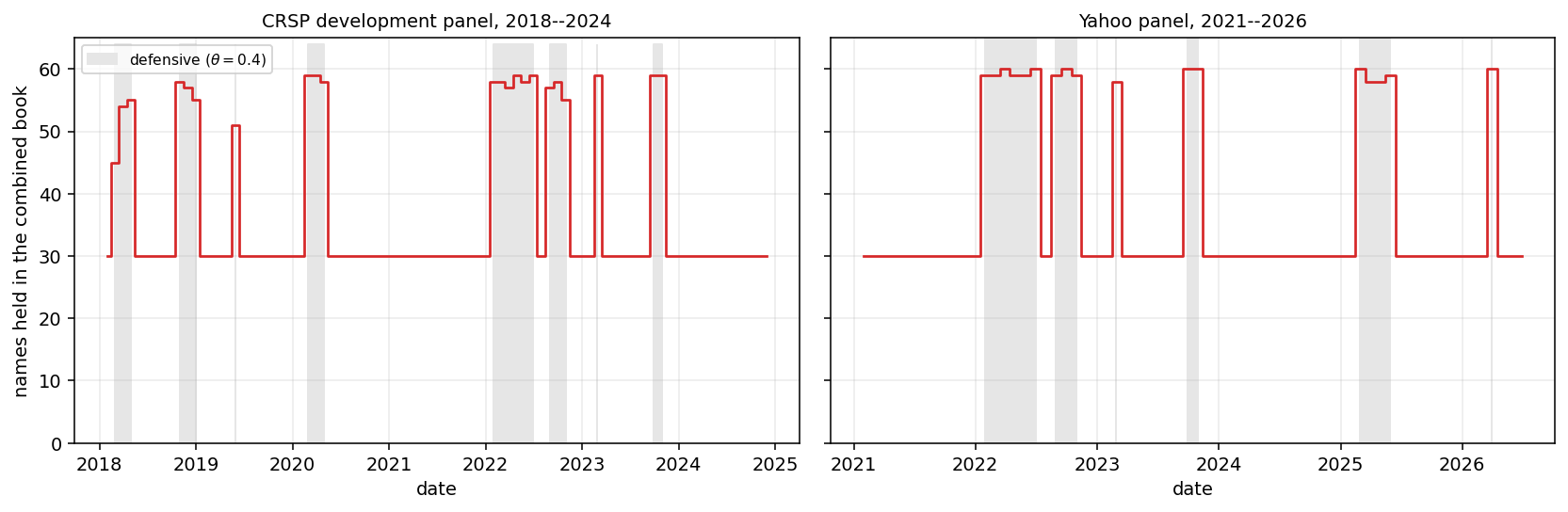}
\caption{Number of names held in the combined book over time, on the CRSP
first test panel (left) and the second test panel (right), with the defensive
periods ($\theta=0.4$) shaded. The book holds the $30$ long-only names in
rising markets and steps up to near sixty when the market-neutral long-short
sleeve switches on in downturns, net of the names the two sleeves share. The
size is piecewise constant at the two regime levels rather than fluctuating.}
\label{fig:holdings}
\end{figure}

The book's size is not constant, but it takes only two values. In a rising
market $\theta$ is one and the book is the $2K=30$ long-only names alone. When
the trailing market turns down $\theta$ drops to $0.4$ and the market-neutral
long-short switches on, adding its $K$ long and $K$ short legs net of the few
they share with the long-only book, so the count steps up to near sixty
(Figure~\ref{fig:holdings}). The book holds thirty names in rising markets and
about fifty-seven to fifty-nine in the defensive periods, a mean near
thirty-seven over each sample, and it switches between the two levels with the
regime rather than drifting. The defensive periods are exactly where the extra
market-neutral names, and the protection they carry, are wanted.

The annualized returns, near $20\%$ on the development window and near $50\%$ on
the recent one, are best read through the Sharpe and against the market of the
same window, which itself returns about $12.5\%$ and about $20\%$ a year, so much
of the level is simply the market the window contains. Whether the excess over
the market is bought with leverage is the operative question, and it is not.

The exposure follows the same two regime levels, and the book does not lever the
market to reach its returns. Leverage is measured by the gross exposure, the sum
of the absolute position weights, the total capital deployed long and short per
unit of equity. The net exposure, their signed sum, is a separate quantity, the
directional bet on the market rather than the leverage. A fully invested
long-only book has both equal to one, and a dollar-neutral long-short is levered
two to one at the gross while its net is zero. In rising markets our book is the
fully invested long-only sleeve, a gross and a net of one. When the
dollar-neutral long-short switches on in a downturn the gross rises to
$2-\theta=1.6$ while the net market exposure falls to $\theta=0.4$, so the book
de-risks its directional exposure exactly when the market drops. Averaged over
the sample the gross is about $1.15$ and the net about $0.85$, and
Tables~\ref{tab:combined} and~\ref{tab:freshoos} give both by test period. The
leverage is conservative by professional standards. A fully invested long-only
fund runs a gross of one, a long-short equity fund commonly runs a gross of
$1.5$ to $2$, and a $130/30$ fund runs $1.6$. Our gross therefore sits below the
long-short norm, reaches a $130/30$ level only in its defensive periods, and
stays well inside the Regulation-T $2\times$ margin bound throughout. The excess
gross is the self-funded long-short overlay rather than borrowed capital, so the
outperformance is a higher Sharpe at a comparable or lower volatility and a
lower average market exposure, which a leverage of the market return, scaling
return and risk together, could not produce.

The blend weight admits other constructions. In place of the hard two-level
step on the sign of the trailing market return, $\theta$ can be a smooth
sigmoid of a standardized driver, and the driver can be a distance-matrix
signal, the change in the participation ratio of $C(t)$, in place of the
return. Chosen on the validation period the trailing-return step is still the
best, with the smooth participation-ratio weight the runner-up just below it,
and a mix of the two reaching a higher out-of-sample Sharpe but a lower
validation one, so it cannot be selected. A smooth weight also holds a steadier
book, near fifty names throughout in place of the step between thirty and near
sixty. On this 2022--2024 period the participation-ratio weight matches the step on
the base book and edges it with the diversified sleeve, a combined Sharpe of
$1.05$ and $1.14$ against the step's $1.06$ and $1.08$, but the edge does not survive the clean test of
Section~\ref{sec:freshoos}, so we keep the step.

The protective sleeve in this book is the bare long-short, which shields
the portfolio by holding no net market exposure. That protection is close
to linear in the market, a neutral position rather than one that pays in a
sharp move. A second and distinct source of protection sits in the directed
stock-to-stock transfer-entropy network of Section~\ref{sec:mmatrix}, whose
centralities forecast neither chain. The HITS hub score of that network is
high for a name that sends information to the names others follow, a
recursive leader score, and a book that longs the leaders and shorts the
rest is market-neutral and convex in the market return, rising in both
tails rather than tracking either. Overlaying a quarter of this hub book on
the protective sleeve turns the sleeve from a neutral hedge into insurance,
a position that gains when the market moves sharply in either direction. We
re-optimize the long-only-to-protective mix $\theta$ on the validation
period with the overlay in place, and the validation selects the same mix as
before, $\theta=1$ in up markets and $0.4$ in down, so the insurance rides
inside the protective sleeve without disturbing the blend.

Table~\ref{tab:insurance} reads the overlay out of sample. The curvature of
the combined book's payoff in the market, its quadratic-beta coefficient,
roughly doubles, from $+0.2$ to $+0.4$, and the worst drawdown eases from
sixteen to thirteen percent. The out-of-sample Sharpe holds at $1.06$. The
overlay trades return for lower risk rather than lifting the ratio, the
annual return easing from $18.1\%$ to $16.8\%$ and the volatility from
$17.0\%$ to $15.8\%$, the premium paid for the convexity. The hub earns
nothing out of sample, so the overlay buys curvature and not return. It is
insurance, and it is priced as such: a small give-up in return in the
ordinary years against a book that bends up when the market breaks.

\begin{table}[htbp]
\centering
\small
\caption{The combined book with and without the convex insurance overlay, a
quarter weight of the HITS-hub leader long-short added to the protective
sleeve. In-sample (2018--2021, the period on which the mix is chosen) and
out-of-sample (2022--2024): annualized return, annualized volatility, Sharpe
ratio, and worst drawdown, net of five basis points a name traded, together
with the convexity of the book's out-of-sample payoff in the market (its
quadratic-beta coefficient). The overlay roughly doubles the out-of-sample
convexity and trims the drawdown, at a small cost in return and Sharpe. The
capitalization-weighted market is shown for reference, its convexity zero by
construction.}
\label{tab:insurance}
\setlength{\tabcolsep}{4pt}
\begin{tabular}{lcccc c cccc c}
\toprule
& \multicolumn{4}{c}{in-sample (2018--2021)} && \multicolumn{4}{c}{out-of-sample (2022--2024)} & \\
\cmidrule{2-5}\cmidrule{7-10}
book & ret & vol & Sharpe & max DD && ret & vol & Sharpe & max DD & convex. \\
\midrule
combined, no overlay        & $8.1\%$ & $19.7\%$ & $0.49$ & $-22\%$ && $18.1\%$ & $17.0\%$ & $1.06$ & $-16\%$ & $+0.2$ \\
combined, insurance overlay & $9.2\%$ & $19.0\%$ & $0.56$ & $-21\%$ && $16.8\%$ & $15.8\%$ & $1.06$ & $-13\%$  & $+0.4$ \\
market                      & $15.3\%$ & $21.2\%$ & $0.78$ & $-33\%$ && $12.6\%$ & $17.0\%$ & $0.78$ & $-22\%$ & $0$ \\
\bottomrule
\end{tabular}
\end{table}

The long-only sleeve holds the highest-scoring names, and the score
rewards a high return-rank probability and, in downturns, a calm
volatility rank. It says nothing about how the held names sit relative to
one another, so the sleeve can pile into momentum winners that move
together. A book whose names are mutually far apart is better diversified,
and the distance matrix already measures that distance. We read it from the
residual distance matrix, the arccos distance of the trailing return
correlations with the equal-weight market mode regressed out, so that what
remains is the excess comovement beyond the market, the crowding of Baltas
\cite{baltas2019}. The diversification of a held book is then the mean
pairwise residual distance of its names, large when they share little
common idiosyncratic movement. We tilt the selection score toward this
residual distance by a weight $\gamma$, rescaled so that $\gamma=0$
reproduces the current sleeve, and select the tilted top names with the
same no-trade band.

\begin{figure}[htbp]
\centering
\includegraphics[width=0.98\textwidth]{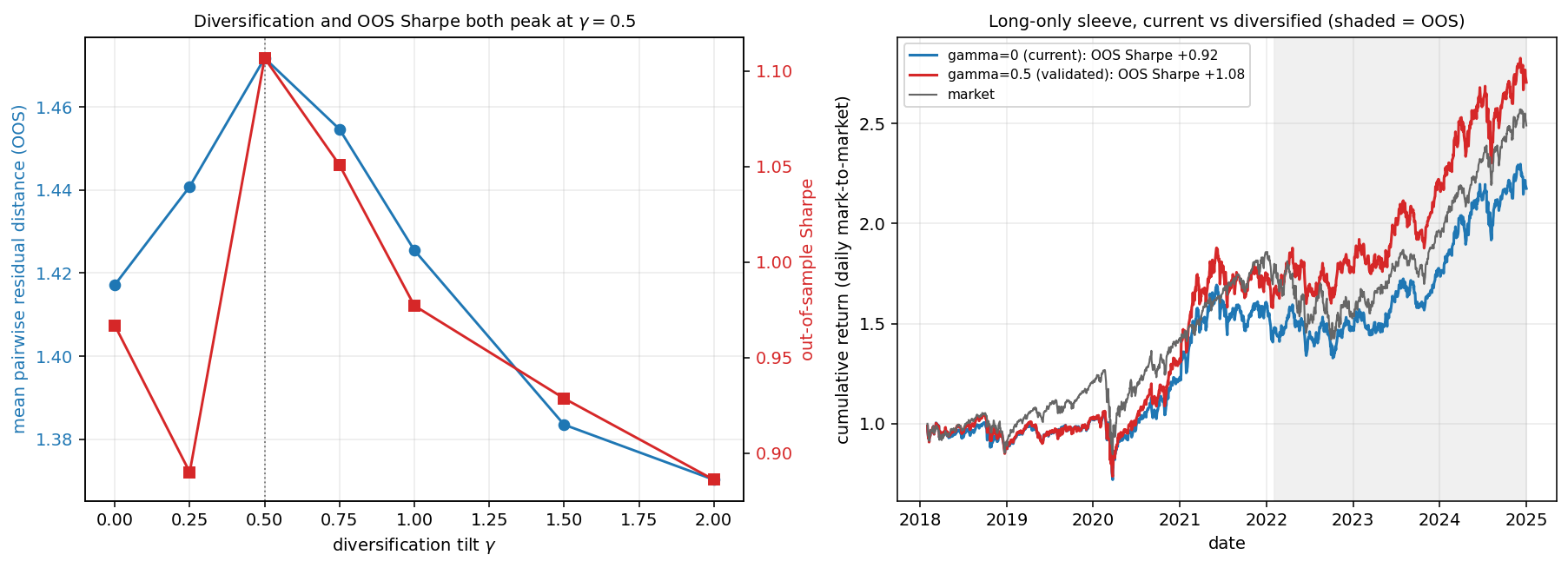}
\caption{Diversifying the long-only sleeve. Left: the held book's realized
diversification, the mean pairwise residual distance of its holdings
(blue), and its out-of-sample Sharpe (red) against the diversification tilt
$\gamma$, both single-peaked at $\gamma=0.5$. A stronger tilt over-selects
a cluster of peripheral outliers, lowering both. Right: cumulative return
of the current sleeve ($\gamma=0$) and the diversified sleeve
($\gamma=0.5$) against the market, out-of-sample period shaded.}
\label{fig:diversify}
\end{figure}

Diversification helps, but only up to a point, and the two statements are
one. Raising $\gamma$ from zero lifts the held book's realized
diversification from $1.42$ to a peak of $1.47$ at $\gamma=0.5$, and its
out-of-sample Sharpe rises with it, from $0.92$ to $1.08$, its annual
return from fourteen to nineteen percent and its worst drawdown two points
lower (Fig.~\ref{fig:diversify}). Past that peak the realized
diversification falls again, because a stronger tilt toward the most
peripheral names selects a cluster of similar oddities, far from the crowd
but close to each other, and the Sharpe falls with it while the volatility
climbs. The out-of-sample Sharpe tracks the achieved diversification, both
single-peaked at the same moderate tilt. Carried into the combined book,
the diversified sleeve raises the out-of-sample Sharpe from $1.06$ to
$1.08$.

The right way to set $\gamma$ is to aim at the diversification itself, not
the return. The realized diversification is known at selection time from
the trailing residual matrix, with no forward information, and choosing
$\gamma$ to maximize it on the training period lands on $\gamma=0.5$, the
out-of-sample peak. Choosing $\gamma$ by validation Sharpe instead
over-tilts to $\gamma=1.5$, since the training bull rewarded the
high-return outliers, and misses the peak. Diversification is a property to
aim at directly, not a by-product of maximizing risk-adjusted return.

This construction is a relative of the classical diversification
portfolios, but not the same object. The global minimum-variance portfolio
minimizes the portfolio variance, with weights $w\propto\Sigma^{-1}\mathbf{1}$,
and the most diversified portfolio of Choueifaty and Coignard
\cite{choueifaty2008} maximizes the ratio of the weighted-average volatility to
the portfolio volatility, with weights $w\propto\Sigma^{-1}\sigma$ for $\sigma$
the vector of individual volatilities. Neither takes any expected return as an
input. They fix the weights, and over the whole universe the names too, from the
second moments of the returns alone, so by construction they keep only the
covariance and discard the first-moment signal, the directional information
about the next ranking that our selection is built on. Both reduce to
inverse-variance and inverse-volatility weighting in the limit of vanishing or
equal correlations. Ours instead equal-weights names chosen to be mutually
distant in the residual correlation, the market mode removed, so it diversifies
the crowding rather than the total risk and leaves the market exposure to the
timing and the long-short sleeve.

The two classical books also carry the cost of their machinery, the same cost
the rank construction was built to avoid. Each needs the full covariance matrix
estimated and inverted. At these sizes the sample covariance is ill-conditioned,
so it must be shrunk toward its diagonal before the inverse is stable, and even
then the inverse amplifies estimation error, the known fragility of variance
optimizers. The raw solutions $\Sigma^{-1}\mathbf{1}$ and $\Sigma^{-1}\sigma$
place large long and short positions, so a long-only book must clip the negative
weights and renormalize, moving the realized portfolio away from the optimum it
solved for. And each is a single-period weighting on one covariance snapshot,
re-solved every month with no dynamics of its own. Our sleeve needs none of
this. It reads the residual distance matrix directly, with no inverse, keeps the
return signal in the selection, and equal-weights the chosen names, so no
covariance is estimated for the weighting, while the ranking chains supply the
dynamics the static optimizer lacks.

Table~\ref{tab:diversify} sets the two side by side. Applying the
minimum-variance or the most-diversified weights to our selected names reaches a
comparable or higher diversification but a lower out-of-sample Sharpe, $0.78$ and
$0.84$ against our $1.08$, because optimizing weights on the total covariance
pulls toward the low-volatility names and gives up the return the score
selected. Built standalone over the whole universe the two classical books are
the most diversified of all, near $1.56$, and the weakest out of sample, Sharpes
of $0.51$ and $0.56$, their diversification bought by discarding the return
signal. More diversification helps only while the return selection is held
fixed. Bought by reweighting toward calm names, it costs more return than risk.

\begin{table}[htbp]
\centering
\small
\caption{The diversified long-only sleeve against the classical
minimum-variance and maximum-diversification portfolios, out-of-sample
(2022--2024). Each book's diversification is the mean pairwise residual
distance of its holdings, the market mode regressed out; annualized return,
annualized volatility, Sharpe, and worst drawdown are net of five basis
points a name traded. The diversified selection, equal-weighted names
tilted toward residual distance, reaches the best Sharpe. The
minimum-variance and most-diversified weights, on the same names or over
the whole universe, reach a higher diversification but a lower Sharpe,
having reweighted toward low-volatility names at the expense of the return
signal. Both take no expected return as an input and weight from an inverted
covariance. The diversified selection does neither, equal-weighting a
return-selected book. The capitalization-weighted market is shown for reference.}
\label{tab:diversify}
\begin{tabular}{lccccc}
\toprule
book & diversification & ann.\ ret & ann.\ vol & Sharpe & max DD \\
\midrule
current sleeve (equal weight)          & $1.42$ & $14.4\%$ & $16.2\%$ & $0.92$ & $-17\%$ \\
diversified selection ($\gamma=0.5$)   & $1.47$ & $18.6\%$ & $17.2\%$ & $\mathbf{1.08}$ & $-15\%$ \\
min-variance weights, our selection    & $1.47$ & $10.7\%$ & $14.3\%$ & $0.78$ & $-20\%$ \\
max-diversification weights, selection & $1.50$ & $11.8\%$ & $14.7\%$ & $0.84$ & $-21\%$ \\
min-variance, whole universe           & $1.56$ & $6.1\%$  & $13.6\%$ & $0.51$ & $-18\%$ \\
max-diversification, whole universe    & $1.56$ & $7.2\%$  & $14.2\%$ & $0.56$ & $-18\%$ \\
market                                 & ---    & $12.6\%$ & $17.0\%$ & $0.78$ & $-22\%$ \\
\bottomrule
\end{tabular}
\end{table}

\begin{figure}[htbp]
\centering
\includegraphics[width=0.98\textwidth]{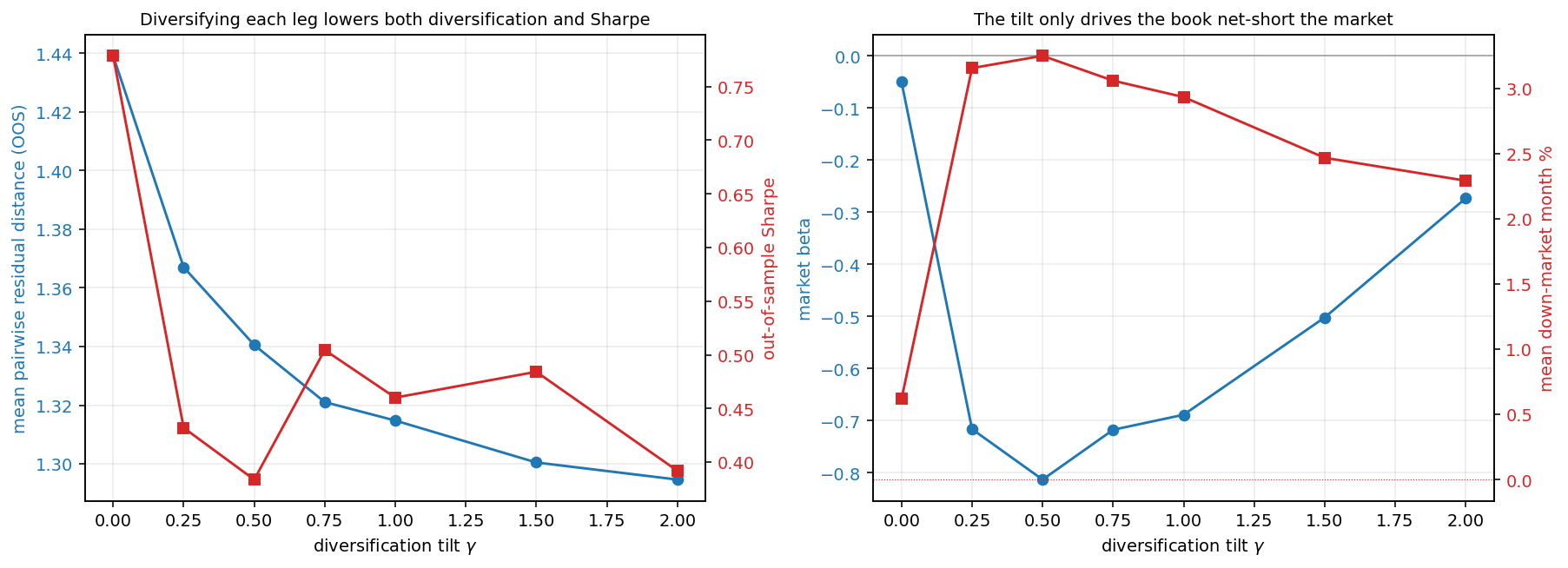}
\caption{The residual-distance tilt does not help the long-short sleeve.
Left: diversifying each leg lowers the realized diversification (blue) as
$\gamma$ rises, and the out-of-sample Sharpe (red) with it. Right: the tilt
drives the book net-short the market, its beta (blue) falling well below
zero, so what down-market gain it adds (red) is a directional short bet
rather than a better-diversified neutral book.}
\label{fig:lsdiv}
\end{figure}

Diversification is a long-only lever. The long-short sleeve is already the
most diversified book in the strategy, its two legs drawn from opposite ends
of the ranking and farther apart in residual space, a mean pairwise
distance of $1.49$, than the long-only sleeve ever reaches. Tilting it
toward residual distance does not help, whichever way it is done. Adding the
tilt to a single score, longing the uncrowded and shorting the crowded,
loads a net crowding factor and breaks the market-neutrality toward a
positive beta, so the apparent Sharpe gain is market exposure in disguise.
Diversifying each leg on its own, longing and shorting the most peripheral
names of each end, instead clumps the legs into the same oddities, lowering
the realized diversification from $1.49$ to $1.34$ and driving the book
net-short the market, beta $-0.8$, at a lower Sharpe
(Fig.~\ref{fig:lsdiv}). The long-short earns its place by staying neutral
and protecting in drawdowns, and neither tilt improves that, so we leave it
undiversified.

Everything above sizes the positions equally, which is the selection
half of portfolio management. The sizing half asks how much of each
name to hold, and the distance matrix offers a natural input, its
fragility. The market-factor share of the trailing correlation matrix
is high when the cross section concentrates onto a single mode, the
fragile, crisis-prone state, and low when it is diverse. We test two
sizing rules against the equal-weight book, each tuned on the
validation period. Conviction sizing weights a held name by its
predicted top-two probability rather than equally. Fragility sizing
sets the adaptive weight $\theta$ from the market-factor share instead
of the trailing return, turning the book market-neutral when the cross
section concentrates. Neither improves the risk-adjusted result.
Conviction sizing nudges the out-of-sample cumulative return up but
leaves the Sharpe essentially unchanged. Fragility sizing is worse on
both the validation and the test periods, since the market-factor share
is a noisier regime signal than the trailing return here. Equal weight
and trailing-return timing remain the better choices, and all the
variants still beat the market out of sample. In this study, then, the edge lives in the
selection and the regime timing, and these sizing refinements do not
add to it, though on a longer or more volatile sample the fragility
measure may yet earn its place.

A final ablation, in the spirit of the fundamentals test of
Appendix~\ref{sec:fundsection}, asks whether the distance matrix earns its keep
as a hedge rather than a forecast, and reaches the same verdict. Zlotnikov
et al.\ \cite{zlotnikov2024crowding} hedge crowding with a fund-holdings
network, long the uncrowded names and short the crowded, a book convex and
negatively correlated with the market. The same centrality already sits in
$M(t)$: the market-loading predictor is a name's eigenvector centrality in
the return-correlation network, and a book long the least-central decile
and short the most-central reproduces that hedge holdings-free, correlating
$-0.50$ with the market against their $-0.47$ and earning $+3.0\%$ in the
average down month. What it misses is the convexity, being concave in the
market ($-2.2$ against the holdings book's $+4.8$), and used as the
market-neutral sleeve in place of the momentum long-short it protects but
does not beat, a daily-marked out-of-sample Sharpe of $0.69$ against the
combined book's $1.06$. Rebuilding the signals from the residual correlations, the
market mode regressed out in the sense of Baltas \cite{baltas2019},
recovers the convex, positively-skewed crowding signature, a quadratic
loading of $+6.8$, but at a positive market correlation. It also forecasts
the ranking worse, so the full matrix stays the predictor and the residual
reads as the cleaner crowding diagnostic. Neither hedge adds to the
momentum-based book, whose convexity the information-leader insurance
overlay already supplies.

\section{The second out-of-sample test set}
\label{sec:freshoos}

The out-of-sample period used so far, 2022--2024, was held out from the
fitting, but it was not left untouched. Every refinement of the book, the
adaptive blend, the insurance overlay, the diversification of the long
sleeve, the sizing rules, and the crowding hedges, was read on that same
period, so it has been looked at many times over. Repeated inspection of one
test set leaks information, however carefully each individual choice is
validated on an earlier period, and the period is no longer perfectly clean.
A stricter test leans on data the study has barely seen. Because the strategy
runs on price and volume data alone, such a test is cheap. We replay it on an
independent data source over a recent window, most of which postdates the
whole study.

We rebuild the entire panel from the independent Yahoo Finance source
for the S\&P 500, the $475$ names with a full history since 2018, and run the
combined book with every hyperparameter frozen at the value chosen in the
earlier study, the book size, the no-trade band, and the $\lambda$ and
$\theta$ timing. The chain coefficients are refit walk-forward, each month on
the history before it, so no future observation enters any forecast. The
headline window is January 2025 to July 2026. It postdates the earlier study,
whose data ended in 2024, and it does not overlap the first test period, so it
is genuinely clean. Marking the book to market daily makes this window a sample
of about $390$ trading days, enough to read on its own, where the eighteen
monthly points of the same span would have been too few.
Table~\ref{tab:freshoos} reports it next to a 2022--2024 slice that replays the
first test period on the independent source, a cross-source check that shares
the first test's calendar but not its data.

\begin{figure}[htbp]
\centering
\includegraphics[width=0.92\textwidth]{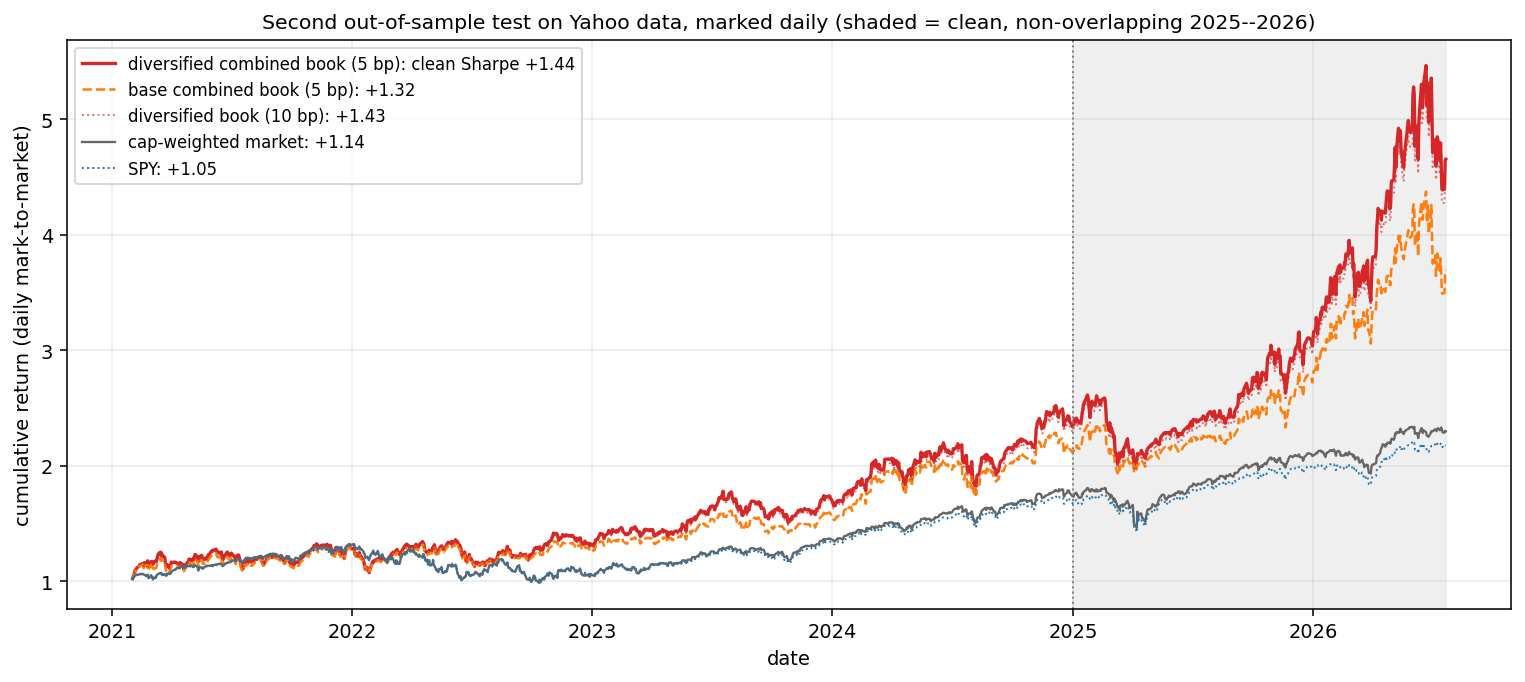}
\caption{The second out-of-sample test set. Cumulative return of
the OMD-Portfolio combined book with its long sleeve diversified by residual
distance (red, five basis points), the same book without the diversification
(orange dashed), the same diversified book at ten basis points (red dotted),
the capitalization-weighted market (grey), and SPY (blue dotted), 2021--2026,
with the clean, non-overlapping 2025--2026 window shaded. Every hyperparameter
is frozen at its earlier-study value and the chain coefficients are refit
walk-forward, so no future data enters any forecast. All curves are marked to
market daily. Both books pull ahead of the market through the window, the
diversified one at a clean-window daily Sharpe $1.44$ against the market's
$1.14$.}
\label{fig:freshoos}
\end{figure}

On the clean window the combined book earns a daily mark-to-market Sharpe of
$1.32$ against the capitalization-weighted market's $1.14$ and the S\&P 500 ETF
SPY's $1.05$, and diversifying its long sleeve by residual distance, the
refinement of Section~\ref{sec:portfolio}, lifts it to $1.44$ at an annualized
return of $56\%$ against the market's $20\%$, well over double
(Fig.~\ref{fig:freshoos}). The 2022--2024 cross-source slice reproduces the
pattern of the first test on the new data, the base book at $0.91$ and the
diversified one at $0.98$ against the market's $0.61$, an edge over the market
of about a third of a point, close to the $1.06$ and $1.08$ against $0.78$ of
the CRSP study. The diversification that helped on the first test period helps
again on the clean one, and by more. The book tracks the market through
2021--2024 and pulls ahead from 2025, its one cost a deeper drawdown than the
strong market of the window, more risk in a rising market for the higher
return. At ten basis points the clean window reads $1.30$ and $1.43$, a hair
below the five-basis-point base case. The runner-up blend weight of
Section~\ref{sec:portfolio}, the smooth participation-ratio construction, tells
the same cautionary story here. It edges the step on the 2022--2024 slice, $0.94$
and $1.00$ against the step's $0.91$ and $0.98$, but gives that edge back on the
clean window, $1.08$ and $1.29$ against $1.32$ and $1.44$, the in-sample gain not
surviving the fresh data, so we keep the step.

A few caveats bound the reading. The universe is the current index membership
with a full price history, which gives the earlier 2022--2024 slice a mild
survivorship tilt, though the clean window, drawn from names that are index
members now, is largely free of it. Market capitalization for the size
covariate is the current share count times the price, a small approximation.
The clean window is eighteen months, informative but not a long record. Within those limits the strategy carries to fresh data and an
independent source with no retuning, the strongest evidence the paper offers
that the edge is real rather than an artifact of repeated looks at one test
period.

\begin{table}[htbp]
\centering
\small
\caption{The second out-of-sample test set, S\&P 500, with every
hyperparameter frozen at its earlier-study value and the chain coefficients
refit walk-forward. Sharpe ratio and annualized return of the base combined
book, the same book with its long sleeve diversified by residual distance, and
the capitalization-weighted market, all marked to market daily and net of five
basis points. The headline window, January 2025 to July 2026, postdates the
whole study and does not overlap the first test period, so it is clean. The
2022--2024 slice replays the first test period on the independent source, a
cross-source check that overlaps the first test in calendar time. At ten basis
points the clean-window Sharpes are $1.30$ (base) and $1.43$ (diversified). The
last two columns give the combined book's gross exposure, its measure of
leverage, and its net exposure, the directional market exposure, averaged over
the period; they are the same for the base and diversified books, and the gross
stays below the long-short norm.}
\label{tab:freshoos}
\setlength{\tabcolsep}{4.5pt}
\begin{tabular}{lccc c ccc c cc}
\toprule
& \multicolumn{3}{c}{Sharpe} && \multicolumn{3}{c}{ann.\ return} && \multicolumn{2}{c}{exposure} \\
\cmidrule{2-4}\cmidrule{6-8}\cmidrule{10-11}
period & base & diversified & market && base & diversified & market && gross & net \\
\midrule
2025--2026 (clean)        & $1.32$ & $1.44$ & $1.14$ && $44\%$ & $56\%$ & $20\%$ && $1.17$ & $0.83$ \\
2022--2024 (cross-source)           & $0.91$ & $0.98$ & $0.61$ && $18\%$ & $22\%$ & $10\%$ && $1.20$ & $0.80$ \\
full 2021--2026                     & $1.09$ & $1.17$ & $0.98$ && $27\%$ & $33\%$ & $16\%$ && $1.15$ & $0.85$ \\
\bottomrule
\end{tabular}
\end{table}

How much of the edge is signal and how much is a short record? We test the
outperformance with a stationary block bootstrap of the daily mark-to-market
excess return, book minus market, resampling geometric-length blocks of about a
month so the serial dependence of the daily returns is respected
(Table~\ref{tab:bootstrap}). The resampling is of the realized daily returns,
not the prices. The signals were computed once, walk-forward and in calendar
order, and are frozen inside the track record, so testing whether its edge
arose by chance re-estimates nothing, and the block structure carries the
serial dependence a mean's significance needs. A model-level resampling of the
prices would recompute the signals but is ill-posed on a temporal strategy
whose walk-forward fit and regime timing depend on calendar order. The return
advantage is significant for the diversified book on the clean 2025--2026 window
itself, an annualized excess of $31\%$ at a one-sided $p$ of $0.04$, and over
the full 2021--2026 record, $15\%$ at $p=0.03$ with a 95\% interval that barely
excludes zero. Daily marking is what buys the significance on the clean window,
its eighteen months being only a handful of monthly points but some $390$ daily
ones, enough for the block bootstrap to separate the edge from noise. The Sharpe
difference reaches significance in no period: eighteen months to five years
cannot separate the book's Sharpe from the market's, which was itself high, so
the advantage the data supports is in the return rather than yet in the
risk-adjusted return. The honest reading is a return edge that is real over the
multi-year record and significant on the clean window on its own.

\begin{table}[htbp]
\centering
\small
\caption{Stationary block bootstrap ($10^4$ resamples, expected block length
about one month, $21$ trading days) of the daily mark-to-market excess return,
combined book minus capitalization-weighted market, on the second test panel,
net of five basis points. Annualized mean excess return with a 95\% bootstrap
interval and a one-sided $p$-value against the null of no outperformance, for
the base book and the residual-distance diversified book.}
\label{tab:bootstrap}
\begin{tabular}{llccc}
\toprule
period & book & excess ann.\ \% & 95\% interval & $p$ \\
\midrule
2025--2026 (clean) & base        & $21.5$ & $[-10.2, 53.8]$ & $0.09$ \\
                   & diversified & $31.4$ & $[-3.7, 66.6]$  & $0.04$ \\
2022--2024         & base        & $8.0$  & $[-9.8, 26.1]$  & $0.19$ \\
                   & diversified & $11.8$ & $[-6.7, 30.8]$  & $0.11$ \\
full 2021--2026    & base        & $10.3$ & $[-3.9, 24.8]$  & $0.08$ \\
                   & diversified & $15.3$ & $[-0.2, 31.4]$  & $0.03$ \\
\bottomrule
\end{tabular}
\end{table}

A second bootstrap asks the complementary question: does the edge rest on a few
names, or is it broad across the cross section? Here we resample the universe
rather than time, drawing a random seventy percent of the names without
replacement, since duplicate names would degenerate the distance geometry, and
re-estimating the entire pipeline on each drawn universe with the calendar order
left intact. Across twenty such universes the diversified book's annualized
excess over the capitalization-weighted market of the same names is positive in
every one, on both periods. It averages $9\%$ over 2022--2024 with a
fifth-to-ninety-fifth-percentile range of $[5, 14]\%$, and $24\%$ over the
clean 2025--2026 window with a range of $[14, 37]\%$, and the daily-marked book
Sharpe runs $[0.75, 1.14]$ over the first period and $[1.20, 1.74]$ over the
clean window, above the market's in every draw. The edge is
thus broad across the cross section, not the product of a handful of names.

\section{Summary and outlook}
\label{sec:summary}

We have set out a Markov-chain framework for dynamic portfolio management, of
which security selection is the first half, extending the ranking-space chains of
the OMD study of the S\&P 500 \cite{halperin2026omdstocks} with covariates. Each
firm characteristic is reduced to a common ten-level ordinal by cross-sectional
rank-bucketing, so the target ranks and the covariates sit on one footing and a
large characteristic library becomes directly usable, and the return and
volatility chains are conditioned on the bucketed covariates through a log-linear
model that nests the unconditional chain. Because the construction stays discrete,
the entropy production and the transfer entropy remain exact plug-in sums, and
they acquire new readings, a covariate-attribution of the arrow of time from the
convexity bound $\sigma_{\mathrm{cond}}\ge\sigma$ and a directional causality
screen from the asymmetry of the transfer entropy.

The empirical picture is a consistent asymmetry between the two chains. The
volatility chain is where the covariates bite: conditioning resolves several times
the pooled entropy production, and its forecast improves out of sample and at
longer horizons. The return chain is close to unforecastable, and its transfer
entropy flags momentum as a follower rather than a leader. Predictors read from
the distance matrix itself, the market-mode loading, the centrality, and a
transfer-entropy lead-lag score, sharpen the volatility chain and lift the return
chain from an out-of-sample overfit to a marginal positive, so the cross-sectional
geometry carries information the individual characteristics miss. A multi-step
check confirms the split, the return chain matching its first-order $P^n$ forecast
over ten months while the volatility chain carries clustering memory beyond one
step. What else predicts the return rank is slow and fundamental, and half of the
informative ratios prove redundant. An out-of-sample ablation adds nothing the
validation period would select, in line with the practitioner view that
backward-looking accounting ratios carry little tradeable signal. Trading volume,
read through the same chains, has a far larger arrow of time than the price ranks,
and its turnover forecasts the next return rank out of sample, real beyond the
training period yet not tradeable once cost is paid.

The forecasts drive a selection rule. The best market-neutral long-short is a
momentum book from the return chain whose gross edge is at first erased by
turnover, until a no-trade band cuts the turnover by more than half and lifts it
to a full-period Sharpe of $0.34$ and an out-of-sample Sharpe near $0.6$, at beta
$-0.02$ with its gains concentrated in the Covid and 2022 drawdowns. Folding in
the more forecastable volatility chain does not help, its forecastability being
about risk measurement rather than selection. Combined with an adaptive-$\lambda$
long sleeve that leans defensive when the market weakens, blended by a
state-dependent weight chosen on the validation period, the book beats the market
out of sample over 2022--2024 at a shallower drawdown. A second test rebuilt from
an independent source, every hyperparameter frozen and the coefficients refit
walk-forward, carries the result to a clean window the study never saw, a
daily-marked Sharpe of $1.32$ against the market's $1.14$ over January 2025 to
July 2026, and $1.44$ once the long sleeve is diversified by residual distance, at
well over double the market's return and significant on a block bootstrap. A
convex insurance overlay from the directed transfer-entropy graph bends the payoff
up in both tails at a small cost, and the residual-distance tilt lifts the
combined book from $1.06$ to $1.08$ while beating the minimum-variance and
most-diversified portfolios, which give up the return signal by optimizing weights
on the total covariance. Position sizing, whether by forecast conviction, a
distance-matrix fragility measure, or a holdings-free crowding hedge
\cite{zlotnikov2024crowding}, does not improve the risk-adjusted result, so the
value stays in the selection and the timing, the tradeable strategy itself a
by-product of the forecasting framework and its diagnostics.

Several checks bear on how much of the design is real rather than fitted. The two
distance-matrix centralities, the full-matrix one used in the book and its
residual alternative, are close to a wash, so the book keeps the simpler
full-matrix one. A modern hyperparameter optimizer run on the validation Sharpe
overfits, raising the validation number and lowering the out-of-sample, so the
conservative hand grid, and the modest book size that is only weakly identified
because the Sharpe rewards diversification without bound, are the point rather
than a limitation. The components also differ sharply under a change of regime,
the market premium the most stable, the market-neutral momentum sleeve the least,
and hedging the market out of the long-only sleeve raises its regime dependence
rather than lowering it. The net-long, regime-timed construction is thus a
compromise built for non-stationary data, and the refinements it invites, a
momentum-crash defence and a worst-regime calibration, each help the regime they
target and fail out of sample, because the momentum-unfriendly and the bear
regimes want opposite things from the long-short.

This returns us to the question the paper opened with. Our result is not,
strictly, a refutation of the efficient-market hypothesis, since what the
hypothesis forbids is an excess return over the correct risk-adjusted benchmark,
and every test of it is a joint test of efficiency and an assumed asset-pricing
model \cite{fama1970}. Beating the capitalization-weighted index is not beating a
risk- and factor-adjusted benchmark, and our edge could be, in part, compensation
for exposures the index does not carry. The construction is built from momentum
and from a volatility ranking, both long-documented anomalies. Relative-strength
momentum was read from the start as evidence against weak-form efficiency
\cite{jegadeesh1993,danielmoskowitz2016}, and the low-volatility and
minimum-variance portfolios beat the index on a risk-adjusted basis
\cite{clarke2011minvar,choueifaty2008}. Such factor premia survive careful
out-of-sample replication \cite{jensen2023replication}. What is new is not that a
systematic S\&P 500 portfolio can outperform the index, but the representation
that produces it, three fixed-size matrices and a pair of discrete chains from
which the momentum and low-volatility edges emerge as readings of one
information-theoretic object rather than as separately engineered factors.

Several extensions follow. The state can carry higher-order memory through lagged
ranks and covariates, which the log-linear form keeps tractable. The return chain
can be extended to the forward-looking earnings estimates it now lacks. The
selection rule can size positions by the full predicted transition distribution
rather than its mean, with the transfer-entropy screen gating which covariates
enter. The out-of-sample volume signal is a candidate stress input for the
regime-timing weight, the one lever within the price representation the book has
not yet spent.

These extensions share one premise, that the state representation is what matters
for decision-making in finance. What we hope to have shown is that the
parsimonious three-matrix formulation is enough, carrying the information that
serves the problem of dynamic portfolio optimization while filtering unhelpful
noise. Modeling multi-period portfolios of many assets is usually held to be
mathematically heavy, which is perhaps why practitioners rarely attempt it. The
framework developed here is instead a minimal \textit{ab initio} formulation that
uses no forward-looking signals and no sophisticated optimization, and does not
even invert a matrix. It still slightly outperforms the market on our tests, a
level best read as a floor, since adding genuinely predictive signals to the same
representation could only raise the performance above it.

\appendix

\section{Fundamentals: information content, lead-lag, and an out-of-sample ablation}
\label{sec:fundsection}

Company fundamentals sit outside the three-matrix information set. We explore
them as potential covariates for our model using a rich dataset of S\&P 500
companies' fundamentals from 1970 to 2025 from Compustat, which lets us ask, over half
a century rather than a single decade, what the fundamentals carry and whether
they earn a place in the tradeable book. We
work on a monthly grid over 1975--2025, restricted at each date to the S\&P 500
members of that month, with thirty financial ratios spanning value,
profitability, leverage, liquidity, and efficiency. Each ratio is carried
point-in-time from the Compustat public date and rank-bucketed into deciles by
Eq.~\eqref{eq:xbucket}, so no future accounting value enters a forecast. This
appendix first reads what information the fundamentals carry about the return
rank and which of them lead it, then ablates them in the portfolio out of
sample.

\subsection{Information content and lead-lag of the fundamentals}
\label{sec:fundamentals}

We treat the monthly return decile as the return-rank state of the chain and
each ratio as a decile-bucketed covariate, and measure three
information-theoretic quantities in nats, pooled across names and months. The
entropy-production gap is the conditioned entropy production of the return chain
minus the pooled value, how much of the chain's directed structure a ratio
resolves. The transfer entropy $\TE(X\!\to\!A)=I(A_{\mathrm{next}};X\mid A)$ and
its reverse $\TE(A\!\to\!X)$ give, through their difference, the lead-lag
direction, positive when the ratio leads the return rank and negative when it
merely follows. The conditional transfer entropy $\TE(X\!\to\!A\mid
\mathrm{size})$ is the same forward information once firm size is controlled
for, the dominant confounder of the cross section.

\begin{figure}[htbp]
\centering
\includegraphics[width=0.98\textwidth]{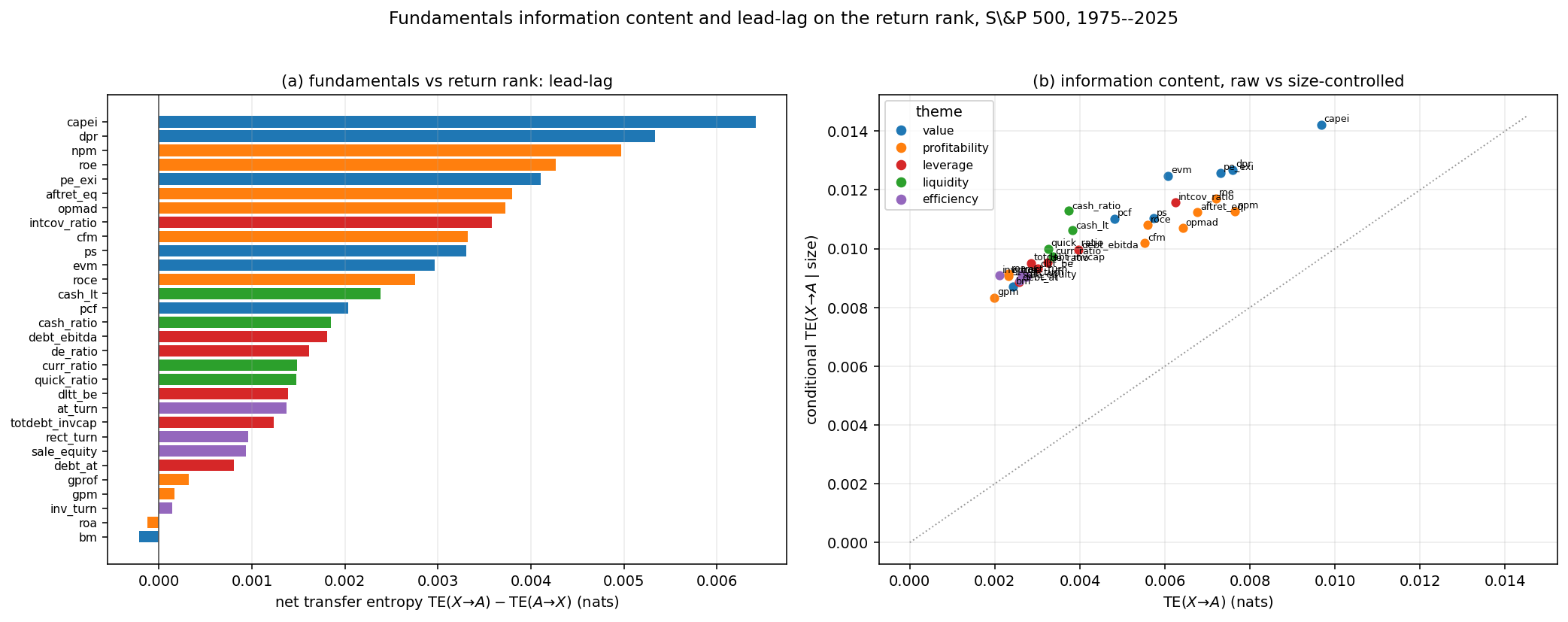}
\caption{Fundamentals information content and lead-lag on the return rank,
S\&P 500, 1975--2025. (a) Net transfer entropy
$\TE(X\!\to\!A)-\TE(A\!\to\!X)$ of each ratio with the return rank, positive
for almost all, so the fundamentals lead the ranking, and largest for the value
and profitability ratios. (b) Raw transfer entropy against the size-controlled
conditional transfer entropy, coloured by theme. Points above the diagonal gain
information once size is held fixed, which is nearly all of them, so the content
is not a size artifact.}
\label{fig:fundwrds}
\end{figure}

The reading is consistent across all three measures, in
Figure~\ref{fig:fundwrds}. The information is concentrated in the value and
profitability ratios. The cyclically adjusted earnings yield leads, with a
transfer entropy into the return rank near $0.010$ nats, followed by the net
profit margin, the dividend payout, the price-to-earnings ratio, and the return
on equity, all near $0.007$. Leverage, liquidity, and efficiency ratios carry
less than half as much. The net transfer entropy is positive for almost every
ratio, so the fundamentals lead the return rank rather than follow it, the
opposite of the momentum covariate of Section~\ref{sec:empirical}, and the
reading expected of exogenous accounting information. Plain book-to-market is
the exception, with a net flow near zero, a reminder that the raw value ratio is
a weaker lead than the earnings-based ones.

Controlling for size sharpens rather than erases the signal. The conditional
transfer entropy exceeds the raw transfer entropy for nearly every ratio
(Figure~\ref{fig:fundwrds}b), so size was masking part of the fundamentals'
information rather than manufacturing it, and the value and profitability ratios
keep their lead once size is held fixed. What emerges is a low-dimensional
value-and-quality direction that carries genuine, size-robust, forward
information about the ranking, though all of it small in absolute terms, a few
thousandths of a nat, the slow payoff characteristic of fundamentals.

A last question is whether it is the level of a fundamental that carries this
information or its change. One expects a firm to move through the ranking when a
fundamental improves or deteriorates, not merely when it is high or low.
Replacing each level by its year-on-year percentage change, decile-bucketed the
same way, raises the mean transfer entropy into the return rank from $0.0045$ to
$0.0053$ nats, and a change carries more than the level for twenty-five of the
thirty ratios (Table~\ref{tab:fundchanges}). The gain is largest among the
profitability ratios. The change in return on assets carries three times the
information of its level, $0.0023$ against $0.0075$ nats, and operating margin,
return on capital, and sales-to-price improve as much. This is the
fundamental-momentum reading, in which an improving balance sheet, not a strong
one, leads the return ranking. The volatility rank behaves oppositely and
prefers the level to either change, so it is the state of the fundamentals that
tracks risk and their trend that tracks return.

\begin{table}[htbp]
\centering
\small
\caption{Level against change as the cross-sectional signal, mean transfer
entropy across the thirty ratios into the return-rank and volatility-rank
chains, S\&P 500 1975--2025, in nats. Ranking the year-on-year percentage change
rather than the level raises the information into the return rank, and does so
for twenty-five of the thirty ratios, while the volatility rank prefers the
level.}
\label{tab:fundchanges}
\begin{tabular}{lcc}
\toprule
cross-sectional signal & TE into return rank & TE into vol rank \\
\midrule
level                          & $0.0045$ & $0.0032$ \\
year-on-year absolute change   & $0.0046$ & $0.0025$ \\
year-on-year percentage change & $0.0053$ & $0.0028$ \\
\bottomrule
\end{tabular}
\end{table}

\subsection{The out-of-sample ablation}
\label{sec:ablation}

Whether that information trades is a separate question, and the long record lets
us settle it with a strict split. The book size and no-trade band are chosen on
the in-sample period 1975--1999, and every number below is read on the
twenty-six-year out-of-sample period 2000--2025. We reuse the conditional-logit
return chain of the manuscript unchanged, with the return-rank class the
trailing six-month return decile and the covariates the monthly analogs of the
price and distance-matrix predictors. Against that baseline we add the eight
fundamentals with the most information content from the previous subsection, and
then the full thirty-ratio library, a broad standard characteristic set in the
spirit of the replication literature \cite{jensen2023replication}. The chain
coefficients are refit walk-forward, each rebalance using only months strictly
before it. Both a market-neutral long-short and a directional long-only are run
at each covariate set.

The two books answer differently, in Table~\ref{tab:fundablation}. The
market-neutral long-short, the sleeve that carries the portfolio's downside
protection, gains nothing from the fundamentals. Its out-of-sample Sharpe is a
weak $0.07$ on the price and distance-matrix predictors alone, a monthly
momentum long-short living through the crashes of 2000, 2009, and 2020, and
adding the eight fundamentals or the full library leaves it at or below zero on
both the validation and the test period. No ex-ante rule adds them to the
market-neutral book. The directional long-only is the one place the fundamentals
help. The informative eight lift its out-of-sample Sharpe from $0.59$ to $0.66$
and its annualized return from $10.5\%$ to $10.9\%$, and the improvement shows
up in sample as well, so a rule that selects on the validation period does
include them. The full thirty-ratio library does no better than the informative
eight, consistent with the low-dimensional value-and-quality direction of the
previous subsection. Ranking the changes of the eight fundamentals rather than
their levels, the representation that carried more information about the return
rank, gives the same out-of-sample edge, a long-only Sharpe of $0.65$ against
$0.66$, so the extra information in the changes does not translate into extra
tradeable performance, and the value-and-quality tilt is present whether the
level or the change is ranked.

What the long history changes is the strength of the negative, not its
direction for the tradeable core. The fundamentals are a small directional tilt,
a value-and-quality lean that adds a few basis points a month to a long book
already carrying the market, not a driver of the market-neutral sleeve that the
combined portfolio leans on in downturns. The mechanism is the one the price
representation already suggests. Markowitz optimization reaches a portfolio
through an expected-return vector estimated from fundamentals, so the
fundamentals enter through the return forecast. Our construction never forms
that vector. It reads the cross section as a rank geometry and a pair of
transition matrices, and conditions on the market-derived state directly. If the
market has already priced the fundamental information into the return
correlations and the rank dynamics, as an efficient market largely should, then
the price representation carries whatever of it is relevant, and the fundamentals
are largely redundant once that representation is in hand. The small long-only
tilt is the residue the price signals do not fully span, a slow value-and-quality
lean, and it does not reach the market-neutral book at all.

The claim is correspondingly narrow. At the monthly horizon on S\&P 500
large-caps, over half a century, the informative fundamentals add a small,
validated tilt to a directional book and nothing to the market-neutral core, so
the tradeable book of Section~\ref{sec:portfolio} is right to run on the price
and distance-matrix predictors alone. Whether a longer holding period, smaller
names, or the forward-looking earnings estimates we lack would change that is
left open, and the discrete framework makes the test easy to repeat.

A final check reruns the ablation on the recent first test dataset, keeping the
isolated-sleeve reading rather than the combined book, since the two sleeves
respond to the different regimes each test period contains. On the in-sample
2018--2021 and out-of-sample 2022--2024, from the daily point-in-time features,
the pattern of the half-century ablation repeats, its Sharpes kept on that
ablation's monthly footing for comparability rather than the daily marking of
the tradeable book. In the directional long-only
the eight fundamentals lift the in-sample Sharpe from $0.65$ to $0.93$ and the
out-of-sample from $0.78$ to $0.81$, a validated gain an ex-ante rule would
take. In the market-neutral long-short they do not, the in-sample Sharpe
slipping from $-0.02$ to $-0.12$ and the out-of-sample from $1.05$ to $0.84$, so
no rule adds them there. The recent period also sharpens what the separate
sleeves show. The long-short carries the book through the 2022 bear at an
out-of-sample Sharpe above one, while the long-only rides the recovery, and the
value-and-quality tilt of the fundamentals reaches only the directional sleeve,
exactly as over the long history.

\begin{table}[htbp]
\centering
\small
\caption{Fundamentals ablation on the long Compustat history, walk-forward out of
sample. Validation-period (1975--1999) and out-of-sample (2000--2025) Sharpe of
the market-neutral long-short (LS) and the directional long-only (LO), with the
long-only annualized return and worst drawdown, the book size and no-trade band
chosen on the validation period. The informative eight are the value and
profitability ratios that lead the return rank in Appendix~\ref{sec:fundamentals}.
Adding fundamentals does nothing for the market-neutral book on either period.
In the directional book the informative eight raise the out-of-sample Sharpe
from $0.59$ to $0.66$, a lift the validation period also shows, and the full
library does no better. The capitalization-weighted market is the directional
benchmark for the long-only. These Sharpes are computed on monthly equity returns,
annualized by $\sqrt{12}$, since daily marks are not available for the
fifty-year fundamentals panel, so they sit on a different footing from the
daily-marked books of the main tests.}
\label{tab:fundablation}
\begin{tabular}{lcccccc}
\toprule
covariate set & LS val & LS OOS & LO val & LO OOS & LO ann\% & LO DD\% \\
\midrule
price $+$ $M$-matrix   & $0.11$  & $0.07$  & $0.74$ & $0.59$ & $10.5$ & $-52$ \\
$+$ 8 fundamentals     & $-0.01$ & $-0.01$ & $0.75$ & $0.66$ & $10.9$ & $-47$ \\
$+$ 30-ratio library   & $-0.14$ & $0.04$  & $0.73$ & $0.65$ & $10.3$ & $-45$ \\
\midrule
market (reference)     & ---     & ---     & ---    & $0.61$ & $8.4$  & --- \\
\bottomrule
\end{tabular}
\end{table}

\begin{figure}[htbp]
\centering
\includegraphics[width=0.98\textwidth]{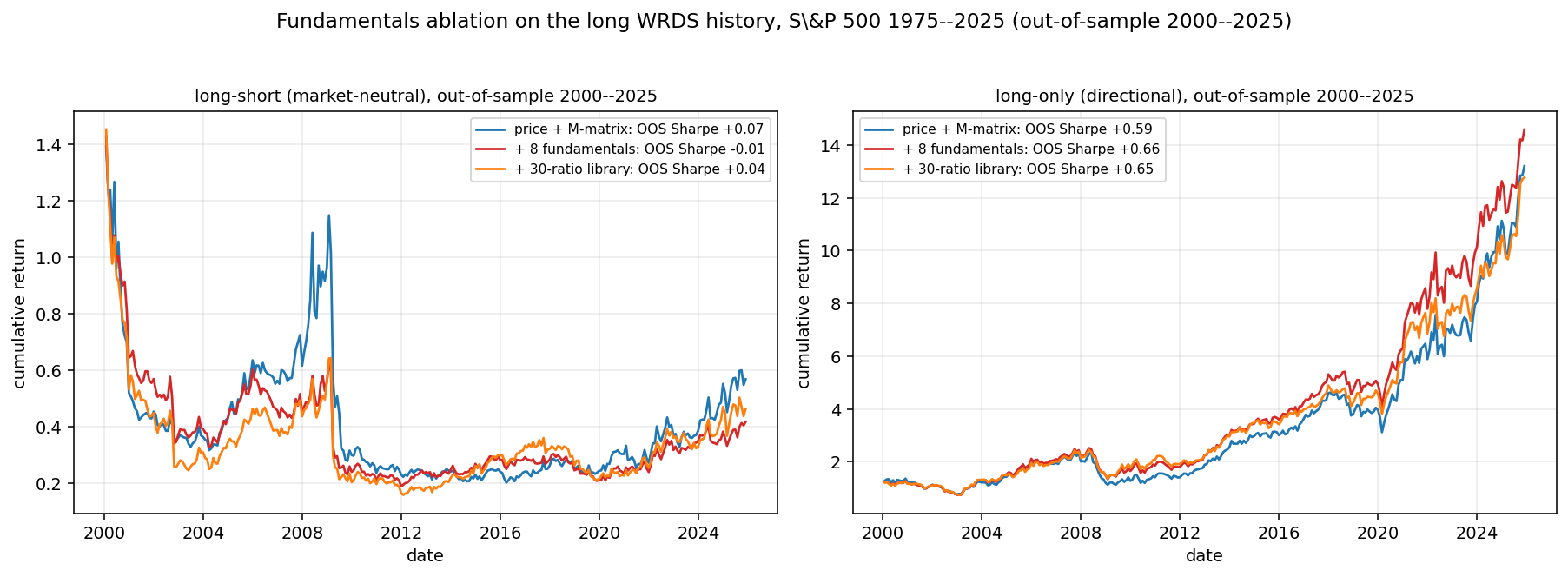}
\caption{Fundamentals ablation on the long Compustat history, out-of-sample
2000--2025. Cumulative return of the market-neutral long-short (left) and the
directional long-only (right), on the price and distance-matrix predictors alone
(blue) and with the informative eight fundamentals (red) or the full
thirty-ratio library (orange) added. The fundamentals leave the market-neutral
book unchanged and give the directional book a small, validated lift.}
\label{fig:fundablation}
\end{figure}

\section{Spectral early-warning signals: a second out-of-sample ablation}
\label{sec:spectralablation}

The blend weight $\theta$ of Section~\ref{sec:portfolio} switches on one bit,
the sign of the trailing market return, opportunistic when the recent market
has risen and defensive when it has fallen. The OMD-Stocks model
\cite{halperin2026omdstocks} and the OMD study \cite{halperin2026omd} read the
dynamics of the eigenvalues and eigenvectors of the rolling return-correlation
matrix as precursors of structural change. A natural question, in the spirit of
the fundamentals ablation of Appendix~\ref{sec:fundsection}, is whether those
spectral observables can time the mix better than the sign of a recent return.
We test it with every choice fixed on an earlier period and the clean recent
window of Section~\ref{sec:freshoos} kept as the arbiter.

From the trailing one-year correlation matrix $C(t)$ we form, each month and
from past data only, the share of variance in its leading eigenvalue, the
participation ratio $(\sum_k\lambda_k)^2/\sum_k\lambda_k^2$ that counts the
effective number of factors, and the mean pairwise correlation, together with a
standardized composite of the three that rises as the cross section collapses
onto the market mode, the fragility score of the OMD-Stocks model. We add the
commutator norm $\|[C(t),C(t\!-\!1\,\text{yr})]\|_F$, which marks a rotation of
the eigenbasis, and the rank-decay slope of the arccos distance matrix that the
OMD study reads as its own precursor.

Used to steer $\theta$ toward the protective sleeve when a signal runs high,
these observables lower the out-of-sample return in every form we tried. The
participation-ratio decline validates better than the trailing-return baseline,
a higher Sharpe on the 2018--2021 period, and yet gives the worst 2022--2024
result among the defensive rules, below the daily-marked baseline of $1.06$. The
mechanism is the one the OMD-Stocks model documents. For the exogenous shocks
and fast rebounds that fill this sample the spectral collapse is coincident with
the drawdown rather than ahead of it, so a defensive tilt taken on the signal is
taken at the bottom and forfeits the recovery.

The more promising use inverts the reading. Instead of turning defensive as
fragility rises, the book stays opportunistic while fragility recedes from a
recent peak, overriding the trailing-return defensive tilt to rejoin the
rebound. Fixed on the 2018--2021 validation period, this override raised the
validation Sharpe above the baseline and lifted the 2022--2024 out-of-sample
Sharpe of the base book from $1.06$ to $1.15$. On an independent decade, the
2005--2016 panel with every parameter frozen, it improved the base book from a
Sharpe of $0.35$ to $0.58$, all sixteen variants of the rule beat the baseline,
and the gain fell in the 2009 and 2015--2016 recoveries that the trailing-return
rule sat out (Table~\ref{tab:spectralablation}).

The clean recent window settles it, and it settles against the rule.
Frozen, the override lowers the base combined book from a Sharpe of $1.32$ to
$1.05$ and its annual return from $44\%$ to $32\%$, and it leaves the
diversified book nearly unchanged, $1.44$ against $1.41$. It fired in the first months of 2026
rather than at any rebound, and re-risking there cost return. A rule that
improved every period used to build it did not survive the one period held out
of its construction.

We therefore keep the trailing-return blend, and the reading matches the
fundamentals ablation of Appendix~\ref{sec:fundsection}. The spectral observables
are a strong contemporaneous read of the cross section, as their originating
studies conclude, and not a timing signal that beats the sign of the recent
market return out of sample. Information that is real in-sample, whether a
backward-looking accounting ratio or an eigenvalue precursor, need not carry
into tradeable out-of-sample structure, and the price representation with a
one-bit reactive timing rule already holds what the blend needs. The exercise
also shows the worth of a period held strictly out of construction, since the
override looked robust on every window that built it and failed on the one that
did not.

\begin{table}[htbp]
\centering
\small
\caption{The contrarian spectral re-risk override of the blend weight $\theta$
against the trailing-return baseline, base combined book, out-of-sample Sharpe
ratio, net of five basis points and marked to market daily. The rule is fixed
on the 2018--2021 validation period and read out of sample everywhere else. The
2005--2016 and 2022--2024 rows are the CRSP study panels on which it was built
and confirmed, the $1.06$ baseline the value of the earlier sections. The
2025--2026 row is the clean second test window of Section~\ref{sec:freshoos},
held out of the rule's construction.}
\label{tab:spectralablation}
\begin{tabular}{lcc}
\toprule
period & baseline $\theta$ & with re-risk override \\
\midrule
2005--2016 (independent decade) & $0.35$ & $0.58$ \\
2022--2024                      & $1.06$ & $1.15$ \\
2025--2026 (clean holdout)      & $1.32$ & $1.05$ \\
\bottomrule
\end{tabular}
\end{table}

\section{Additional signal studies}
\label{sec:signalstudies}

Several further studies test signals that sit outside the tradeable book but
bear on it. Most reach the conclusion of the spectral ablation of
Appendix~\ref{sec:spectralablation}, their information real in sample and not
tradeable out of sample. The last, on trading volume, sharpens the reading,
since its information is real out of sample as well and fails only at the price
of trading it.

\subsection{Transfer-entropy market leaders}
\label{sec:teleaders}

The transfer entropy that conditions the ranking chains in
Section~\ref{sec:analysis} also answers a question of its own, which individual
names lead the market. We work in the discretized setting of the paper. For a
horizon $h$ of twenty-one or thirty trading days we bucket into deciles the
market's forward $h$-day return $F_t$, the market's trailing $h$-day return
$m_t$, and each stock's trailing $h$-day return $x_t$, where the market is the
equal-weight index of the universe. The leadership score of a stock is the
transfer entropy to the market,
\begin{equation}
\TE(x \to \text{market};\,h) = I(F_t;\,x_t \mid m_t),
\end{equation}
the conditional mutual information that measures how much the stock's current
state says about the market's next $h$-day return beyond what the market's own
current state already says \cite{schreiber2000}. Conditioning on $m_t$ removes
market momentum, so the score isolates lead information. We rank the names on
each of three long periods, the 1996--2005 dot-com era, the 2003--2012 window
around the financial crisis, and the 2015--2024 window around the pandemic, and
correct the finite-sample bias of the plug-in estimate by subtracting a
shuffled-surrogate null.

\begin{figure}[htbp]
\centering
\includegraphics[width=0.9\textwidth]{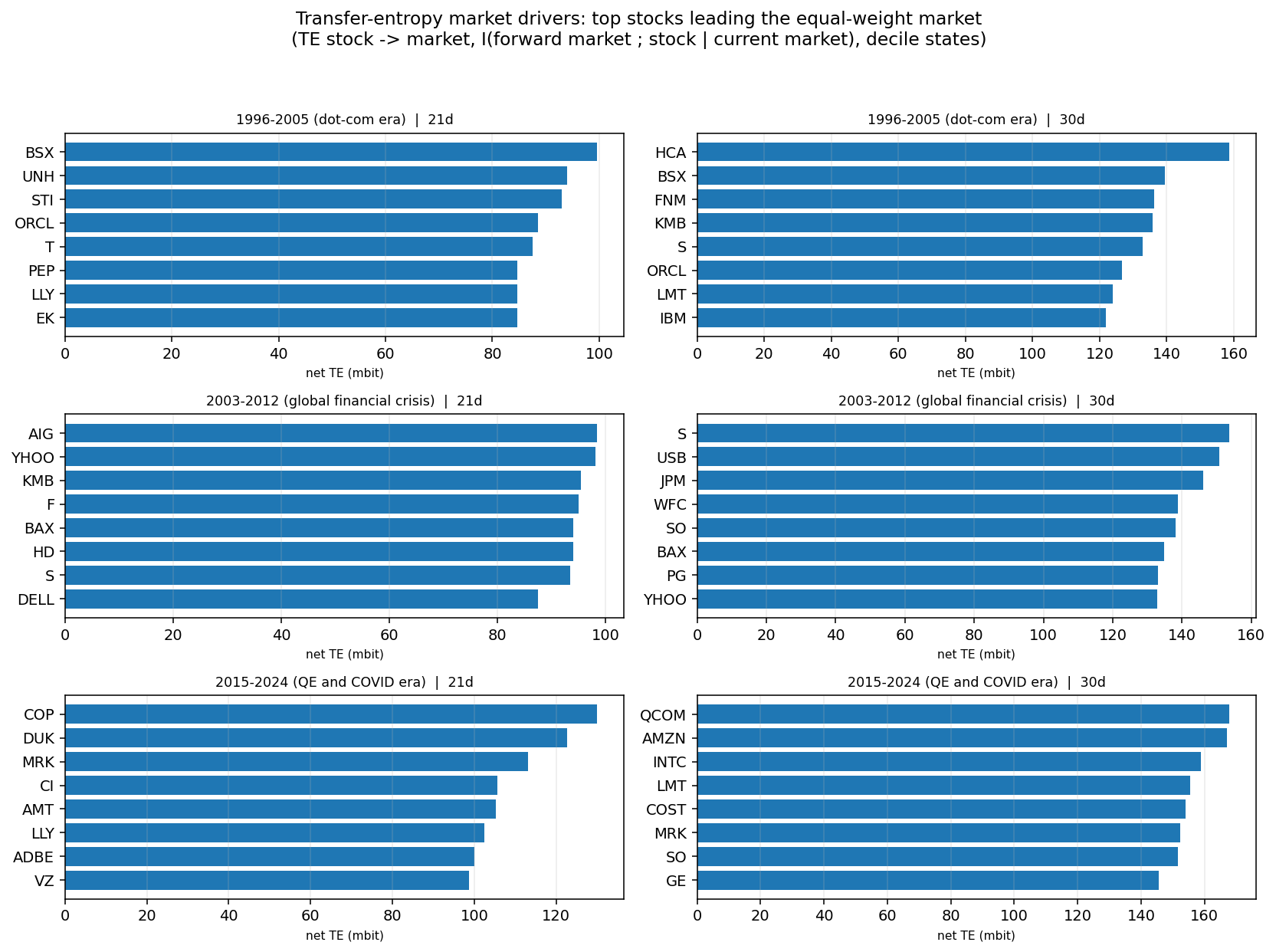}
\caption{Top transfer-entropy market leaders by period and horizon, ranked by
net transfer entropy to the equal-weight market. The financials lead the crisis
window and the mega-cap technology names lead the recent window.}
\label{fig:teleaders}
\end{figure}

The leaders are economically legible (Table~\ref{tab:teleaders} and
Figure~\ref{fig:teleaders}). The thirty-day leaders of the crisis window are
dominated by financials, with Sprint, U.S.\ Bancorp, JPMorgan, and Wells Fargo
at the top, and the recent window is led by the mega-cap technology names,
Qualcomm, Amazon, and Intel. All the top scores are many surrogate standard
deviations above the null.

\begin{table}[htbp]
\centering\small
\caption{Top five transfer-entropy market leaders,
$\TE(\text{stock}\!\to\!\text{market})$, by period and horizon.}
\label{tab:teleaders}
\begin{tabular}{lll}
\toprule
period & horizon & top five leaders (by net $\TE$) \\
\midrule
1996--2005 & 21d & BSX, UNH, STI, ORCL, T \\
           & 30d & HCA, BSX, FNM, KMB, S \\
2003--2012 & 21d & AIG, YHOO, KMB, F, BAX \\
           & 30d & S, USB, JPM, WFC, SO \\
2015--2024 & 21d & COP, DUK, MRK, CI, AMT \\
           & 30d & QCOM, AMZN, INTC, LMT, COST \\
\bottomrule
\end{tabular}
\end{table}

\begin{figure}[htbp]
\centering
\includegraphics[width=0.95\textwidth]{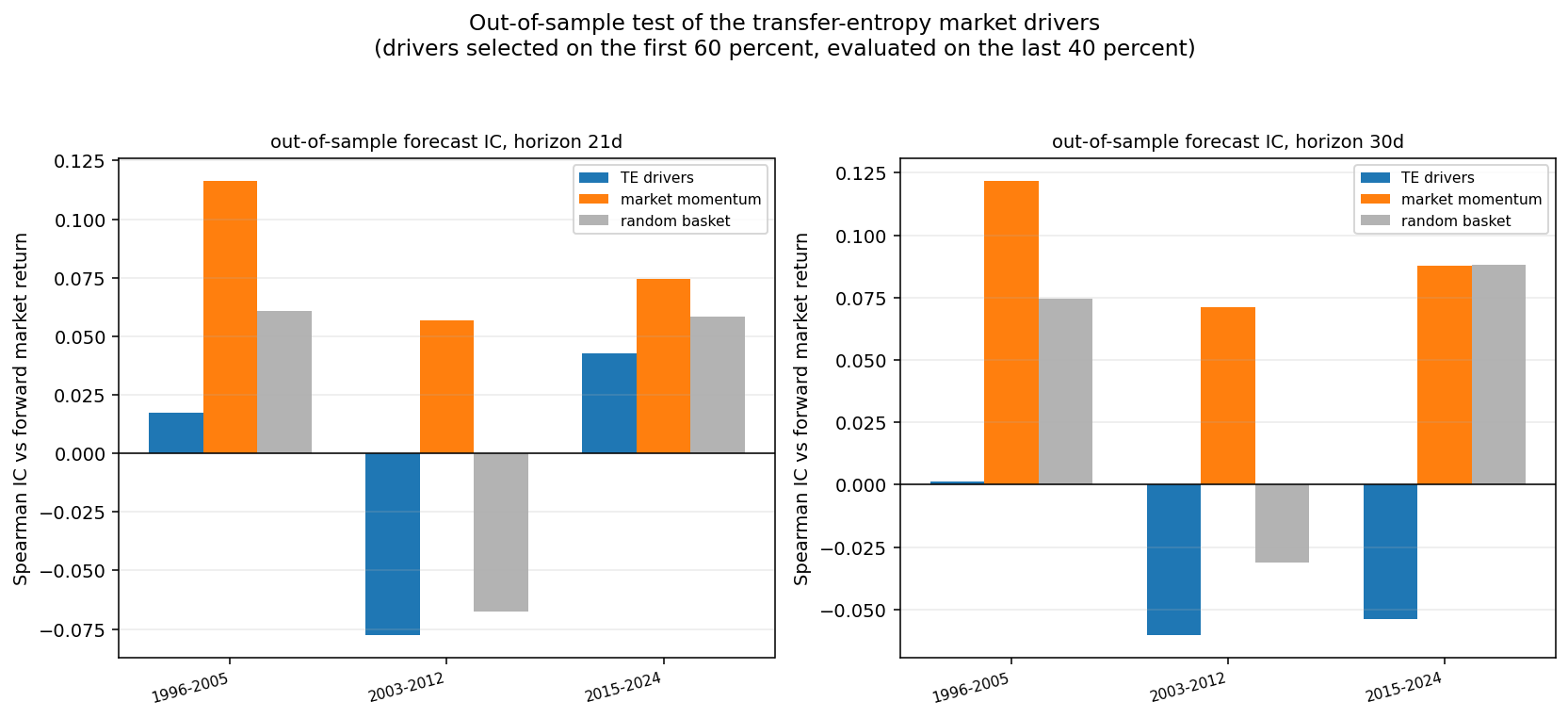}
\caption{Out-of-sample test of the transfer-entropy leaders, selected on the
first sixty percent of each period and read on the last forty. A timing signal
from the leaders (blue) does not beat the market's own momentum (orange) or the
random basket (grey) on the held-out period.}
\label{fig:teleadersoos}
\end{figure}

An out-of-sample test selects the leaders on the first sixty percent of each
period and evaluates them on the last forty (Figure~\ref{fig:teleadersoos}). The
selection is persistent. On the held-out period the train-selected leaders keep
a higher transfer entropy to the market than randomly chosen baskets in five of
the six period-and-horizon cases, so the leadership is a stable property rather
than a fit to the training noise. A forecast built from the leaders is not.
Learning each leader's average forward market return in each of its decile
states on the training period, and summing these responses into one
market-timing signal, we find that out of sample the signal does not beat the
market's own trailing return, whose rank correlation with the forward market is
the higher of the two in every period. The leaders carry genuine and persistent
information about where the market goes next, and that information is nonlinear
and state-conditional rather than a simple directional alpha.

Whether to add these signals to the portfolio then has the same answer as the
spectral and fundamental ablations. As a naive linear market-timing overlay they
are not worth adding, since the one-bit trailing-return rule of
Section~\ref{sec:portfolio} already captures the tradeable part and the leaders'
linear read does not beat it out of sample. The persistence of the leadership
does point to two uses worth a separate study, reading a leader's decile state
as a conditioning covariate in the ranking chains, which the framework already
admits, and extracting the state-conditional information with a nonlinear rather
than a mean-response readout. We leave both to future work and keep the tradeable
book on market data with the reactive blend.

\begin{figure}[H]
\centering
\includegraphics[width=0.95\textwidth]{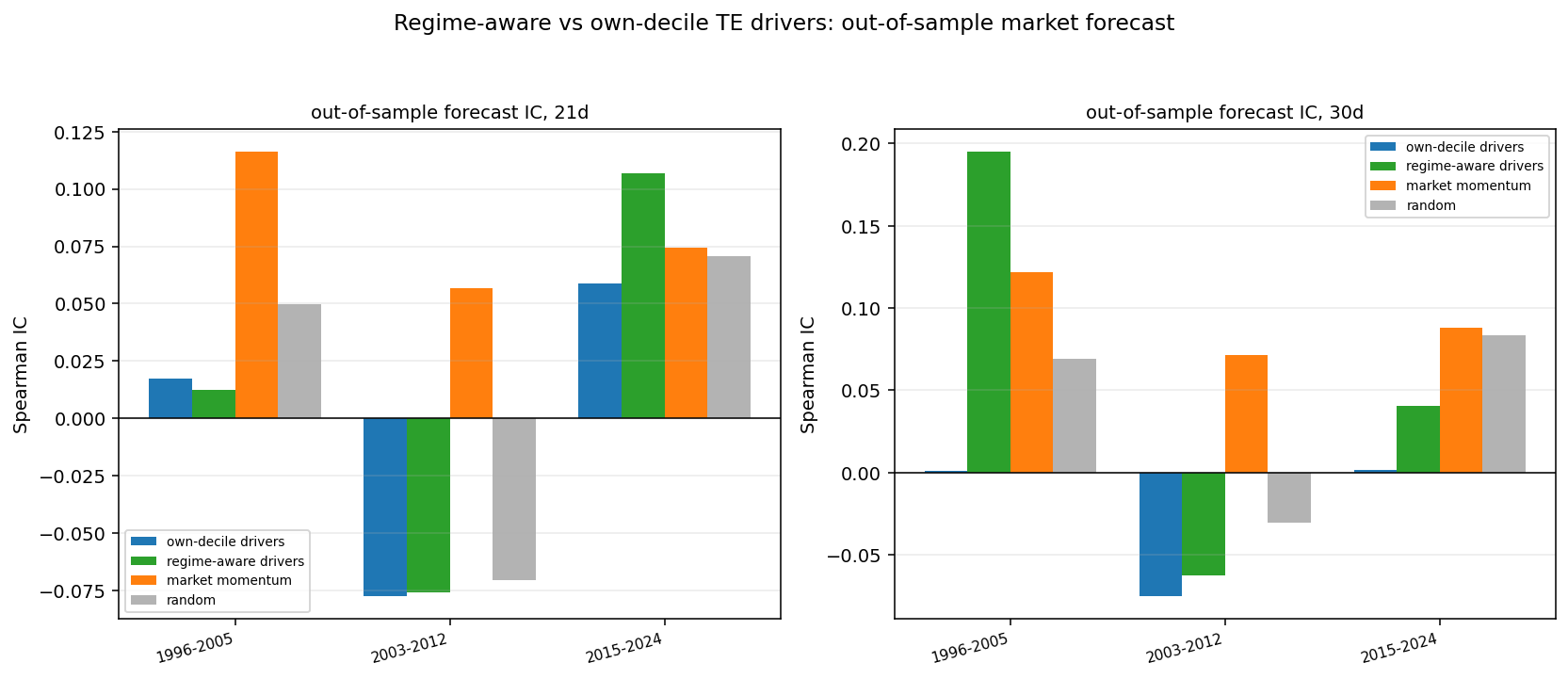}
\caption{Out-of-sample market-timing forecast of the transfer-entropy leaders,
regime-aware buckets against per-stock own deciles, at the twenty-one and thirty
day horizons. The regime-aware leaders (green) beat the market momentum baseline
(orange) on the dot-com and the recent windows, where the own-decile leaders
(blue) do not; the financial-crisis window is adverse for both.}
\label{fig:teregime}
\end{figure}

The regime-aware chains of Section~\ref{sec:regimeaware} change this reading in
part. Ranking the same drivers with the fixed regime-aware thresholds rather than
each name's own deciles leaves the leaders themselves almost unchanged, so the
leadership is a property of the market and not of the discretization. The
out-of-sample forecast they carry, however, improves. A market-timing signal built
from the regime-aware leaders beats the market's own momentum on the dot-com and
the recent windows, a rank correlation with the forward market of $0.20$ and
$0.11$ against the momentum baseline's $0.12$ and $0.08$, where the own-decile
signal never beat momentum (Figure~\ref{fig:teregime}). By making the leaders'
states comparable across names and by carrying the market regime, the regime-aware
discretization extracts a tradeable piece the market-neutral reading could not,
though the financial-crisis window stays adverse for both. The signal is not yet
strong or stable enough across periods to enter the tradeable book, and it is the
most promising of the covariate overlays we have found and the natural first
candidate for the conditioning use noted above.

\subsection{Alternative early-warning predictors}
\label{sec:altews}

\begin{figure}[htbp]
\centering
\includegraphics[width=0.82\textwidth]{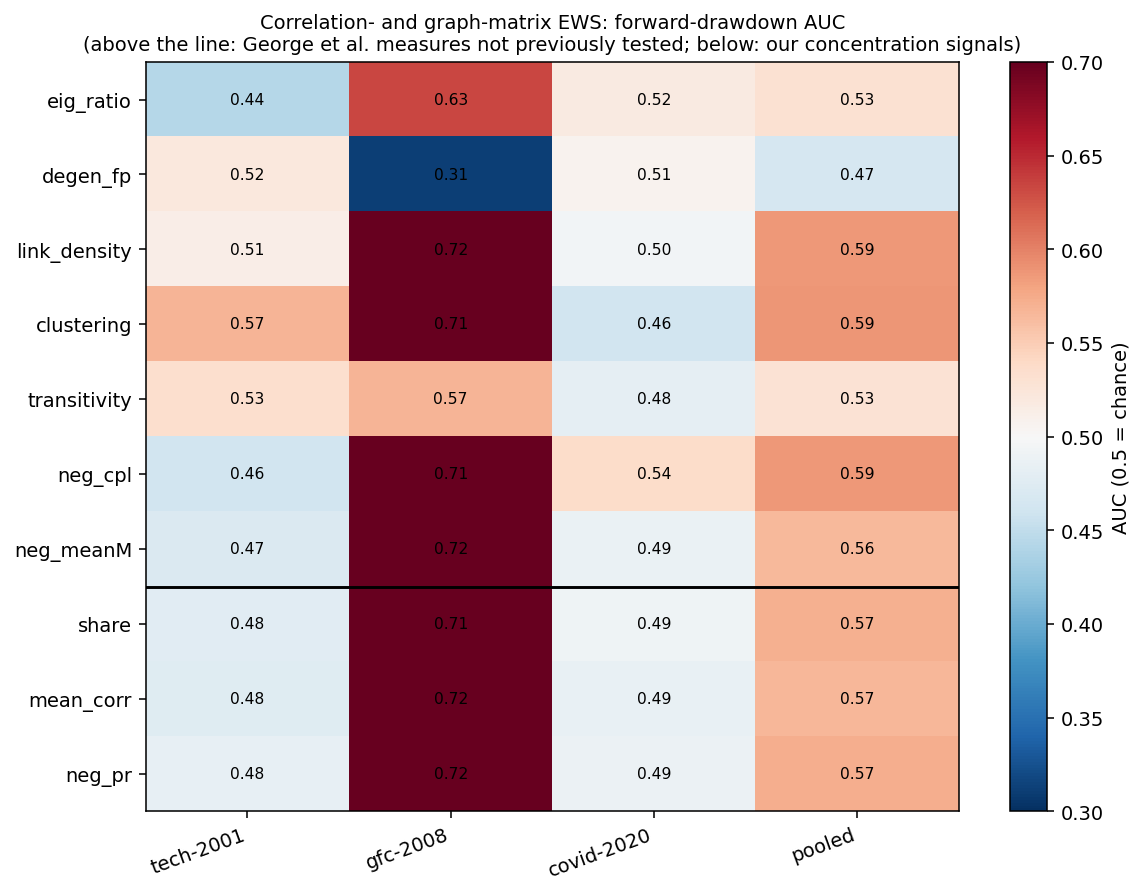}
\caption{Forward-drawdown area under the ROC curve for the correlation- and
graph-matrix early-warning signals of the transition literature (above the line)
and for the concentration signals already in use (below the line), per crisis
and pooled. The network measures reproduce the 2008 skill, the temporal
slowing-down measures do not.}
\label{fig:graphews}
\end{figure}

The spectral ablation of Appendix~\ref{sec:spectralablation} asked whether the
eigenvalue precursors of the OMD studies time the blend. We ran the same test on
the broader set of correlation- and network-matrix early-warning signals
collected in a recent review of critical transitions \cite{george2023}, and on
the information dissipation length introduced for the 2008 collapse
\cite{quax2013}, to see whether any of them beats the reactive rule. Scored by
the causal forward-drawdown area under the ROC curve on the three crises, the
network-connectivity measures, the link density, the clustering coefficient, the
characteristic path length, and the mean geodesic distance of the arccos matrix,
reproduce the read of the spectral signals exactly. They forecast the endogenous
2008 drawdown at an area near $0.72$ and sit at chance for the exogenous 2020
shock, and they add no skill beyond the eigenvalue share and the mean
correlation already tested, because at the level of the correlation matrix
connectivity and concentration are one phenomenon (Figure~\ref{fig:graphews}).
The classical multivariate slowing-down measures, degenerate fingerprinting and
the eigenvalue-gap ratio, fail outright on equities, the former scoring below
chance before 2008.

\begin{figure}[htbp]
\centering
\includegraphics[width=0.88\textwidth]{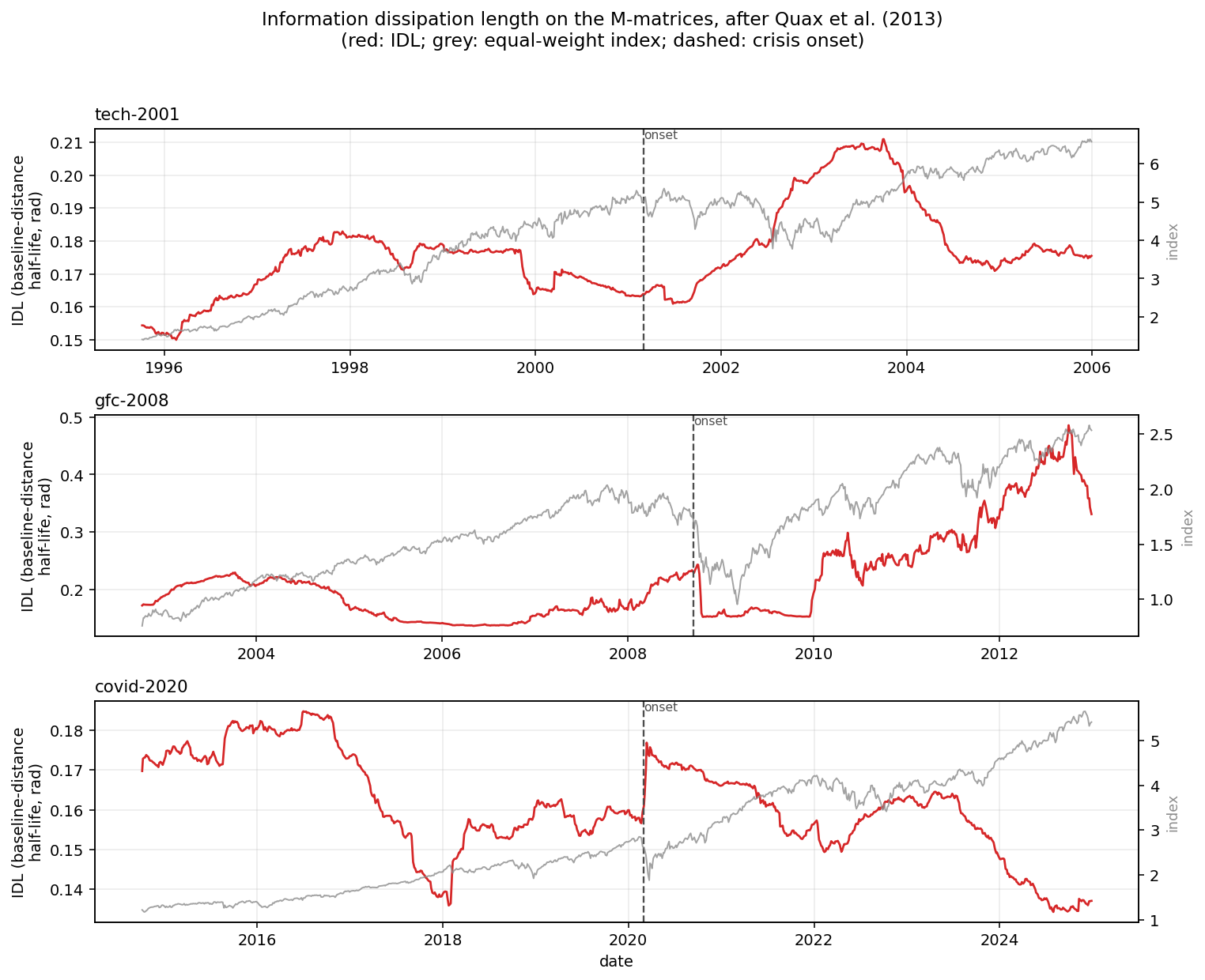}
\caption{The information dissipation length on the arccos distance matrices
through the three crises. It climbs into and peaks at the 2008 bankruptcy, does
not build up before the 2020 shock, and peaks at the 2000 top before the
dispersed 2001 unwind. Grey is the equal-weight index, the dashed line the
crisis onset.}
\label{fig:idlport}
\end{figure}

The information dissipation length measures how far information about one name
reaches across the others before it dissipates, fitting the decay of the mutual
information with a logical distance and reporting its half-life \cite{quax2013}.
The arccos distance matrix supplies the logical distance that the original
construction took from bond maturities. Computed this way the length climbs into
and peaks at the 2008 bankruptcy and then falls, reproducing the original
finding, and it does not build up before the 2020 shock
(Figure~\ref{fig:idlport}). Its forward-drawdown skill is modest and below that
of the concentration signals. The reading is again that of
Appendix~\ref{sec:spectralablation}. These measures are a strong contemporaneous
description of the cross section and are useful for understanding a crisis, and
they are not timing signals that beat the sign of the recent market return out of
sample.

\subsection{Regime-aware signals for the book}
\label{sec:regimebook}

The regime-aware chains of Section~\ref{sec:regimeaware} read the market state
well, as Section~\ref{sec:regimediag} shows, so the natural question is whether
they improve the tradeable book. We test both levers, timing and selection, with
everything chosen on the validation period through 2021 and read out of sample
over 2022--2024, reusing the sleeves and the market so that only the regime rule
changes. The daily regime-aware signals drive the monthly rebalance.

On the timing side we steer the blend weight $\theta$ with the regime-aware
signals, the cross-sectional mean bucket as a risk-on/off state and the
entropy production of the chains as the arrow of time. The defensive use fails.
The rule chosen on the validation period, the arrow of time at a daily Sharpe of
$0.57$ against the baseline's $0.49$, gives an out-of-sample Sharpe of $0.72$
against the baseline's $1.06$, for the reason the spectral ablation of
Appendix~\ref{sec:spectralablation} already gave. The arrow of time is coincident
with the drawdown rather than ahead of it, so a defensive tilt taken on it is
taken at the bottom. The contrarian re-risk use, which helped the spectral signal
inside the 2022--2024 window, gives a grid of configurations centered on the
baseline out of sample, a median Sharpe of $1.05$ against $1.06$ with half the
configurations above it, so there is no structural edge.

On the selection side we use the absolute-momentum regime bucket to choose the
book, a long-only sleeve that holds only the names beating the market and so
de-risks itself when few do. It has the highest validation Sharpe of any variant
we tried, $0.59$ against the baseline's $0.49$, and yet an out-of-sample number
below the baseline, $0.84$, a clean case of validation overfitting. In the 2022 bear it held
as few as four names, so the de-risking is real, but concentrating the book into
them is not rewarded.

The reading is the same as the fundamental and the spectral ablations. The
regime-aware chains earn their place as diagnostics of the market state, the
arrow of time, and the endogenous build-up, and not as a new source of tradeable
alpha.

\subsection{Information in trading volume}
\label{sec:volume}

The daily panel carries trading volume alongside price, and volume reads with the
same chains as any other characteristic. We rank-bucket the cross section into
ten deciles each day, so dollar volume, its daily log-change, share turnover, the
dollar volume over market capitalization, and the turnover change each become a
ten-state process on the footing of the return and volatility ranks. Three
questions follow, in the tools of Sections~\ref{sec:condep} and~\ref{sec:te}:
whether the volume process has an arrow of time of its own, whether it carries
directional information about the return and volatility ranks, and whether that
information is real and tradeable out of sample.

The volume process is strongly irreversible. The entropy production of its own
decile chain, in Table~\ref{tab:volume}, is far above that of the return or
volatility rank, two orders of magnitude for the daily change, and many standard
deviations above a reversible surrogate. The direction is the asymmetry of a
spike, a jump in volume followed by a decay rather than the reverse, so the
change chain carries a large circulating flow. This is a property of volume
itself, not yet of its link to price.

That link is directional, and it is turnover, not raw volume, that carries it.
The surrogate-corrected transfer entropy from turnover to the next return rank
exceeds the reverse, so turnover leads the return rank. Raw dollar volume does
the opposite, the return rank leading it, because dollar volume is dominated by
size, a slow and persistent cross-section. Only after dividing by capitalization
does the return-leading information appear. The daily changes of volume and
turnover do not lead the return rank in either form. Conditioning the return
chain on turnover resolves a large arrow of time it had hidden, a gap
$\Delta\sigma$ two orders of magnitude above the pooled value, so the
near-reversible return rank becomes strongly irreversible once turnover is known.

The predictive content survives out of sample, where most signals in this paper
do not. Fitting the one-step return-rank transition with and without the
covariate on the training period through 2021, and reading the held-out
2022--2024, turnover raises the mean predictive log-likelihood of the next rank,
a strictly proper score on which an overfit covariate would lose, and its
transfer entropy stays positive within the held-out period. The volume and
turnover changes, informative in sample, turn negative out of sample, the
signature of overfitting.

What does not survive is tradeability. The long-short the signal implies, long
the high-turnover decile and short the low, earns a held-out gross Sharpe of only
$0.25$ at a daily holding horizon, and the cost of re-forming it every day, an
annualized turnover near four hundred, drives the net Sharpe to $-0.91$. Held to
the next month-end instead, the fast signal has decayed, a gross Sharpe of $0.09$
and a net of zero. Turnover carries genuine information about the next day's
return ranking, real out of sample, but at a horizon too fast to monetize once
trading is paid for, the same wall the daily-rebalancing test of
Section~\ref{sec:portfolio} met. It is the cleanest case in the paper of
information that is real, and real out of sample, without being a tradeable edge.

\begin{table}[htbp]
\centering
\small
\caption{Information in trading volume, CRSP daily panel 2015--2024, the cross
section rank-bucketed into ten deciles each day. Own-chain entropy production
$\sigma$ (the arrow of time of the covariate's own decile chain), the
surrogate-corrected net transfer entropy to the next return rank
$\TE(X\!\to\!A)-\TE(A\!\to\!X)$ (positive when the covariate leads), and the
out-of-sample gain in mean predictive log-likelihood of the next return rank from
adding the covariate, fit on the training period through 2021 and read on the
held-out 2022--2024. All in units of $10^{-3}$ nats. The return and volatility
ranks are shown for scale. Turnover, the dollar volume over market
capitalization, is the only covariate that both leads the return rank and keeps a
positive out-of-sample gain, and its long-short is nonetheless not tradeable net
of cost.}
\label{tab:volume}
\begin{tabular}{lccc}
\toprule
covariate & own-chain $\sigma$ & net $\TE\to$ return & OOS log-lik.\ gain \\
\midrule
return rank (reference)     & $0.15$ & ---              & --- \\
volatility rank (reference) & $0.20$ & ---              & --- \\
dollar volume, level        & $5.2$  & $-7.7$           & $+0.8$ \\
dollar volume, change       & $37.7$ & $-5.7$           & $-0.3$ \\
turnover, level             & $5.8$  & $\mathbf{+15.8}$ & $\mathbf{+14.2}$ \\
turnover, change            & $37.7$ & $-5.8$           & $-0.2$ \\
\bottomrule
\end{tabular}
\end{table}

\end{document}